\documentclass[preprint,12pt]{emulateapj-rtx4}
\usepackage[varg]{txfonts}
\usepackage{graphicx}
\usepackage{booktabs}
\usepackage{afterpage}
\usepackage{changepage}
\usepackage{chngcntr}
\bibpunct{(}{)}{;}{a}{}{,} 
\usepackage{color}
\usepackage[caption=false]{subfig}
\usepackage[colorlinks=true,citecolor=blue,urlcolor=green]{hyperref}

\usepackage{longtable}

\newlength{\offsetpage}
\newenvironment{widepage}{\begin{adjustwidth}{-\offsetpage}{-\offsetpage}%
    \addtolength{\textwidth}{2\offsetpage}}%
{\end{adjustwidth}}

\begin{document}

\title{Eccentricity from transit photometry: \\
small planets in Kepler multi-planet systems have low eccentricities}

\hyphenation{Kepler}

\author{Vincent~Van~Eylen$^{1,2}$ and Simon~Albrecht$^1$}

\affil{$^1$Stellar Astrophysics Centre, Department of Physics and Astronomy, Aarhus University, Ny Munkegade 120, \\
DK-8000 Aarhus C, Denmark \\
$^2$MIT Kavli Institute for Astrophysics and Space Research, 70 Vassar St., Cambridge, MA 02139}

\email{vincent@phys.au.dk}

\shorttitle{Small planets in Kepler multi-planet systems have low eccentricities}
\shortauthors{Van Eylen and Albrecht}

\received{receipt date}
\revised{revision date}

\begin{abstract}
{Solar system planets move on almost circular orbits. In strong contrast, many massive gas giant exoplanets travel on highly elliptical orbits, whereas the shape of the orbits of smaller, more terrestrial, exoplanets remained largely elusive. Knowing the eccentricity distribution in systems of small planets would be important as it holds information about the planet's formation and evolution, and influences its habitability. We make these measurements using photometry from the \textit{Kepler} satellite and utilizing a method relying on Kepler's second law, which relates the duration of a planetary transit to its orbital eccentricity, if the stellar density is known. Our sample consists of 28 bright stars with precise asteroseismic density measurements. These stars host 74 planets with an average radius of $2.6~R_\oplus$. We find that the eccentricity of planets in \textit{Kepler} multi-planet systems is low and can be described by a Rayleigh distribution with $\sigma = 0.049 \pm 0.013$. This is in full 
agreement with solar system eccentricities, but in contrast to the eccentricity distributions previously derived for exoplanets from radial velocity studies. Our findings are helpful in identifying which planets are habitable because the location of the habitable zone depends on eccentricity, and to determine occurrence rates inferred for these planets because planets on circular orbits are less likely to transit. For measuring eccentricity it is crucial to detect and remove Transit Timing Variations (TTVs), and we present some previously unreported TTVs. Finally transit durations help distinguish between false positives and true planets and we use our measurements to confirm six new exoplanets.}
\end{abstract}


\maketitle
 
\section{Introduction}

In the solar system, the orbit of Mercury has the highest ellipticity with an eccentricity ($e$) of 0.21, where an eccentricity of 0 indicates a circular orbit, whereas the mean orbital eccentricity of the other seven planets is 0.04. In contrast, Radial Velocity (RV) measurements revealed a wide range of eccentricities for gas giant planets \citep{butler2006}, where HD~80606b is the current record holder with an eccentricity of 0.927 \citep{naef2001}. RV surveys also found evidence that orbital eccentricities for sub-Jovian planets reach up to 0.45 \citep{wright2009,mayor2011}. For Earth-sized planets and Super-Earths, RV detections of eccentricities are typically not feasible, even with modern instruments, because of the small orbital RV signal amplitude $K$ \citep{marcy2014}, and the fact that the amplitude of the eccentricity scales with $e \times K$ \citep[see e.g.][]{lucy2005}. One alternative way to measure orbital eccentricities relies on the timing of secondary transits (eclipses), but this method 
is limited to the hottest and closest-in exoplanets. In some systems with multiple transiting planets Transit Timing Variations (TTVs) can be used to infer planetary mass ratios and orbital eccentricities. While these two parameters are often correlated, sometimes eccentricity information can nevertheless be inferred using statistical arguments \citep[e.g.][]{lithwick2012,wu2013}, or from the ``chopping'' effect \citep[e.g.][]{deck2015}. Low-eccentricity as well as some higher eccentricity systems have been found \citep{hadden2014}. Unfortunately, TTVs are only detected in a subset of all transiting multiple systems, and the interpretations of the results is complex as systems with TTVs are typically found near resonances, and it's unclear if such systems have undergone the same evolution as systems without such resonances. 

\begin{figure*}[ht]
\centering
\resizebox{\hsize}{!}{\includegraphics{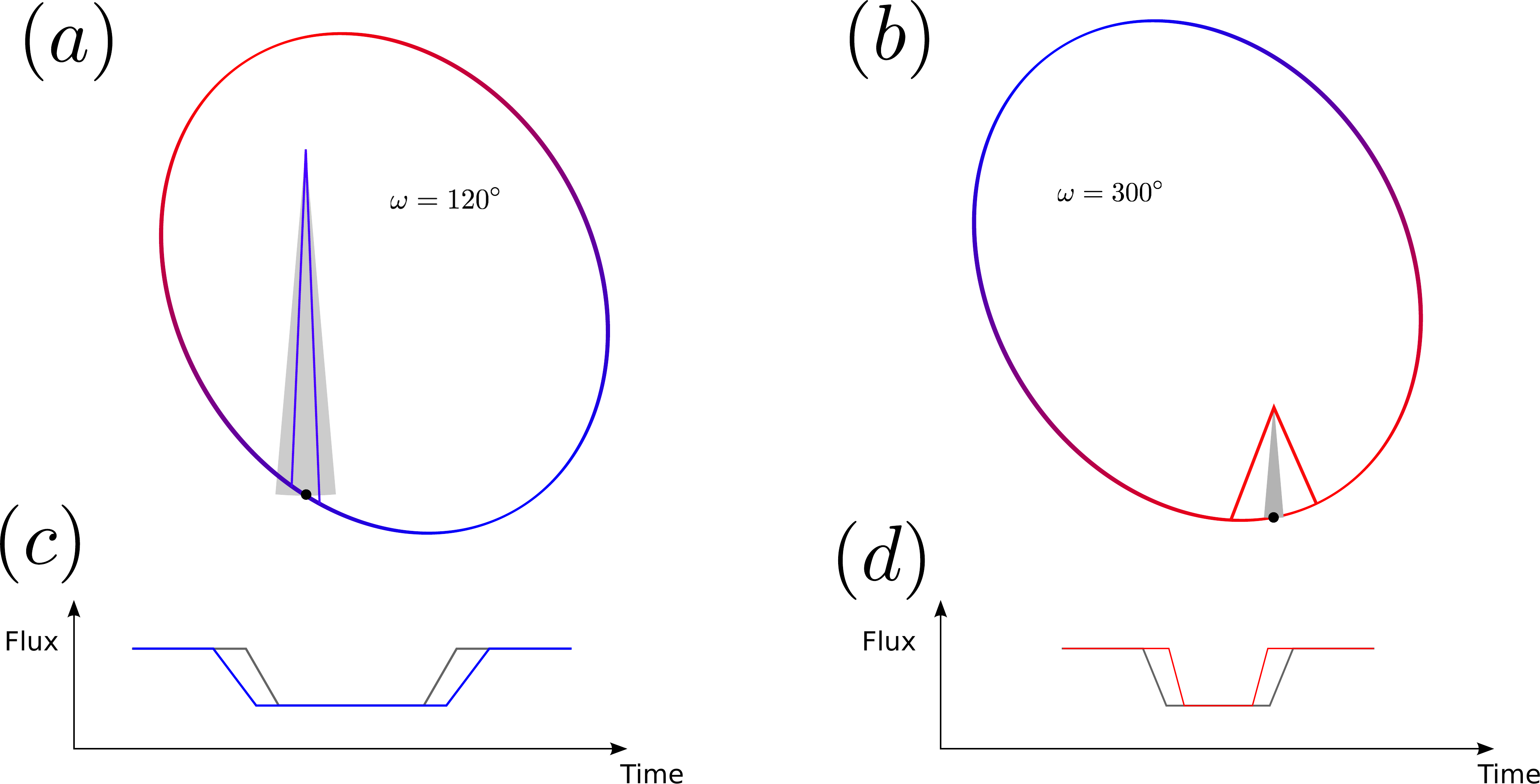}}
\caption{The left top panel (a) pictures an orbit with $e$ = 0.6 and an argument of periastron of $120^\circ$. The observer is located below the figure. Panel (b) shows the same orbit, now with  $\omega = 300^\circ$. The pies outlined with blue and red lines in the two panels encompass the same surface areas and the corresponding arcs are traveled by the planet during $1/36$ of their orbital period. These times are centered around the time of planetary transit. The gray filled pies correspond to the surface areas covered by a planet traveling on a circular orbit with the same apparent $a/R_\star$ ratio. According to Kepler's second law the transit in the eccentric system in panel (a) lasts longer than in the system with the circular orbit. The reverse is true for panel (b). Corresponding schematic light curves are shown in panels (c) and (d).\label{fig:eccentricity_illustration}}
\end{figure*}

Here we determine orbital eccentricities of planets making use of Kepler's second law, which states that eccentric planets vary their velocity throughout their orbit. This results in a different duration for their transits relative to the circular case: transits can last longer or shorter depending on the orientation of the orbit in its own plane, the argument of periastron~($\omega$). This is illustrated in Figure~\ref{fig:eccentricity_illustration}. Transit durations for circular orbits are governed by the mean stellar density \citep{seager2003}. Therefore if the stellar density is known from an independent source then a comparison between these two values constrains the orbital eccentricity of a transiting planet independently of its mass \citep{ford2008,tingley2011}.

Using this technique, individual measurements of eccentric orbits were made successfully, making use of high-quality \textit{Kepler} transit observations. For highly eccentric Jupiters, the technique is powerful even when only loose constraints on the `true' stellar density are available, as shown for Kepler-419 \citep{dawson2012} and later confirmed by radial velocity observations \citep{dawson2014}. \cite{kipping2012} suggested that multiple planets in the same system can be compared to constrain the sum of eccentricities in cases where the stellar density is not known. For close-in hot Jupiters where the orbits are assumed to be circular due to tidal forces, the technique provides stellar densities which rival the accuracy provided by other methods such as asteroseismology, and good agreement is typically found \citep[e.g.\ HAT-P-7b,][]{vaneylen2013}. For Kepler-410b, a Super-Earth, a small but significantly non-zero eccentricity (0.17$^{+0.07}_{-0.06}$) was measured, thanks to an accurately 
determined stellar density from asteroseismology and the brightness of the star \citep[\textit{Kepler} magnitude 9.4,][]{vaneylen2014}. The orbits of both Kepler-10b ($1.4~R_\oplus$) and Kepler-10c ($2.4~R_\oplus$) were found to be consistent with circularity \citep{fogtmannschulz2014}.

An ensemble study, based on early \textit{Kepler} catalog data and averaging over impact parameters, found the eccentricity distribution of large planet candidates ($\geq 8~R_\oplus$) to be consistent with the RV eccentricity distribution, with some evidence that sub-Neptune planets had lower average eccentricities \citep{kane2012}. However, subsequent ensemble studies have revealed a range of complications, such as a correlation with the transit impact parameter \citep{huber2013}, the influence of planetary false positives \citep{sliski2014} and uncertainties or biases in stellar parameters \citep{plavchan2014,rowe2014}. \cite{price2015} recently investigated the feasibility of such studies for the smallest planets.\footnote{We note that the authors made use of \textit{Kepler} 30-minute integration time data in their study, while the data used in this work has a one-minute (short cadence) sampling, which complicates a direct comparison (see also Section~\ref{sec:results_b}).} \cite{kipping2014} identified a 
number of other mechanisms that influence transit durations, e.g.\ TTVs. We approach these complications in two ways.

Firstly, we design a data analysis pipeline that allows us to identify and remove TTVs, measure transit parameters and their correlations, and insert and recover artificial transits to test our methods. Secondly, we focus on a sample of 28 bright stars observed by \textit{Kepler} \citep{borucki2010}: the brightest host star has a \textit{Kepler} magnitude 8.7 and all but one are brighter than magnitude 13. They have all been observed in short-cadence mode with a one-minute integration time. Their mean stellar density is constrained through asteroseismology. The 17 brightest of these stars were analyzed in \cite{silvaaguirre2015} and the average accuracy of their mean density measurements is $1.7\%$. The other 11 stars were previously modeled by \cite{huber2013} and the average uncertainty on the mean stellar density of these objects is $6.7\%$. All 28 stars also have separate mass and radius measurements, while the detailed modeling of individual frequencies by \cite{silvaaguirre2015} also provides stellar 
ages with a median uncertainty of $14\%$. They all contain multiple planets (74 in total) and all but three contain confirmed planets. The planets are small with an average radius of $2.8~R_\oplus$ and have orbital periods ranging from 0.8 to 180 days.

In Section~\ref{sec:methods} we describe our analysis methods. We present the pipeline developed to model the planetary transits and discuss several important parameter correlations. Our main results are presented in Section~\ref{sec:results}. We present the eccentricity distribution of our sample of planets, as well as homogeneous planetary parameters and several previously unreported transit timing variations. We also validate several previously unconfirmed exoplanets. In Section~\ref{sec:discussion} we discuss the implications of our findings in the context of planetary habitability and planetary occurrence rates. Our conclusions are presented in Section~\ref{sec:conclusions} In Appendix \ref{sec:individual_posterior_discussion} we present the eccentricities of individual exoplanet systems.

\section{Methods}
\label{sec:methods}

We built a customary data reduction and analysis pipeline to measure all transit parameters and their correlations. This also allows us to do transit insertion and recovery tests. In Section~\ref{sec:pipeline} we describe the pipeline and how we extract the relevant parameters. In Section~\ref{sec:results_correlations} we discuss parameter correlations. In Section~\ref{sec:transit_insertion}, we present the results of modeling artificial transits that we inserted in the data.

\subsection{Pipeline}
\label{sec:pipeline}

The pipeline performed the following main steps:

\begin{enumerate}
 \item \textit{Kepler} data reduction and normalisation
 \item Period determination and Transit Timing Variation (TTV) assessment; data folding
 \item MCMC transit fit module
\end{enumerate}

We now describe each step in more detail.

\subsubsection{Data reduction}

The first part of our pipeline is responsible for reducing and normalising \textit{Kepler} light curves. For a given \textit{Kepler} object of interest (KOI), the pipeline searches for observations in any quarter (Q), between Q0-Q17. Only the quarters which contain short cadence observations are downloaded (in fits-file format), because the one minute sampling is required to resolve the planetary ingress and egress (see Section~\ref{sec:results_b}). Our analysis starts with the Presearch Data Conditioning (PDC) version of the data \citep{smith2012}.

In the following, we only focus on data directly before, during or after the transits (typically encompassing about 5-10 hours before and 5-10 hours after a given transit). An initial estimate of the transit times is calculated with the ephemeris available at the \textit{Kepler} database\footnote{http://exoplanetarchive.ipac.caltech.edu/}. From the same source a value for the transit duration is obtained and used to determine the in-transit data points. By default the transit duration is increased by three hours to make sure no in-transit data points are erroneously used for the data normalisation. In case of (previously known or subsequently detected) TTVs, the transit duration is further increased to catch all in-transit data points. The data before and after the transits are then fitted by a second order polynomial which is used to normalise the data.

In a final step, all transits are visually inspected. In some cases, (instrumental or astrophysical) data jumps or gaps can cause the transit fits to fail or the true transit to be poorly determined. These transits are manually removed. Similarly, when multiple transits happen simultaneously, these data points are removed to avoid biasing the transit measurement. 

\subsubsection{Period and TTV determination}
\label{sec:period_ttv_finder}

This part of the pipeline measures times of individual transits and uses them to find the orbital period, as well as detect any TTVs. The measurement of an individual transit time is done by fitting the best transit model to the individual transits, keeping all transit parameters fixed except for the transit mid-time. During the first iteration, the model is based on the parameters extracted from the \textit{Kepler} database, afterwards the best model from the MCMC analysis in Step 3 (transit fit module) is used, a procedure which is repeated until convergence is reached. The uncertainty of each transit-mid time is calculated by first subtracting the best fitting transit model from the original light curve, bootstrapping the residuals with replacement, injecting the best fitting transit model and fitting this new light curve. The steps after and including the permutation of the residuals are repeated 200 times for each transit, to calculate the mid-time uncertainty from the spread in these fits. 

Now the planetary period is obtained by (weighted) fitting for a linear ephemeris to the individual transit times. From this we calculate the observed minus calculated (O-C) transit times. Next we refit, this time ignoring $3-\sigma$ outliers (as determined by the standard deviation around the linear ephemeris), and repeat until convergence is reached (no more outliers are removed).

Once the linear ephemeris has been determined we perform a search for TTVs as these might cause biases in the eccentricity calculations, as explained below. For this a sinusoidal model is fitted to the O-C diagram. A list of the systems where TTVs were included is given in Table~\ref{tab:ttv_table}. The transits are subsequently folded based on their period and TTVs if present. The folded transit curve is binned to contain a maximum of 6000 data points, which even for the longest transits implies more than 10 data points per minute, which is an oversampling compared to the original one minute \textit{Kepler} sampling.

\subsubsection{Transit fit module}
\label{sec:transit_fit_module}

This part of the pipeline consists of a transit fitting module, which makes use of a Markov Chain Monte Carlo (MCMC) algorithm. We choose to employ an Affine-Invariant Ensemble Sampler \citep{goodman2010} as implemented in the Python module $emcee$ \citep{foremanmackey2013}. Planetary transits are modeled analytically \citep{mandel2002}\footnote{We gratefully acknowledge the implementation of planetary transit equations into Python by Ian J. M. Crossfield, upon which our code was based; see http://www.lpl.arizona.edu/~ianc/python/transit.html.}.

For each planet in the system, we sample five parameters: the impact parameter $b$, relative planetary radius $R_\textrm{p}/R_\star$, $\sqrt{e}\cos \omega$, $\sqrt{e} \sin \omega$, mid-transit time $T_0$ and flux offset $F$. In addition, two stellar limb darkening parameters are adjusted. These are common for all planets in one system, leading to $6n+2$ parameters per planetary system, where $n$ is the amount of planets in the system. The MCMC chains were run using 200 walkers, each producing a chain of 500\,000 steps, after a burn-in phase of 150\,000 steps was completed.

We sample uniformly in $R_\textrm{p}/R_\star$ and place a uniform prior on $T_0$ and $b$, where the latter is sampled between $-2$ and $2$ to allow grazing orbits and avoid border effects around $0$. We do not sample directly in $e$ and $\omega$, as this biases the eccentricity results for nearly circular orbits due to the boundary at zero \citep{lucy1971,eastman2013}. Instead we sample uniformly in $\sqrt{e}\cos \omega$ and $\sqrt{e} \sin \omega$ (both between $-1$ and $1$), which corresponds to a uniform sampling in $e \in [0,1]$ and $\omega \in [0,360]^\circ$ after conversion and rejection of values corresponding to $e>1$. The conversion between $e$ and $\omega$ and the stellar density ratio is given by \citep{kipping2010,moorhead2011,tingley2011,dawson2012}

\begin{equation}
\label{eq:eccentricity_periastron}
\frac{\rho_{\star}}{\rho_{\star,\textrm{transit}}} = \frac{(1-e^2)^{3/2}}{(1+e\sin \omega)^3},	
\end{equation}

and this can be further converted into the ratio of semi-major axis to stellar radius $R_\star /a$ using \citep{seager2003}

\begin{equation}\label{eq:stellar_density}
 \rho_{\star,\textrm{transit}} = \frac{3 \pi}{G P^2} \left( \frac{a}{R_\star}\right)^3.
\end{equation}

Here $G$ represents the gravitational constant. It is $R_\star /a$ which is used in the analytical transit model \citep{mandel2002}. For circular orbits, $R_\star /a$ directly constrains the stellar density ($\rho_{\star,\textrm{transit}} = \rho_{\star}$). In general, when $\rho_{\star}$ is known (e.g.\ from asteroseismology \citep{huber2013,silvaaguirre2015}), $R_\star /a$ constrains the combination of $e$ and $\omega$ given by the right-hand side of Equation~\ref{eq:eccentricity_periastron}. We note that it is possible to sample directly from the stellar density ratio (or from $R_\star /a$) \citep{dawson2012,vaneylen2014}, since the data always constrains a combination of $e$ and $\omega$ simultaneously, but doing so makes it more complicated to achieve an uninformative flat prior in $e$ and $\omega$.

Multiple planets around the same star are modeled simultaneously using the same limb darkening parameters. We use a quadratic limb darkening law with parameters $u_1$ and $u_2$ ($I(\mu)/I(1) = 1 - u_1(1-\mu) + u_2(1-\mu)^2$, where $I(1)$ represents the specific intensity at the centre of the disc and $\mu$ the cosine of the angle between the line of sight and the emergent intensity) and place a Gaussian prior with a standard deviation of 0.1 on each parameter, centered on predicted values interpolated for a Kurucz atmosphere \citep{claret2011}. This is a compromise to avoid fixing the parameters entirely, while still making use of the detailed stellar parameters available for the stars in our sample.

The final part of this module of the pipeline consists of the processing of the MCMC chains. Convergence is checked by visually inspecting traceplots, checking that an increase in burn-in time does not influence the posteriors, and confirming that MCMC chains initialized with different starting conditions give equivalent results. Transit fits for the final parameters are produced. All parameter distributions and their mutual correlations are plotted and visually inspected. A range of statistics, such as the mean, median, mode and confidence intervals are calculated for each parameter.

The results for our combined sample are presented in terms of the stellar density ratio in Section~\ref{sec:multis_loweccentricity}. The results for all individual systems and parameters are presented in Table~\ref{tab:parameter_table} and the eccentricity posterior distributions are shown in the Appendix.

\subsection{Parameter correlations}
\label{sec:results_correlations}

\begin{figure}[ht]
\centering
\resizebox{\hsize}{!}{\includegraphics{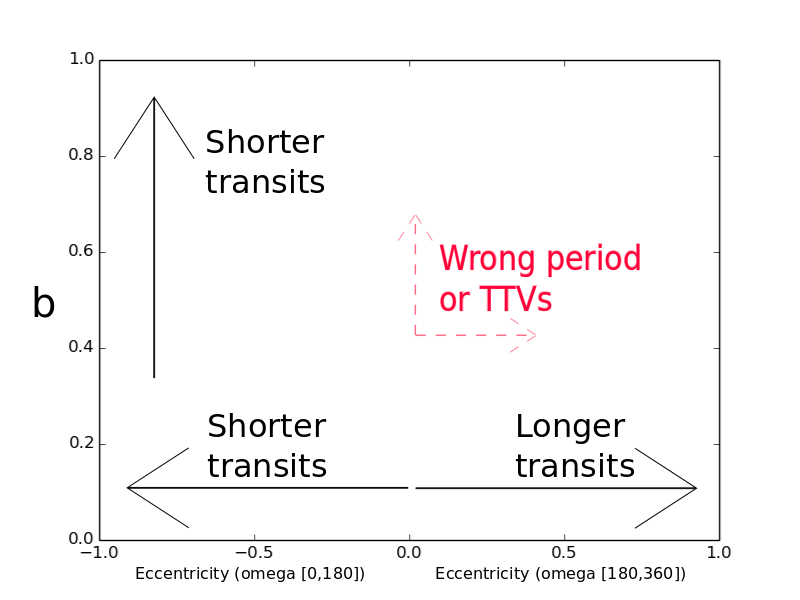}}
\caption{Illustration showing the influence of impact parameter $b$ and eccentricity $e$ on the transit duration. Misidentified periods or inadequately removed TTV signals cause a bias in $b$ and $e$.\label{fig:illustration_b_ecc}}
\end{figure}

There are several correlations between eccentricity and other parameters which are addressed here. The most important correlation occurs between eccentricity and angle of periastron $\omega$ and was already reported above (Equation~\ref{eq:eccentricity_periastron}). We explain how this complication can be overcome for a \textit{sample} of systems, by directly using the relative density instead, as well as its influence on eccentricity estimates for individual systems. Another important correlation occurs with impact parameter $b$. The influence of TTVs is also discussed. The effect of $\omega$, $b$ and TTVs on the eccentricity is summarized in Figure~\ref{fig:illustration_b_ecc}. We briefly discuss other commonly anticipated complications. 

\subsubsection{Correlation with angle of periastron}
\label{sec:results_omega}

\begin{figure}[ht]
\centering
\resizebox{\hsize}{!}{\includegraphics{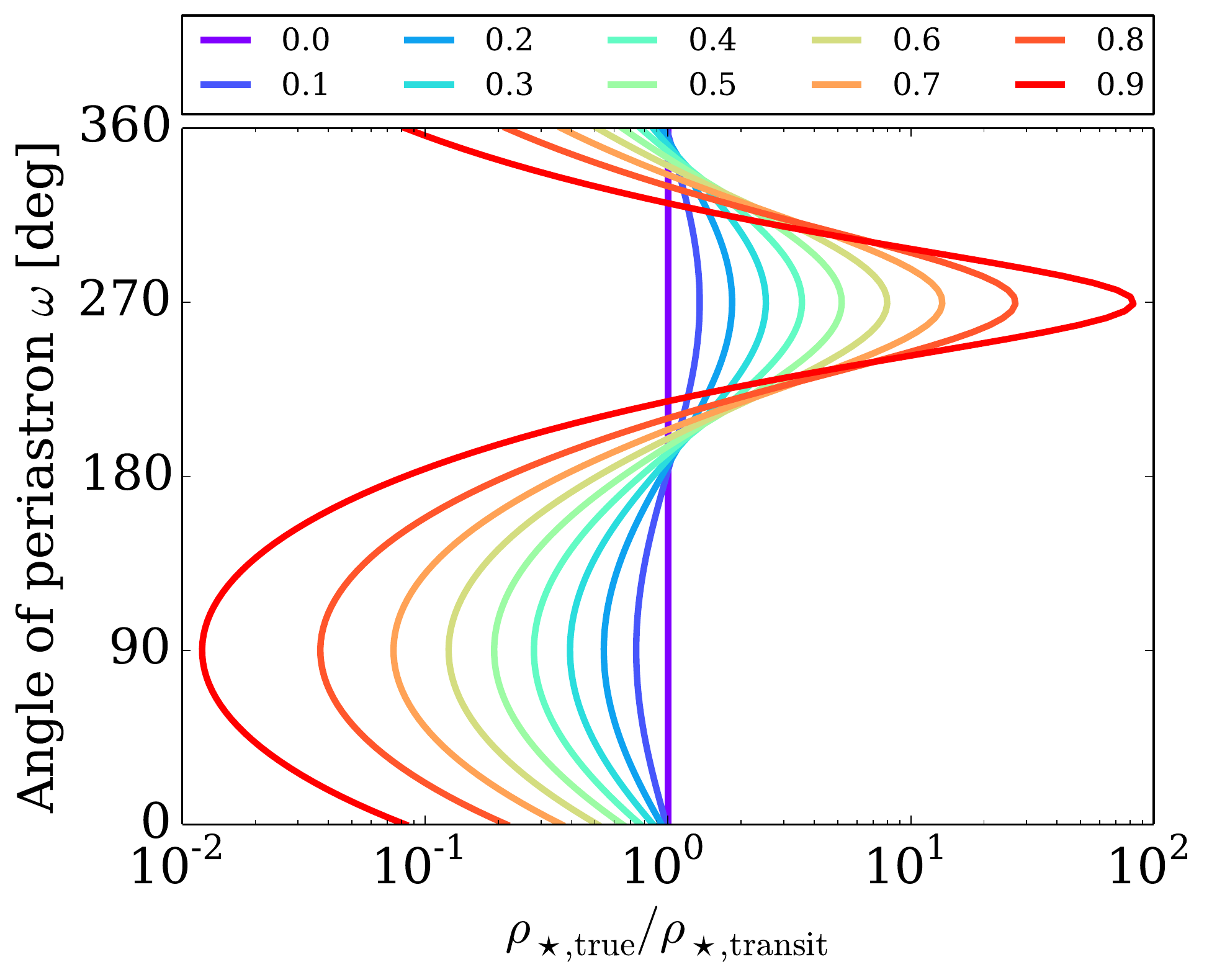}}
\caption{The influence of $e$ and $\omega$ on the relative density measured. The colored lines indicate different eccentricities ranging from 0 (inner) to 0.9 (outer). Different combinations of $e$ and $\omega$ can correspond to the same relative density.\label{fig:illustration_relativedensity}}
\end{figure}

When measuring transits, a combination of eccentricity and angle of periastron is constrained, as given by Equation~\ref{eq:eccentricity_periastron}. The combined influence of $e$ and $\omega$ is illustrated in Figure~\ref{fig:illustration_relativedensity}. For $\omega \in [0,180]^\circ$, eccentric orbits lead to shorter transits, while for $\omega \in [180,360]^\circ$, eccentricity increases the transit duration (see Figure~\ref{fig:illustration_b_ecc}). The left-hand side of the equation (the relative density $\rho_\textrm{circ.}/{\rho_\textrm{transit}}$) is the observable property, i.e.\ it is used to fit transits. Each relative density corresponds to a given eccentricity but also depends on the angle $\omega$, which is illustrated in Figure~\ref{fig:illustration_relativedensity}. 

When looking at an ensemble of systems, this complication can be avoided by reporting the measured relative densities, which is what we do in Section~\ref{sec:multis_loweccentricity}. This is the true observable (i.e.\ it influences the transit model), and it holds information on both $e$ and $\omega$ in a way that is defined by Equation~\ref{eq:eccentricity_periastron}. For an ensemble of systems, $\omega$ is expected to be randomly distributed\footnote{In general, the transit probability depends itself on $\omega$ for eccentric orbits, but given the low eccentricity orbits we find in our sample $\omega$ can be assumed to be randomly distributed.} so that the distribution of relative densities can be directly compared to any anticipated eccentricity distributions.

Note that for individual systems information on $e$ and $\omega$ can still be separately extracted, although the incomplete knowledge of $\omega$ increases the uncertainty of $e$. We discuss individual systems in Appendix~\ref{sec:individual_posterior_discussion} and report eccentricity modal values and highest probability density intervals which represent 68\% confidence in Table~\ref{tab:parameter_table}. We also show full posterior distributions of eccentricity (see Appendix). 

\subsubsection{Correlation with impact parameter}
\label{sec:results_b}

\begin{figure}[ht]
\centering
\resizebox{0.9\hsize}{!}{\includegraphics{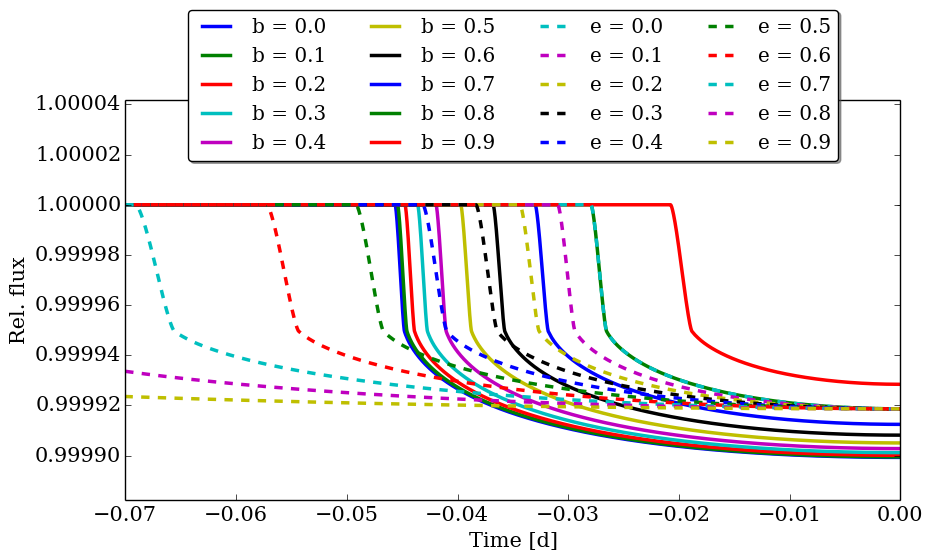}}
\resizebox{0.9\hsize}{!}{\includegraphics{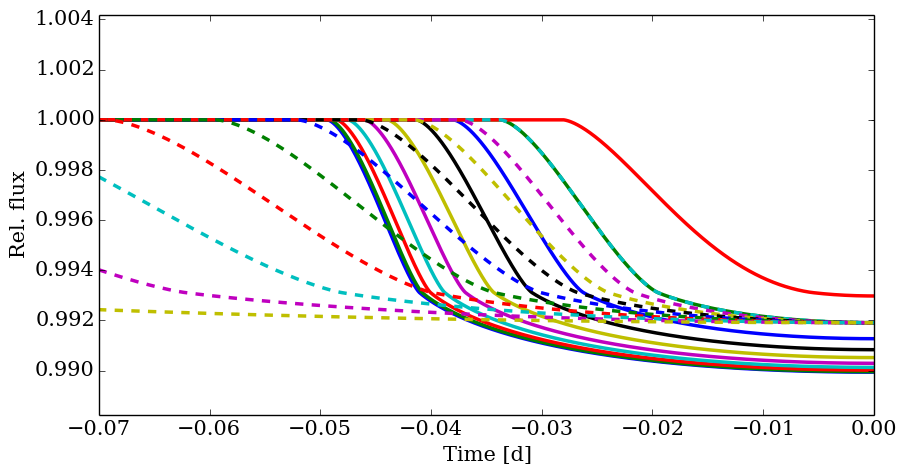}}
\caption{ \textbf{Top}: Earth-sized planet, \textbf{bottom:} Jupiter-sized planet. Solid lines show transits for different impact parameter $b$ and $e = 0$, dotted lines show transits for different $e$ and $b = 0.8$. All angle of periastron $\omega$ are taken to be $270^\circ$. In the Earth-size planet case, high $b$ and medium eccentricity look very similar to zero $b$ and zero $e$, while in the Jupiter-size case, there is much less degeneracy.
\label{fig:analytical_transits}}
\end{figure}

Eccentricity can be correlated with the transit impact parameter $b$. This can be understood by looking at Figure~\ref{fig:analytical_transits}, in which the effect of changing impact parameters and eccentricities is plotted for two analytically generated transit curves. While eccentric orbits change transit durations (increasing or decreasing it depending on the angle of periastron), increasing the impact parameters also shortens transits since a smaller part of the stellar disk is being crossed. Fortunately, changing the impact parameter also has the effect of \textit{deforming} the planetary transit. This is caused by the ingress and egress taking up more of the total transit time and leads to the typical V-shaped transits for high impact parameters. However, for smaller planets, ingress/egress times are intrinsically very short and the deformation of the transit shape is therefore far more limited, causing $b$ and $e$ to be more degenerate for smaller planets than for larger planets (see also \citealt{
ford2008}, and Figure~5 therein). This is why the availability of short cadence observations with a one minute sampling is crucial. Long cadence data, with an integration time of 30 minutes, smears out the ingress and egress of the planet. Therefore measuring eccentricities for small planets is more complicated for two reasons: transits of smaller planets require higher accuracy light curves to obtain the same signal-to-noise ratio in the light curve than needed for larger planets, and for small planets eccentricity and impact parameter are more degenerate. The effect of $b$ and $e$ on the transit duration is illustrated in Figure~\ref{fig:illustration_b_ecc}. Therefore apart from reporting eccentricity confidence intervals we also present two-dimensional histograms which show the posterior distribution in the $e - b$ plane (see Appendix~\ref{sec:individual_posterior_discussion}). In a few cases (see Table~\ref{tab:parameter_table}) the correlation between $b$ and $e$ caused the eccentricity range to be 
uninformative (here defined as an $1-\sigma$ interval larger than $0.4$). These 8 systems were excluded from the sample presented in Section~\ref{sec:multis_loweccentricity} as they do not present any additional information \cite[see e.g.][]{price2015}. 

\subsubsection{The influence of TTVs}
\label{sec:results_ttvissues}

Transit timing variations have the potential to influence eccentricity measurements. Contrary to what one might expect, the major issue with TTVs is not that they cause the total transit duration to be mismeasured, but rather that TTVs can cause the impact parameter to be measured wrongly \citep{kipping2014}. When combining multiple transits which are not correctly aligned, the best-fit model transit will be more V-shaped (higher impact parameter) than the original transit. As high impact parameters typically have shorter transit durations, this bias in $b$ can then be `compensated' by a higher eccentricity (and an angle of periastron within $[180,360]^\circ$). Consequently, when TTVs are not properly taken into account, a bias occurs towards the top right on the illustration in Figure~\ref{fig:illustration_b_ecc}. This bias due to TTVs can be quite large. For example, we inserted an artificial planet on a circular orbit into the \textit{Kepler} observations and added a sinusoidal TTV signal with an 
amplitude of 20 minutes and a period of 250 days. An eccentricity of 0.7 was recovered (with small formal uncertainty), while for the same case without TTVs the correct circular orbit was recovered. However, these clear cases of TTVs can easily be measured and removed, which we do in our pipeline. 

Smaller TTV signals can be more difficult to detect and adequately remove. With smaller planetary radii (smaller transit depths), the ability to measure individual transit times decreases and therefore also our ability to detect a TTV signal. On the other hand, for the smallest planets, the impact parameter is typically poorly constrained, making (small) TTVs less important relative to other sources of uncertainties (see Section~\ref{sec:transit_insertion}). It is not always straightforward to determine whether TTVs should be included in the modeling. We found that classic tests such as the likelihood ratio tests or the Bayesian Information Criterium sometimes favor the inclusion of a TTV signal for the smallest planets, on artificial transits inserted without TTVs into real \textit{Kepler} observations. This could be caused by an underestimate of the errors on the transit times for very small planets, or the influence of light curve inperfections (instrumental or astrophysical, e.g. star spots). 

In our final analysis we include only clearly detectable sinusoidal TTV signals, after confirming that in cases where there was doubt, the decision to include TTVs or not did not influence the eccentricity measurement (see also Section~\ref{sec:transit_insertion}). A list of systems with included TTVs is given in Table~\ref{tab:ttv_table} and for Kepler-103, Kepler-126, Kepler-130 and Kepler-278, these TTVs have not been previously reported. Four systems were excluded from our initial sample because their TTVs could not be adequately removed using a sinusoidal model; they are discussed in Appendix~\ref{sec:excluded_systems}.

\subsubsection{Other potential complications}
\label{sec:results_othercomplications}

We briefly discuss several other issues that have been previously identified as potential sources of error for measurements of eccentricities from transit photometry.

\textit{False positives} can complicate eccentricity measurements. When a planetary transit's host star is misidentified, the true stellar density can differ significantly from the one used to calculate the eccentricity \citep{sliski2014}. In our sample, all but three systems (KOI-5, KOI-270 and KOI-279) contain planets which were previously confirmed or validated as true exoplanets. Kepler-92 contains two confirmed planets and one additional candidate. We discuss the planetary nature of these planet candidates in Section~\ref{sec:validation}. Therefore our sample is not biased due to false positives.

Similar to planetary false positives is the issue of \textit{light curve dilution}. Here, the planet orbits around its host star, but third light dilutes the light curve, causing the transit depth to be reduced. This results not only in a biased planet radius \citep{ciardi2015}, but also in a biased impact parameter, which in turn can cause the eccentricity to be wrongly measured. However, most of the targets from our selected sample of bright stars have been followed up with adaptive optics \citep{adams2012} and Speckle images \citep{howell2011}. No significant sources of dilution have been found for any of our confirmed planets. The reported light curve contamination for KOI-5, KOI-270 and KOI-279 is taken into account prior to the modeling. Quarter to quarter variations in the light curve owing to pixel sensitivity are of the percentage order \citep{vaneylen2013} and do not affect our eccentricity measurement.

Stellar \textit{limb darkening} is another potential source of complication. Visual inspection yields no evidence of a correlation with eccentricity \citep[see also][]{ford2008}. We use a prior on the limb darkening based on stellar atmosphere models \citep{claret2011} to speed up MCMC convergeance, but nevertheless allow the limb darkening parameters to vary to avoid this source of complications. 

Another potential influence on eccentricity measurements would be a bias in the \textit{stellar densities} determined from asteroseismology. Part of the values from our sample are taken from \cite{silvaaguirre2015}, and are based on individual frequency modeling using several different stellar evolution codes. The remaining densities are taken from \citep{huber2013} and are based on scaling relations. Such relations have been proven accurate and unbiased for dwarfs and subdwarfs, such as the stars considered in this study \citep{huber2012,silvaaguirre2012,silvaaguirre2015}.

Finally we note that the \textit{uncertainty in the folded light curve} could be of potential concern. Ideally, all individual transits would be normalised and modeled simultaneously, while also fitting for the period and any potential TTVs and modeling the correlated noise. However, such an approach is computationally unfeasible. Consequentially, these errors are not fully propogated and the resulting uncertainties could be underestimated. In most cases many transits are available, causing the period to be very well determined. Of bigger concern are TTVs, but tests with artificial planetary transits (see Section~\ref{sec:transit_insertion}) show no evidence of any bias or underestimated error bars. 

\subsection{Transit insertion tests}
\label{sec:transit_insertion}

We have inserted artificial transits into the data to test the performance of our pipeline. The procedure we used is as follows. First, an artificial planetary transit was generated, and inserted into the light curve that has been observed for one of the stars analyzed in our sample. The lightcurves in which we inserted artificial transits were chosen randomly from our sample of stars with two or three transiting planets (stars with more planets were not chosen to avoid 'crowding' due to the pre-existing planets). Subsequently the procedure described in Section~\ref{sec:period_ttv_finder} was followed to find the orbital period and potential TTVs and fold the data. The period and ephemeris information of the (genuine) planets already present in the light curve was used to remove overlapping transits, as is done for genuine planets. Finally the folded light curve is modeled as described in Section~\ref{sec:transit_fit_module}.

The aim of these tests is not to be complete in covering the full parameter space, which is indeed challenging as it spans different stellar and planetary parameters, periods and eccentricities, as well as amplitudes and periods of TTVs, while transit insertion tests are computationally expensive. Rather, the purpose is to evaluate representative cases to understand the performance of our pipeline and judge any potential limitations. A total of 141 artificial transits have been generated, inserted in real \textit{Kepler} data, and modeled. We now describe a few cases in more detail.

In a first number of tests, we generated planets with radii and periods representative for our sample, and assigned a random eccentricity, uniform between $0$ and $1$, and a random angle of periastron $\omega$. We were able to recover the correct eccentricities within the uncertainties. In another set of tests, we attempted to reproduce our sample of planets more closely. The light curves in which the transits were inserted were drawn randomly from the light curves in our sample. The periods and planetary radii were drawn randomly from our sample of planets (Table~\ref{tab:parameter_table}). The impact parameters were chosen uniformly between 0 and 1 for outer planets, and uniformly within a $1.6^\circ$ spread for inner planets \citep{fabrycky2014}. The eccentricities were typically recovered within the uncertainties.

We have also tested the influence of TTVs by adding sinusoidal TTV signals to the inserted transits. The influence of TTVs depends not only on the TTV amplitude, but also on the size of the planet. For example, for a $3.5~R_\oplus$ planet on a 15 day orbit, a 20 minute TTV signal can have a large influence on the derived eccentricity (see Section~\ref{sec:results_ttvissues}), but the TTV signal is easily recovered and after removal, the correct eccentricity is determined within the uncertainty (and without bias). For smaller planets, it can be difficult to adequately remove the TTV signal, and it can escape detection entirely. However, we find that in these cases, the influence of TTVs on the eccentricity determination is small because other uncertainties dominate. For example, when inserting a TTV signal with an amplitude of 15 minutes, for a planet of $1.5~R_\oplus$ with an orbital period of 8 days, we did not recover the TTV signal but were nevertheless able to retrieve the correct eccentricities. 
Other, similar TTV tests revealed similar results, and we also obtained a similar result when modeling genuine planets: when there was significant doubt about the TTV signal, the decision to include it or not did not influence the outcome of our eccentricity measurement.

\section{Results}
\label{sec:results}

Here we present the results of our analysis. In Section~\ref{sec:multis_loweccentricity} we report the distribution of eccentricities for the planets in our sample (eccentricities for individual planets are discussed in Appendix~\ref{sec:individual_posterior_discussion}). In Section~\ref{sec:other_parameters} we present the other parameters that result from our analysis, such as homegeneous planetary parameters, a distribution in impact parameters and new and updated TTVs. Finally in Section~\ref{sec:validation} we discuss the systems with unconfirmed planetary candidates and validate six new planets.

\subsection{Multi-planet systems with small planets have low eccentricities}
\label{sec:multis_loweccentricity}

\begin{figure*}[!htbp]
\centering
\resizebox{0.9\hsize}{!}{\includegraphics{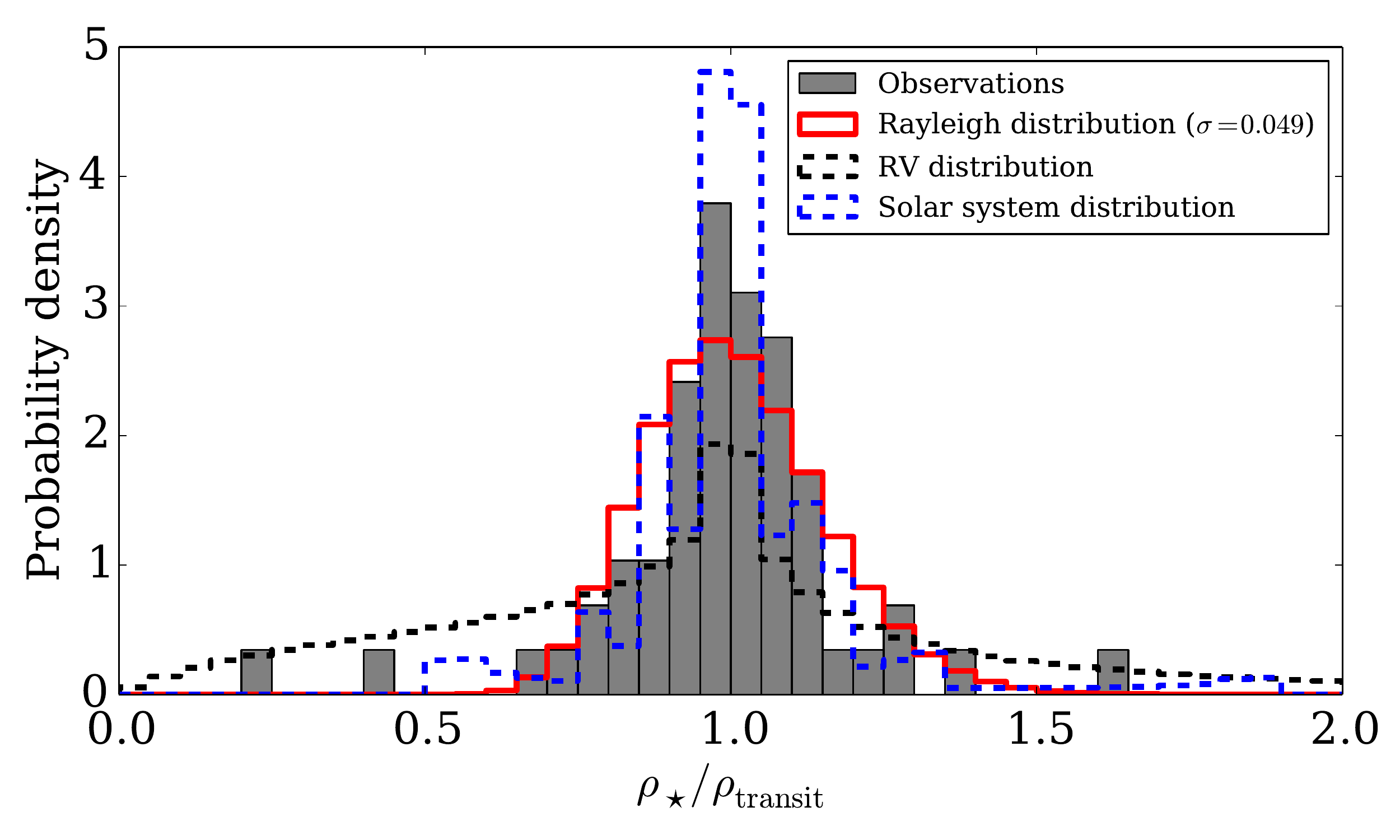}}
\caption{The stellar density determined from asteroseismology divided by the stellar density determined from the planetary transit if the orbit was circular. Values much smaller than one indicate eccentric orbits with $\omega \in [0,180]^\circ$ and short transits, while values much larger indicate $\omega \in [180,360]^\circ$ and longer transits. The best fitted Rayleigh distribution is overplotted and has $\sigma = 0.049$. For illustration the densities which would be observed assuming the RV eccentricity distribution are also indicated, as well as the distribution derived from the Solar System planet's eccentricities.\label{fig:relativedensities}}
\end{figure*}
%
The stellar density encompasses the combined influence of the orbital eccentricity and angle of periastron on the transit duration as described in Section~\ref{sec:results_correlations}. In Figure~\ref{fig:relativedensities} we show a histogram of the ratio of the densities derived from asteroseismology to the densities derived from the transit fits. In this figure large eccentricities would be revealed as very large or small density ratios, depending on the argument of periastron. The absence of such ratios already indicates that low eccentricities are common.

To quantitatively constrain the eccentricity distribution we now assume a Rayleigh distribution for the eccentricities, which provides a best fit to the data for $\sigma = 0.049 \pm 0.013$. The resulting distribution of density ratios is shown in Figure~\ref{fig:relativedensities}. The Rayleigh distribution has the additional advantage that it can be directly compared to some other eccentricity determinations, such as $\sigma = 0.018$ found for some TTV systems \citep{hadden2014}. \cite{kipping2013} suggests the use of a Beta distribution, which has the advantage of being convenient to use as a prior for transit fits. Using this distribution to model our results we find a good fit to our data with Beta parameters $\alpha = 1.03 \pm 0.15$ and $\beta = 13.6 \pm 2.1$. The best-fit values are calculated by drawing random eccentricity values from the chosen distribution (Rayleigh distribution or Beta distribution) and assigning a random angle of periastron to calculate the corresponding density ratio. The 
distribution of density ratios is then compared to the observed density ratio distribution, by minimizing the $\chi^2$ when comparing the cumulative density functions, to avoid a dependency of the fit on binning of the data \citep[see e.g.][]{kipping2013}. The uncertainty on the parameters is calculating by bootstrapping the observed density ratios (with replacement) and repeating the procedure, and calculating the scatter in the best-fit parameters. Individual systems are discussed in Appendix~\ref{sec:individual_posterior_discussion}.

\begin{figure*}[!htbp]
\centering
\resizebox{0.9\hsize}{!}{\includegraphics{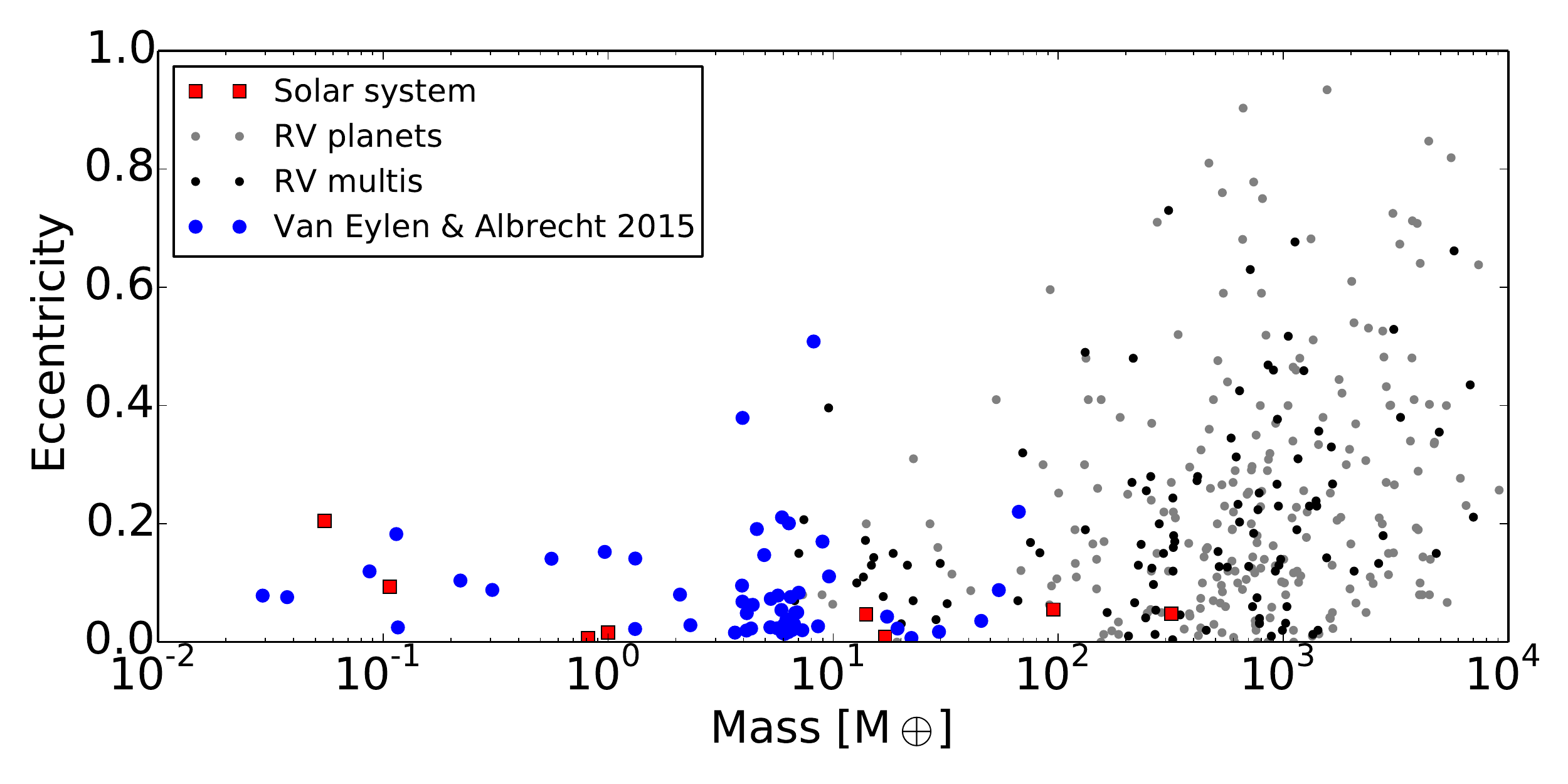}}
\caption{The eccentricity and mass measurements for exoplanets are plotted as taken from exoplanets.org on 27 April 2015, for planets where both values are determined. Planets which are flagged as multi-planet systems are highlighted. For comparison, the solar system is shown. The eccentricities of the planets in our sample are plotted with their mass estimated based on radius \citep{weiss2013,weiss2014}. Error bars are omitted for clarity.\label{fig:masseccentricity}}
\end{figure*}

The distribution is similar to that of the Solar System which is plotted in the same figure for comparison (integrated over different angles). In contrast we also plot the relative densities that would have been observed if our sample had the same eccentricity distribution as measured for RV planets \citep{shen2008} in Figure~\ref{fig:relativedensities}. Figure~\ref{fig:masseccentricity} compares the eccentricities in our sample with the solar system planets and the exoplanets with RV observations. The RV observations were taken from exoplanets.org (27 April 2015) and include all data points where the eccentricity was measured (not fixed to zero), and the RV amplitude ('K') divided by its uncertainty ('UK') is greater than ten. The masses for the planets in our sample were estimated based on the radius, using \cite{weiss2013} for planets with $R \geq 4~R_\oplus$ and following \cite{weiss2014} for planets where $R \leq 4~R_\oplus$.

Our sample differs in two important ways from the RV sample: planetary size and planetary multiplicity. These properties are not independent since smaller planets are frequently found in multiple planet systems \citep{latham2011}. A hint towards smaller eccentricities for smaller/less massive planets and higher multiplicity has already been observed in RV systems. In systems with sub-Jovian mass planets and systems with multiple planets, eccentricities are limited to 0-0.45 \citep{wright2009,mayor2011}. Even so the eccentricities observed in our sample have a much narrower range, possibly because the average size of the planets is much smaller even when 
compared to the sub-Jovian RV sample \citep[most planets in our sample cannot be detected with RV measurements, and even when RV mass measurements are possible eccentricity determinations are not feasible,][]{marcy2014}.

Analyzing TTV signals for \textit{Kepler} planets, \cite{hadden2014} find an rms eccentricity of $0.018^{+0.005}_{-0.004}$. They further note that eccentricities of planets smaller than $2.5~R_\oplus$ are about twice as large as those larger than this limit, although they caution a TTV detection bias may influence this result. We have compared our eccentricity measurements with the planetary radii in Figure~\ref{fig:parameter_comparison} (see also Section~\ref{sec:other_parameters}) and found no evidence for a correlation. However, the difference between the rms eccentricity for planets smaller and larger than $2.5~R_\oplus$ is only 0.009 \citep{hadden2014}, which would likely not be detectable in our sample.

Planet-planet interactions have been brought forward as a mechanism to explain the observed eccentricities in massive planets \citep{fabrycky2007,chatterjee2008,nagasawa2008,ford2008_planetplanet,juric2008}. In this picture gravitational interactions lead to high eccentricities and planetary migration. However, despite finding a small anti-correlation between mass and eccentricity for massive planets, \cite{chatterjee2008} suggested that damping from residual gas or planetesimals could more effectively reduce the eccentricities of low-mass planets after scattering. Furthermore, it has been suggested that there may exist a dependence of eccentricity on the orbital semi-major axis, because the mean eccentricity depends on the velocity dispersion scaled by the Keplerian velocity \citep[see e.g.][]{ida2013,petrovich2014}. Consequentially the eccentricity may be proportional to the square root of the semi-major axis \citep{ida2013}. The majority of the planets in our sample have orbital distances that are 
unlikely to be affected by tidal circularisation, but it was suggested very recently that tidal effects in compact multi-planet systems may propagate further than for single planet systems \citep{hansen2015}

The observed low eccentricities could be related to the planet multiplicity, which was also observed by \cite{limbach2014}. Highly eccentric planets in multi-planet systems are also less likely to be stable over longer timescales, which could lead to lower observed eccentricities in compact systems because systems with more eccentric systems would not survive. \cite{pu2015} found that planets with circular orbits can be more tightly packed than systems with eccentric planets. The systems in our sample have between 2 and 5 transiting planets but the true multiplicity could be underestimated if additional non-transiting planet are present.

\subsection{Homogeneous stellar and planetary parameters and new TTVs}
\label{sec:other_parameters}

Next to orbital eccentricities our analysis also yields a homegeneous set of planetary parameters. They are not only derived from homegeneous transit modeling but also from a homegemeous set of stellar parameters, which were all derived from asteroseismology \citep{huber2013,silvaaguirre2015}. We report the eccentricities and the planetary radii, as well as the stellar masses and radii upon which they were based \citep{huber2013,silvaaguirre2015} in Table~\ref{tab:parameter_table}. The modes and 68\% highest probability density intervals are quoted for all values. The full posterior distributions, including the correlations between parameters, are available upon request. 

\begin{figure}[ht]
\centering
\resizebox{0.9\hsize}{!}{\includegraphics{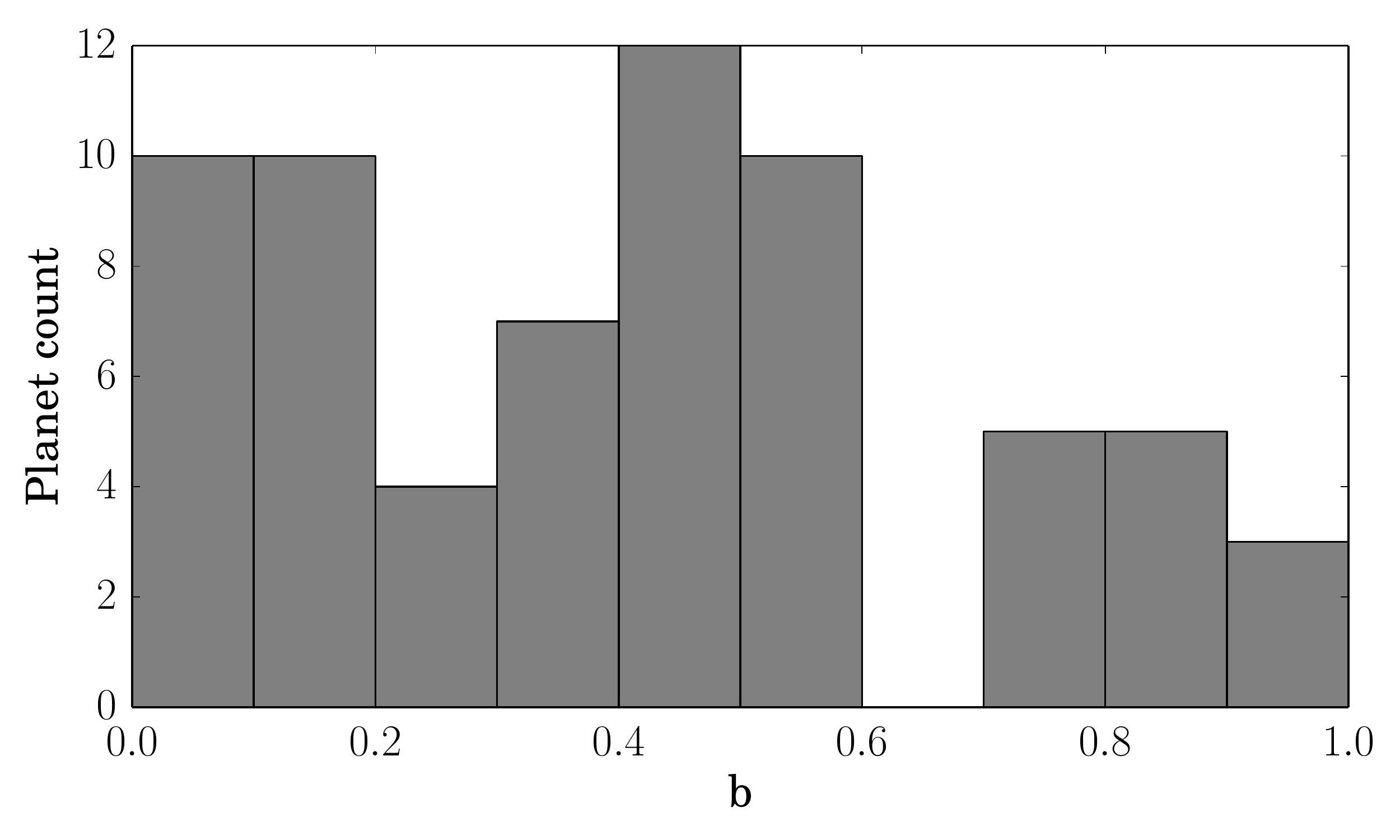}}
\resizebox{0.9\hsize}{!}{\includegraphics{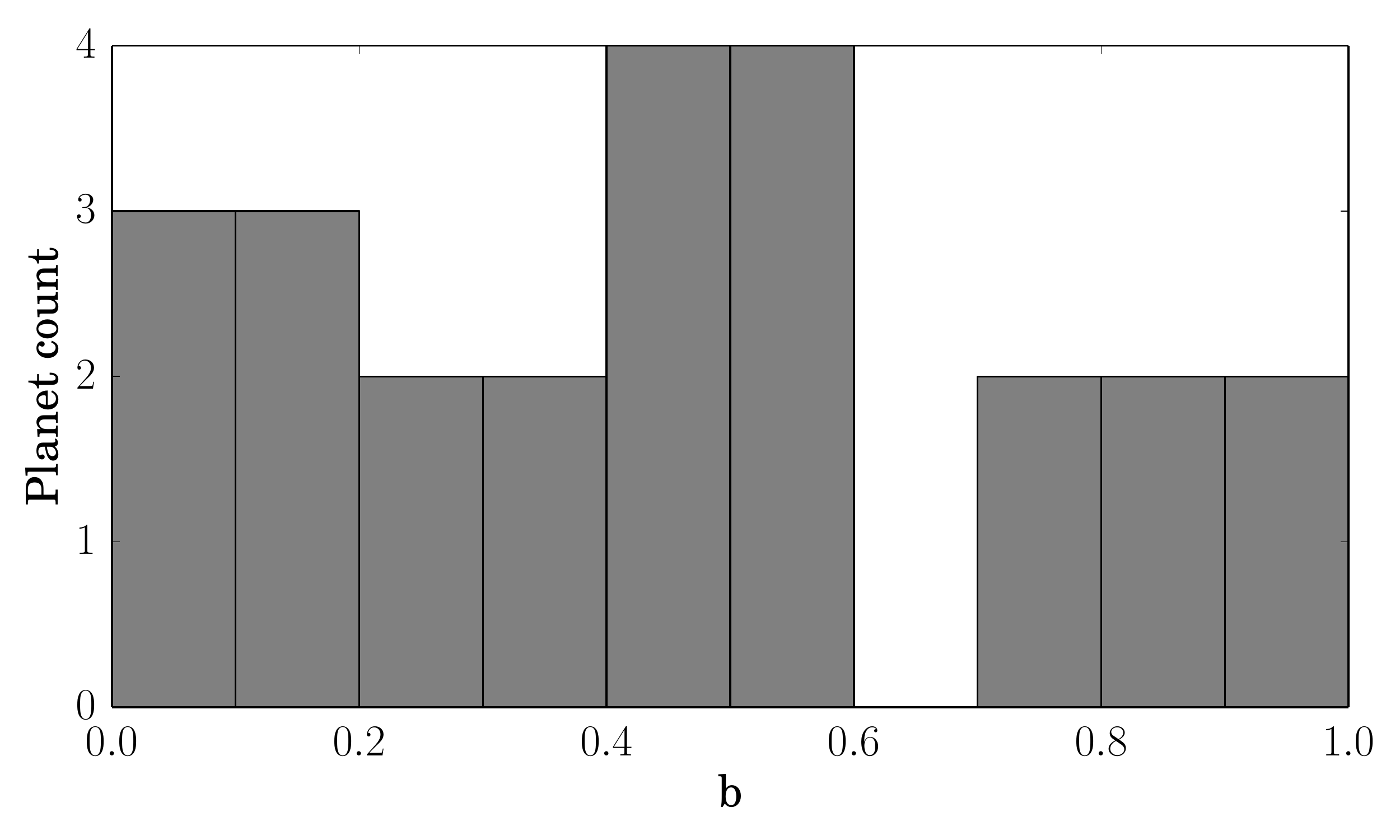}}
\resizebox{0.9\hsize}{!}{\includegraphics{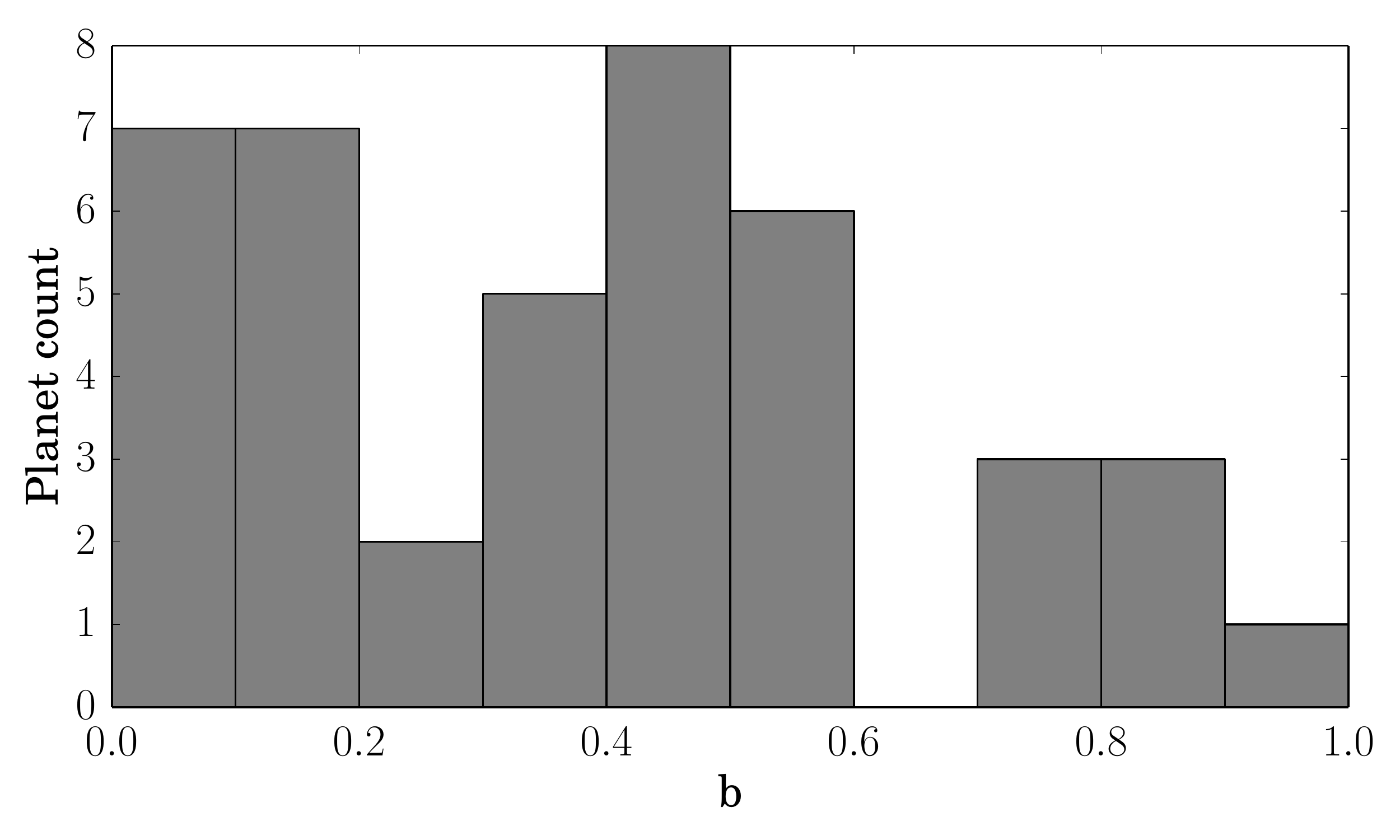}}
\caption{Histogram of the modes of the impact parameters for individual planets. \textbf{Top:} all planets, \textbf{middle:} only outer planets, \textbf{bottom:} planets that are not the outer transiting planet.\label{fig:impactparameter_histogram}}
\end{figure}

We checked the distribution of transit impact parameters and show a histogram in Figure~\ref{fig:impactparameter_histogram}. Because we are dealing with multi-transiting systems a bias towards lower impact parameters is expected since such systems are more likely to have multiple planets transiting. When we plot the impact parameter of all planets, low impact parameter values indeed appear favored and the distribution is inconsistent with a homegeneous one between 0 and 1 (KS-test with p-value of 0.003). If we only plot the impact parameters of the outer planet (the longest period) in each system, a distribution which appears uniform in impact parameter is observed (KS-test with p-value of 0.86, see Figure~\ref{fig:impactparameter_histogram}). That planets on shorter orbital periods have lower impact parameter than the outer planets in the same system shows that most systems in our sample have very low mutual inclinations, consistent with earlier work \citep{fabrycky2014}.

\begin{centering}
\begin{table*}[ht]
\scriptsize
\centering
\begin{tabular}{llllll|llllllllllll}
	  &		&	$e$ (mode)	&$e$ (68\%)	&$R_\textrm{p}$ [$R_\oplus$]	&Period [d]		&Ref.	& $M_\star$ [$M_\odot$]		& $R_\star$ [$R_\odot$]		&Density [g/cm3]&\\
\hline \hline
Kepler-10b&	KOI-72.01&	$0.06$ &	$[0, 0.22]$	&$1.473 \pm 0.026$	&$0.83749026 (29)$	&(2)	&$0.920^{-0.020}_{0.010}$	&$1.0662^{-0.0075}_{0.0069}$	&$1.0679^{-0.012}_{0.0072}$	&\\
Kepler-10c&	KOI-72.02&	$0.05$ &	$[0, 0.25]$	&$2.323 \pm 0.028$	&$45.294292 (97)$	&(2)	&$0.920^{-0.020}_{0.010}$	&$1.0662^{-0.0075}_{0.0069}$	&$1.0679^{-0.012}_{0.0072}$	&\\
Kepler-23b&	KOI-168.03&	$0.06$ &	$[0, 0.32]$	&$1.694 \pm 0.076$	&$7.106995 (73)$	&(1)	&$1.078^{-0.077}_{0.077}$	&$1.548^{-0.048}_{0.048}$	&$0.410^{-0.023}_{0.023}$	&\\
Kepler-23c&	KOI-168.01&	$0.02$ &	$[0, 0.41]$	&$3.12 \pm 0.10$	&$10.742434 (39)$	&(1)	&$1.078^{-0.077}_{0.077}$	&$1.548^{-0.048}_{0.048}$	&$0.410^{-0.023}_{0.023}$	&\\
Kepler-23d&	KOI-168.02&	$0.08$ &	$[0, 0.32]$	&$2.235 \pm 0.088$	&$15.27429 (17)$	&(1)	&$1.078^{-0.077}_{0.077}$	&$1.548^{-0.048}_{0.048}$	&$0.410^{-0.023}_{0.023}$	&\\
Kepler-25b&	KOI-244.02&	$0.05$ &	$[0, 0.16]$	&$2.702 \pm 0.037$	&$6.2385369 (33)$	&(2)	&$1.160^{-0.050}_{0.040}$	&$1.299^{-0.016}_{0.015}$	&$0.7454^{-0.0098}_{0.0093}$	&\\
Kepler-25c&	KOI-244.01&	$0.01$ &	$[0, 0.08]$	&$5.154 \pm 0.060$	&$12.7203678 (35)$	&(2)	&$1.160^{-0.050}_{0.040}$	&$1.299^{-0.016}_{0.015}$	&$0.7454^{-0.0098}_{0.0093}$	&\\
Kepler-37b&	KOI-245.03&	$0.08$ &	$[0, 0.29]$	&$0.354 \pm 0.014$	&$13.36805 (38)$	&(2)	&$0.810^{-0.010}_{0.020}$	&$0.7725^{-0.0063}_{0.0051}$	&$2.486^{-0.025}_{0.022}$	&\\
Kepler-37c&	KOI-245.02&	$0.09$ &	$[0, 0.27]$	&$0.705 \pm 0.012$	&$21.302071 (92)$	&(2)	&$0.810^{-0.010}_{0.020}$	&$0.7725^{-0.0063}_{0.0051}$	&$2.486^{-0.025}_{0.022}$	&\\
Kepler-37d&	KOI-245.01&	$0.15$ &	$[0.05, 0.22]$	&$1.922 \pm 0.024$	&$39.792232 (54)$	&(2)	&$0.810^{-0.010}_{0.020}$	&$0.7725^{-0.0063}_{0.0051}$	&$2.486^{-0.025}_{0.022}$	&\\
Kepler-65b&	KOI-85.02&	$0.02$ &	$[0, 0.19]$	&$1.409 \pm 0.017$	&$2.1549156 (25)$	&(2)	&$1.199^{-0.030}_{0.030}$	&$1.401^{-0.014}_{0.013}$	&$0.6158^{-0.0079}_{0.0071}$	&\\
Kepler-65c&	KOI-85.01&	$0.08$ &	$[0, 0.2]$	&$2.571 \pm 0.033$	&$5.8599408 (23)$	&(2)	&$1.199^{-0.030}_{0.030}$	&$1.401^{-0.014}_{0.013}$	&$0.6158^{-0.0079}_{0.0071}$	&\\
Kepler-65d&	KOI-85.03&	$0.10$ &	$[0, 0.33]$	&$1.506 \pm 0.040$	&$8.131231 (21)$	&(2)	&$1.199^{-0.030}_{0.030}$	&$1.401^{-0.014}_{0.013}$	&$0.6158^{-0.0079}_{0.0071}$	&\\
Kepler-68b&	KOI-246.01&	$0.02$ &	$[0, 0.15]$	&$2.354 \pm 0.020$	&$5.3987533 (13)$	&(2)	&$1.070^{-0.020}_{0.010}$	&$1.2379^{-0.0051}_{0.0067}$	&$0.7949^{-0.011}_{0.0052}$	&\\
Kepler-68c&	KOI-246.02&	$0.42$ &	$[0.32, 0.83]$	&$0.927 \pm 0.025$	&$9.604979 (45)$	&(2)	&$1.070^{-0.020}_{0.010}$	&$1.2379^{-0.0051}_{0.0067}$	&$0.7949^{-0.011}_{0.0052}$	&\\
Kepler-92b&	KOI-285.01&	$0.17$ &	$[0, 0.27]$	&$3.65 \pm 0.13$	&$13.748933 (75)$	&(2)	&$1.209^{-0.030}_{0.020}$	&$1.719^{-0.013}_{0.011}$	&$0.3355^{-0.0040}_{0.0044}$	&\\
Kepler-92c&	KOI-285.02&	$0.04$ &	$[0, 0.26]$	&$2.455 \pm 0.053$	&$26.72311 (19)$	&(2)	&$1.209^{-0.030}_{0.020}$	&$1.719^{-0.013}_{0.011}$	&$0.3355^{-0.0040}_{0.0044}$	&\\
Kepler-92d&	KOI-285.03&	$0.07$ &	$[0.03, 0.41]$	&$2.067 \pm 0.056$	&$49.3568 (24)$		&(2)	&$1.209^{-0.030}_{0.020}$	&$1.719^{-0.013}_{0.011}$	&$0.3355^{-0.0040}_{0.0044}$	&\\
Kepler-100b&	KOI-41.02&	$0.13$ &	$[0, 0.40]$	&$1.305 \pm 0.030$	&$6.887037 (47)$	&(2)	&$1.109^{-0.020}_{0.020}$	&$1.5131^{-0.011}_{0.0093}$	&$0.4542^{-0.0058}_{0.0043}$	&\\
Kepler-100c&	KOI-41.01&	$0.02$ &	$[0.01, 0.17]$	&$2.221 \pm 0.022$	&$12.815909 (26)$	&(2)	&$1.109^{-0.020}_{0.020}$	&$1.5131^{-0.011}_{0.0093}$	&$0.4542^{-0.0058}_{0.0043}$	&\\
Kepler-100d&	KOI-41.03&	$0.38$ &	$[0.22, 0.50]$	&$1.514 \pm 0.034$	&$35.33313 (43)$	&(2)	&$1.109^{-0.020}_{0.020}$	&$1.5131^{-0.011}_{0.0093}$	&$0.4542^{-0.00579}_{0.00431}$	&\\
Kepler-103b&	KOI-108.01&	$0.03$ &	$[0, 0.23]$	&$3.476 \pm 0.039$	&$15.965316 (18)$	&(2)	&$1.099^{-0.030}_{0.019}$	&$1.455^{-0.013}_{0.024}$	&$0.5070^{-0.0050}_{0.0050}$	&\\
Kepler-103c&	KOI-108.02&	$0.02$ &	$[0, 0.21]$	&$5.319 \pm 0.052$	&$179.6133 (47)$	&(2)	&$1.099^{-0.030}_{0.019}$	&$1.450^{-0.009}_{0.009}$	&$0.5070^{-0.0050}_{0.0050}$	&\\
Kepler-107b&	KOI-117.03&	$0.02$ &	$[0, 0.22]$	&$1.581 \pm 0.056$	&$3.180026 (12)$	&(1)	&$1.142^{-0.068}_{0.068}$	&$1.411^{-0.047}_{0.047}$	&$0.581^{-0.049}_{0.049}$	&\\
Kepler-107c&	KOI-117.02&	$0.02$ &	$[0, 0.28]$	&$1.664 \pm 0.065$	&$4.901441 (30)$	&(1)	&$1.142^{-0.068}_{0.068}$	&$1.411^{-0.047}_{0.047}$	&$0.581^{-0.049}_{0.049}$	&\\
Kepler-107d&	KOI-117.04&	$0.14$ &	$[0, 0.39]$	&$1.064 \pm 0.062$	&$7.95825 (11)$		&(1)	&$1.142^{-0.068}_{0.068}$	&$1.411^{-0.047}_{0.047}$	&$0.581^{-0.049}_{0.049}$	&\\
Kepler-107e&	KOI-117.01&	$0.02$ &	$[0, 0.20]$	&$2.92 \pm 0.10$	&$14.749176 (34)$	&(1)	&$1.142^{-0.068}_{0.068}$	&$1.411^{-0.047}_{0.047}$	&$0.581^{-0.049}_{0.049}$	&\\
Kepler-108b&	KOI-119.01&	$0.22$ &	$[0.10, 0.41]$	&$9.56 \pm 0.53$	&$49.18354 (18)$	&(1)	&$1.377^{-0.089}_{0.089}$	&$2.19^{-0.12}_{0.12}$		&$0.188^{-0.024}_{0.024}$	&\\
Kepler-108c&	KOI-119.02&	$0.04$ &	$[0, 0.23]$	&$8.23 \pm 0.47$	&$190.3214$ (n/a)	&(1)	&$1.377^{-0.089}_{0.089}$	&$2.19^{-0.12}_{0.12}$		&$0.188^{-0.024}_{0.024}$	&\\
Kepler-109b&	KOI-123.01&	$0.21$ &	$[0, 0.30]$	&$2.338 \pm 0.034$	&$6.4816370 (80)$	&(2)	&$1.069^{-0.040}_{0.040}$	&$1.339^{-0.015}_{0.017}$	&$0.6278^{-0.0068}_{0.0076}$	&\\
Kepler-109c&	KOI-123.02&	$0.03$ &	$[0, 0.22]$	&$2.634 \pm 0.043$	&$21.222620 (30)$	&(2)	&$1.069^{-0.040}_{0.040}$	&$1.339^{-0.015}_{0.017}$	&$0.6278^{-0.0068}_{0.0076}$	&\\
Kepler-126b&	KOI-260.01&	$0.07$ &	$[0, 0.17]$	&$1.439 \pm 0.020$	&$10.495634 (30)$	&(2)	&$1.148^{-0.049}_{0.051}$	&$1.345^{-0.018}_{0.015}$	&$0.666^{-0.010}_{0.010}$	&\\
Kepler-126c&	KOI-260.03&	$0.19$ &	$[0, 0.37]$	&$1.498 \pm 0.062$	&$21.86964 (10)$	&(2)	&$1.148^{-0.049}_{0.051}$	&$1.345^{-0.018}_{0.015}$	&$0.666^{-0.010}_{0.010}$	&\\
Kepler-126d&	KOI-260.02&	$0.02$ &	$[0, 0.11]$	&$2.513 \pm 0.031$	&$100.28208 (41)$	&(2)	&$1.148^{-0.049}_{0.051}$	&$1.345^{-0.018}_{0.015}$	&$0.666^{-0.010}_{0.010}$	&\\
Kepler-127b&	KOI-271.03&	$0.47$ &	$[0.08, 0.51]$	&$1.52 \pm 0.13$	&$14.43577 (10)$	&(1)	&$1.240^{-0.086}_{0.086}$	&$1.359^{-0.035}_{0.035}$	&$0.697^{-0.023}_{0.023}$	&\\
Kepler-127c&	KOI-271.02&	$0.03$ &	$[0, 0.17]$	&$2.389 \pm 0.067$	&$29.39344 (17)$	&(1)	&$1.240^{-0.086}_{0.086}$	&$1.359^{-0.035}_{0.035}$	&$0.697^{-0.023}_{0.023}$	&\\
Kepler-127d&	KOI-271.01&	$0.03$ &	$[0, 0.31]$	&$2.668 \pm 0.084$	&$48.62997 (57)$	&(1)	&$1.240^{-0.086}_{0.086}$	&$1.359^{-0.035}_{0.035}$	&$0.697^{-0.023}_{0.023}$	&\\
Kepler-129b&	KOI-275.01&	$0.01$ &	$[0, 0.25]$	&$2.409 \pm 0.040$	&$15.791619 (53)$	&(2)	&$1.159^{-0.030}_{0.030}$	&$1.649^{-0.014}_{0.012}$	&$0.3659^{-0.0037}_{0.0042}$	&\\
Kepler-129c&	KOI-275.02&	$0.20$ &	$[0, 0.35]$	&$2.522 \pm 0.066$	&$82.1908$ (n/a)	&(2)	&$1.159^{-0.030}_{0.030}$	&$1.649^{-0.014}_{0.012}$	&$0.3659^{-0.0037}_{0.0042}$	&\\
Kepler-130b&	KOI-282.02&	$0.15$ &	$[0, 0.29]$	&$0.976 \pm 0.045$	&$8.45725 (11)$		&(1)	&$0.934^{-0.059}_{0.059}$	&$1.127^{-0.033}_{0.033}$	&$0.927^{-0.053}_{0.053}$	&\\
Kepler-130c&	KOI-282.01&	$0.08$ &	$[0, 0.23]$	&$2.811 \pm 0.084$	&$27.508686 (37)$	&(1)	&$0.934^{-0.059}_{0.059}$	&$1.127^{-0.033}_{0.033}$	&$0.927^{-0.053}_{0.053}$	&\\
Kepler-130d&	KOI-282.03&	$0.80$ &	$[0.40, 0.89]$	&$1.31 \pm 0.13$	&$87.5211 (24)$		&(1)	&$0.934^{-0.059}_{0.059}$	&$1.127^{-0.033}_{0.033}$	&$0.927^{-0.053}_{0.053}$	&\\
Kepler-145b&	KOI-370.02&	$0.43$ &	$[0.18, 0.61]$	&$2.56 \pm 0.28$	&$22.95102 (23)$	&(2)	&$1.419^{-0.030}_{0.030}$	&$1.887^{-0.014}_{0.012}$	&$0.2976^{-0.0038}_{0.0045}$	&\\
Kepler-145c&	KOI-370.01&	$0.11$ &	$[0, 0.22]$	&$3.92 \pm 0.11$	&$42.88254 (15)$	&(2)	&$1.419^{-0.030}_{0.030}$	&$1.887^{-0.014}_{0.012}$	&$0.2976^{-0.0038}_{0.0045}$	&\\
Kepler-197b&	KOI-623.03&	$0.02$ &	$[0, 0.25]$	&$1.064 \pm 0.038$	&$5.599293 (39)$	&(1)	&$0.922^{-0.059}_{0.059}$	&$1.120^{-0.033}_{0.033}$	&$0.907^{-0.052}_{0.052}$	&\\
Kepler-197c&	KOI-623.01&	$0.08$ &	$[0, 0.29]$	&$1.208 \pm 0.048$	&$10.349711 (54)$	&(1)	&$0.922^{-0.059}_{0.059}$	&$1.120^{-0.033}_{0.033}$	&$0.907^{-0.052}_{0.052}$	&\\
Kepler-197d&	KOI-623.02&	$0.03$ &	$[0, 0.23]$	&$1.244 \pm 0.049$	&$15.67787 (13)$	&(1)	&$0.922^{-0.059}_{0.059}$	&$1.120^{-0.033}_{0.033}$	&$0.907^{-0.052}_{0.052}$	&\\
Kepler-197e&	KOI-623.04&	$0.38$ &	$[0.21, 0.63]$	&$0.983 \pm 0.048$	&$25.2097 (14)$		&(1)	&$0.922^{-0.059}_{0.059}$	&$1.120^{-0.033}_{0.033}$	&$0.907^{-0.052}_{0.052}$	&\\
Kepler-278b&	KOI-1221.01&	$0.04$ &	$[0, 0.37]$	&$4.59 \pm 0.26$	&$30.15856 (91)$	&(1)	&$1.298^{-0.076}_{0.076}$	&$2.935^{-0.066}_{0.066}$	&$0.07240^{-0.00094}_{0.00094}$	&\\
Kepler-278c&	KOI-1221.02&	$0.51$ &	$[0.39, 0.70]$	&$3.31 \pm 0.12$	&$51.0851 (35)$		&(1)	&$1.298^{-0.076}_{0.076}$	&$2.935^{-0.066}_{0.066}$	&$0.07240^{-0.00094}_{0.00094}$	&\\
Kepler-338b&	KOI-1930.01&	$0.04$ &	$[0, 0.31]$	&$2.58 \pm 0.13$	&$13.72699 (47)$	&(1)	&$1.142^{-0.084}_{0.084}$	&$1.735^{-0.082}_{0.082}$	&$0.309^{-0.034}_{0.034}$	&\\
Kepler-338c&	KOI-1930.02&	$0.03$ &	$[0, 0.27]$	&$2.48 \pm 0.14$	&$24.31168 (87)$	&(1)	&$1.142^{-0.084}_{0.084}$	&$1.735^{-0.082}_{0.082}$	&$0.309^{-0.034}_{0.034}$	&\\
Kepler-338d&	KOI-1930.03&	$0.03$ &	$[0, 0.25]$	&$2.66 \pm 0.15$	&$44.4287 (16)$		&(1)	&$1.142^{-0.084}_{0.084}$	&$1.735^{-0.082}_{0.082}$	&$0.309^{-0.034}_{0.034}$	&\\
Kepler-338e&	KOI-1930.04&	$0.05$ &	$[0, 0.28]$	&$1.587 \pm 0.083$	&$9.34149 (40)$		&(1)	&$1.142^{-0.084}_{0.084}$	&$1.735^{-0.082}_{0.082}$	&$0.309^{-0.034}_{0.034}$	&\\
Kepler-444b   &	KOI-3158.01&	$0.08$ &	$[0, 0.30]$	&$0.381 \pm 0.021$	&$3.600125 (28)$	&(2)	&$0.740^{-0.010}_{0.010}$	&$0.7492^{-0.0040}_{0.0046}$	&$2.498^{-0.025}_{0.018}$	&\\
Kepler-444c&	KOI-3158.02&	$0.12$ &	$[0, 0.29]$	&$0.490 \pm 0.024$	&$4.545817 (44)$	&(2)	&$0.740^{-0.010}_{0.010}$	&$0.7492^{-0.0040}_{0.0046}$	&$2.498^{-0.025}_{0.018}$	&\\
Kepler-444d &	KOI-3158.03&	$0.18$ &	$[0, 0.34]$	&$0.530 \pm 0.025$	&$6.189512 (54)$	&(2)	&$0.740^{-0.010}_{0.010}$	&$0.7492^{-0.0040}_{0.0046}$	&$2.498^{-0.025}_{0.018}$	&\\
Kepler-444e &	KOI-3158.04&	$0.02$ &	$[0, 0.29]$	&$0.533 \pm 0.019$	&$7.74350 (10)$		&(2)	&$0.740^{-0.010}_{0.010}$	&$0.7492^{-0.0040}_{0.0046}$	&$2.498^{-0.025}_{0.018}$	&\\
Kepler-444f &	KOI-3158.05&	$0.58$ &	$[0.21, 0.70]$	&$0.679 \pm 0.008$	&$9.740529 (36)$	&(2)	&$0.740^{-0.010}_{0.010}$	&$0.7492^{-0.0040}_{0.0046}$	&$2.498^{-0.025}_{0.018}$	&\\	    
Kepler-449b&	KOI-270.01&	$0.03$ &	$[0, 0.31]$	&$2.056 \pm 0.069$	&$12.58242 (27)$	&(1)	&$0.969^{-0.053}_{0.053}$	&$1.467^{-0.033}_{0.033}$	&$0.439^{-0.016}_{0.016}$	&\\
Kepler-449c&	KOI-270.02&	$0.05$ &	$[0, 0.29]$	&$2.764 \pm 0.086$	&$33.6727 (10)$		&(1)	&$0.969^{-0.053}_{0.053}$	&$1.467^{-0.033}_{0.033}$	&$0.439^{-0.016}_{0.016}$	&\\
Kepler-450b   &	KOI-279.01&	$0.02$ &	$[0, 0.16]$	&$6.14 \pm 0.33$	&$28.454851 (25)$	&(1)	&$1.346^{-0.084}_{0.084}$	&$1.570^{-0.085}_{0.085}$	&$0.478^{-0.064}_{0.064}$	&\\
Kepler-450c   &	KOI-279.02&	$0.02$ &	$[0, 0.19]$	&$2.62 \pm 0.14$	&$15.413135 (85)$	&(1)	&$1.346^{-0.084}_{0.084}$	&$1.570^{-0.085}_{0.085}$	&$0.478^{-0.064}_{0.064}$	&\\
Kepler-450d   &	KOI-279.03&	$0.14$ &	$[0, 0.38]$	&$0.837 \pm 0.068$	&$7.51464 (23)$		&(1)	&$1.346^{-0.084}_{0.084}$	&$1.570^{-0.085}_{0.085}$	&$0.478^{-0.064}_{0.064}$	&\\
	    &	KOI-5.01&	$0.09$ &	$[0, 0.27]$	&$7.87 \pm 0.14$	&$4.78032767 (84)$	&(2)	&$1.199^{-0.030}_{0.020}$	&$1.795^{-0.014}_{0.015}$	&$0.2920^{-0.0034}_{0.0027}$	&\\
	    &	KOI-5.02&	$0.10$ 	&	$[0, 0.40]$	&$0.642 \pm 0.061$	&$7.05174 (13)$		&(2)	&$1.199^{-0.030}_{0.020}$	&$1.795^{-0.014}_{0.015}$	&$0.2920^{-0.0034}_{0.0027}$	&\\
\end{tabular}
\caption{ Planetary and stellar parameters for all planets analyzed. The source of the stellar parameters is indicated in the ref. column: (1) \cite{huber2013}; (2) \cite{silvaaguirre2015}. Individual systems are discussed in Appendix~\ref{sec:individual_posterior_discussion}. \label{tab:parameter_table}}
\end{table*}
\end{centering}

\begin{figure*}
\begin{center}
\centering
\resizebox{0.49\hsize}{!}{\includegraphics{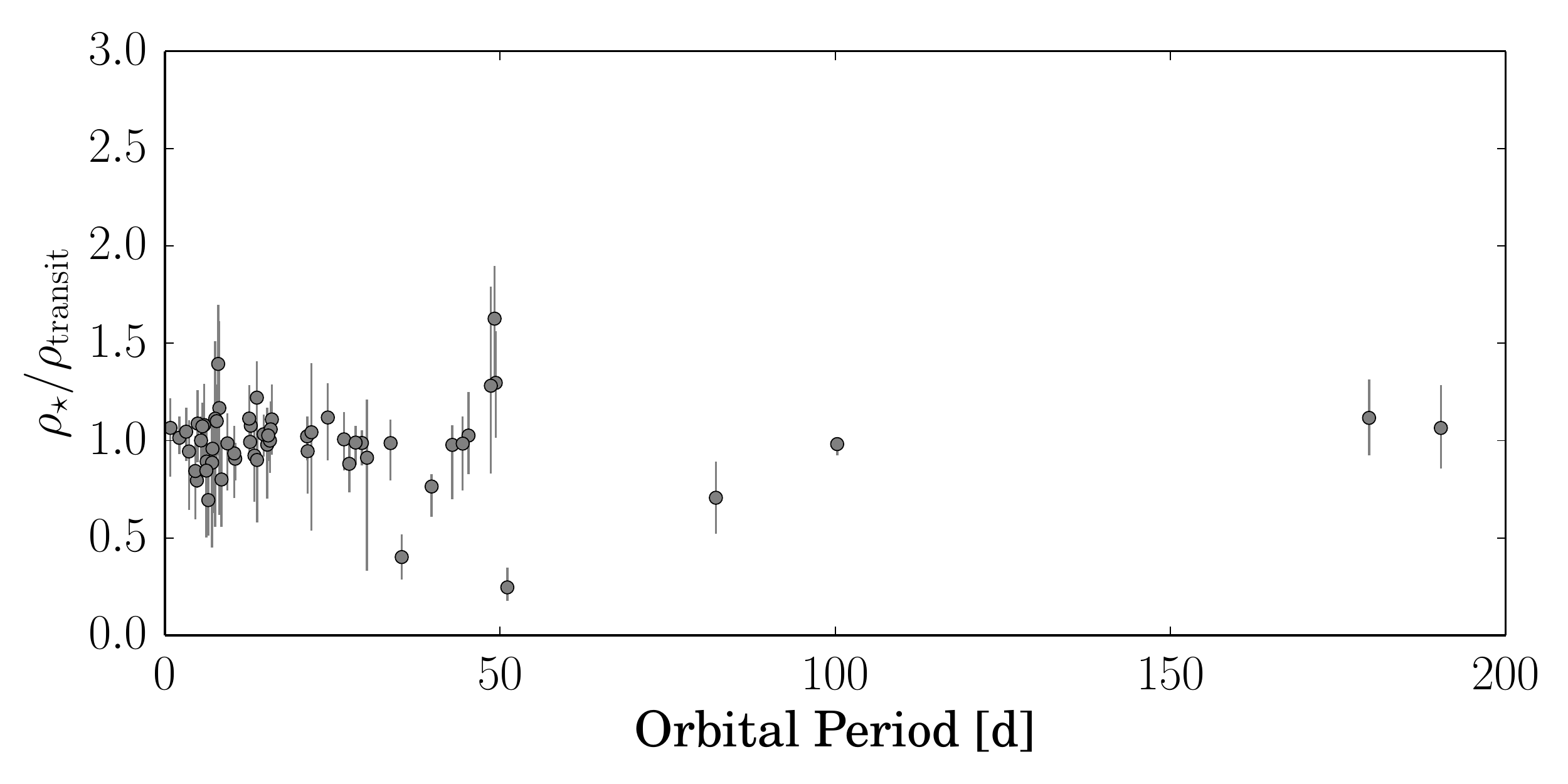}}
\resizebox{0.49\hsize}{!}{\includegraphics{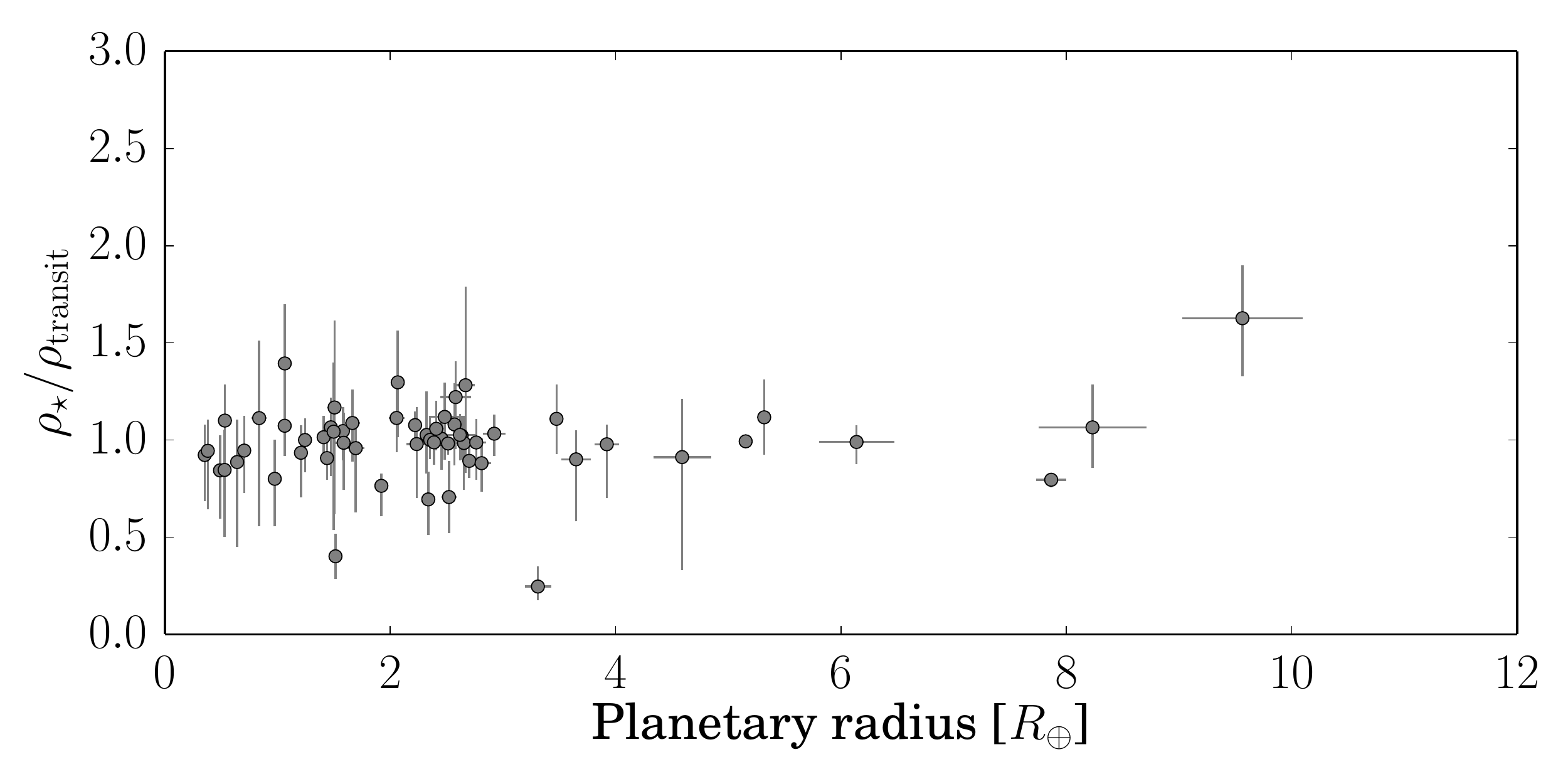}}
\centering
\resizebox{0.49\hsize}{!}{\includegraphics{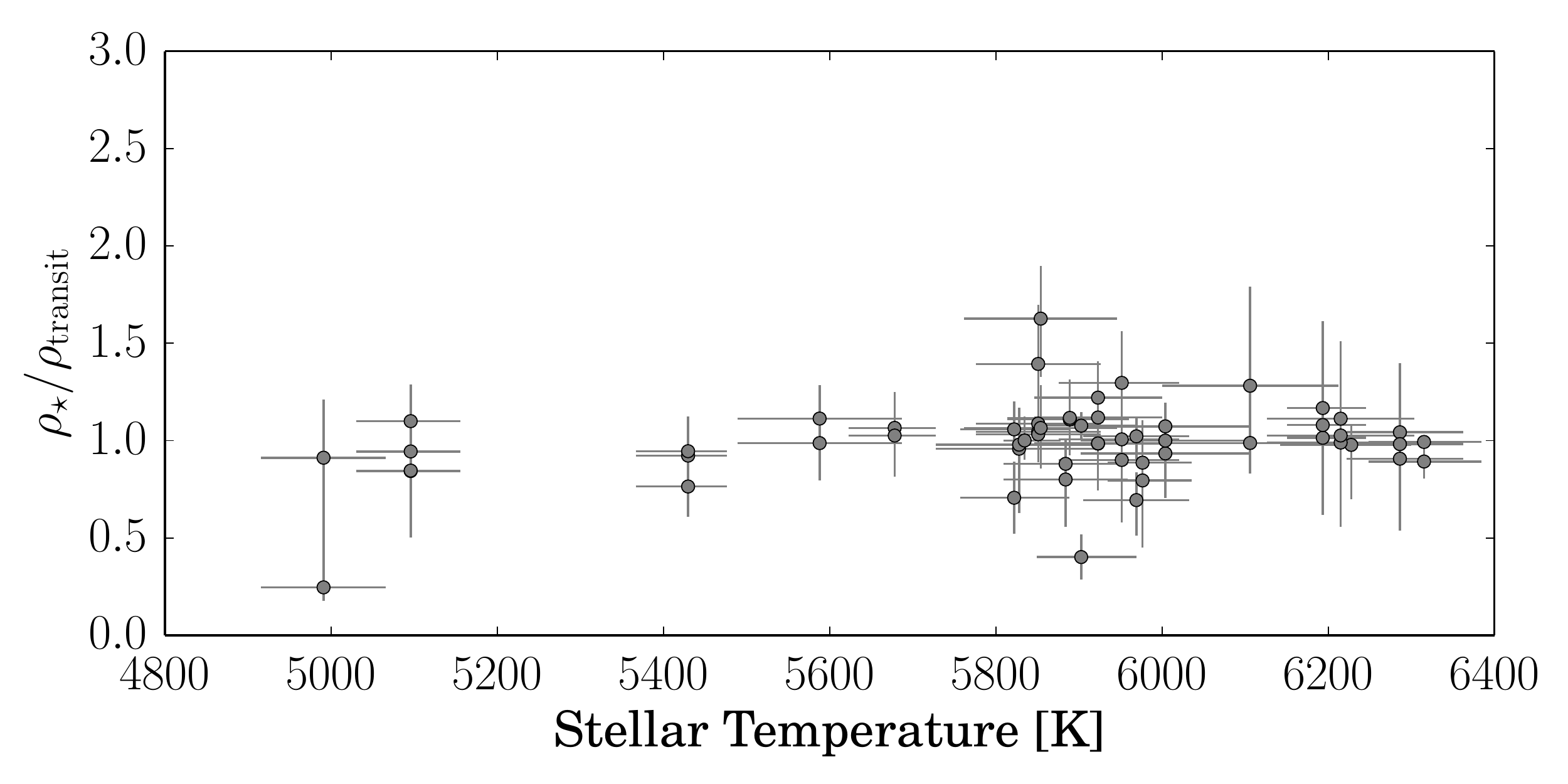}}
\resizebox{0.49\hsize}{!}{\includegraphics{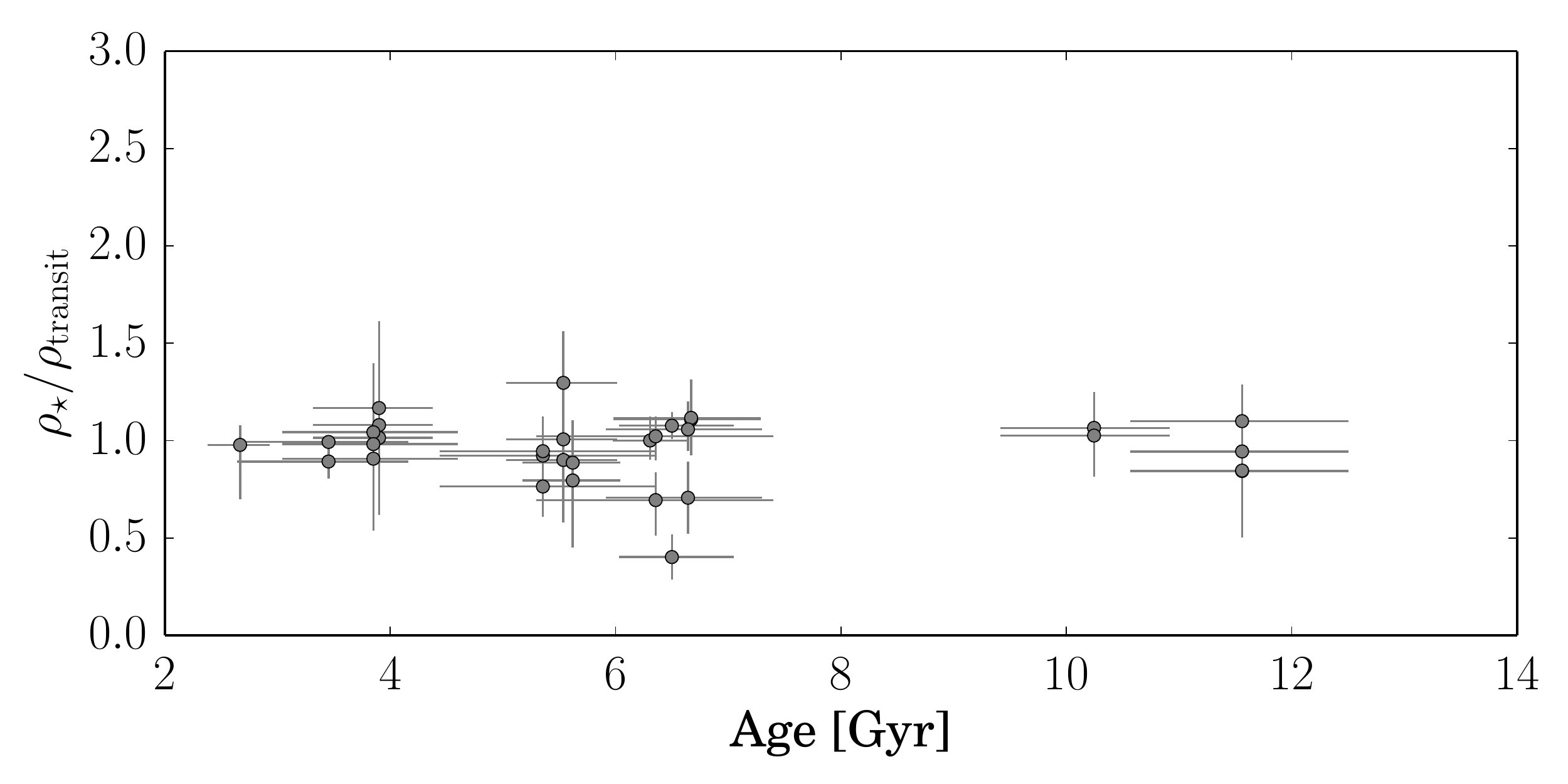}}
\caption{The orbital period and planetary radius of planets in our sample and the stellar temperature and age, plotted versus the measured relative density (where one indicates a circular orbit). \label{fig:parameter_comparison}}
\end{center}
\end{figure*}

We furthermore compared the eccentricity to other parameters and found no correlation (see Figure~\ref{fig:parameter_comparison}). We plot the eccentricity versus the orbital period and planetary radius. We also compare the eccentricity to stellar temperature and stellar age, two parameters which might influence tidal circularisation. We note that ages are only available for part of our sample \citep{silvaaguirre2015}. We see no correlations.

We have determined transit times and (re)derived orbital periods in a way that is robust to outliers (see Section~\ref{sec:methods}). In several cases, we found clear evidence of TTVs. The TTV periods and amplitudes that were included in our analysis are listed in Table~\ref{tab:ttv_table}. For Kepler-103, Kepler-126, Kepler-130 and Kepler-278, these TTVs have not been previously reported. In some cases, hints of small TTVs were found, in which cases we have checked that the decision whether or not to include them had no significant influence on the derived eccentricity, and ultimately did not include any TTVs in the final analysis. All measured times of individual transits are available upon request.
\begin{center} \scriptsize
\begin{table}[ht]
\begin{tabular}{lcccccccc}
 	& TTV period [d]	& TTV amplitude [min]\\
 \hline \hline
Kepler-23b&	433&	21.8\\
Kepler-23c&	472&	23.0\\
Kepler-23d&	362&	22.3\\
Kepler-25b&	327&	3.8\\
Kepler-25c&	348&	1.1\\
Kepler-36b&	449&	166.5\\
Kepler-36c&	446&	116.2\\
Kepler-50b&	2127&	61.0\\
Kepler-50c&	739&	8.7\\
Kepler-103b&	264&	2.7\\
Kepler-103c&	514&	22.2\\
Kepler-126b&	2052&	9.4\\
Kepler-126c&	372&	8.0\\
Kepler-126d&	1052&	6.4\\
Kepler-128b&	413&	55.2\\
Kepler-128c&	355&	103.7\\
Kepler-130b&	2043&	53.8\\
Kepler-130c&	491&	2.8\\
Kepler-278c&	464&	88.5\\
KOI-279.01&	1008&	2.0\\
\end{tabular}
\caption{Overview of the period and amplitude of sinusoidal transit timing variations which were included in the modeling. The transit times and the best model fits are shown in Figure~\ref{fig:oc_partone}. \label{tab:ttv_table}}
\end{table}
\end{center}
%
%

\subsection{Planetary valdidation}
\label{sec:validation}

Multi-planet systems can often be confirmed based on statistical grounds because their multiplicity makes false positive scenarios very unlikely \citep{rowe2014,lissauer2014}. However, this is no longer generally true if the light curve consists of two or more blended stars of different magnitudes, because it can be difficult to tell at which object the transits occur \citep[e.g.][]{vaneylen2014}. 

Transit durations can be used to confirm the planetary nature of transiting candidates when the stellar density of the suspected host star is well known \citep{tingley2005}. However, because eccentricity also influences the transit duration, in general it is difficult to distinguish between eccentric planets and false positives \citep{sliski2014}. Because we find that eccentricities are very small for multi-planet systems, this complication does not arise for these systems and transit durations can be readily used to assess the validity of transit signals in these systems. The transit duration provides a direct estimate of the stellar density, which can be compared to an independent measurement of the stellar density of the stars to determine which of the stars in the aperture hosts the transiting planet(s).

Here we compare the stellar density estimates from the planetary candidates with the asteroseismic density of the brightest star in the system. Any mismatch would be a strong indication that the star is not the true host. A clear agreement is strong evidence the star is the true host, especially if the other star in the system has a very different density. In KOI-5, we cannot draw a clear conclusion because only one of the planets provides meaningful constraints. For KOI-270, we confirm that the transits are caused by true planets which could orbit either KOI-270A or KOI-270B, two stars which are very similar. We confirm that the three planet candidates for KOI-279 are genuine planets orbiting KOI-279A, and finally we also confirm a third planet orbiting Kepler-92 (two other planets were previously confirmed). We discuss these systems in more detail below.

\subsubsection{KOI-5} 

KOI-5 contains two transiting planet candidates which have not been validated or confirmed as true planets. The inner planet candidate has an orbital period of 4.8 days and a 7.9 $R_\oplus$ radius, while the second planet candidate orbits in 7 days and is much smaller ($0.6~R_\oplus$). The reason the candidates have not been validated is the presence of a second, fainter companion star which is physically associated \citep{wang2014,kolbl2015}. We refer to it here as KOI-5B. 

We take a 6\% flux dilution \citep{wang2014,kolbl2015} caused by KOI-5B into account before modeling the planet candidates assuming they orbit the bright star (KOI-5A). The posterior distribution for the inner planet is shown in Figure~\subref{fig:koi5b} and its eccentricity is tightly constrained ($[0.05,0.27]$ within $1-\sigma$). Even within $2-\sigma$, the lower eccentricity bound is 0.04. An alternative way to present this is the relative density, for which a 95\% confidence interval is $[0.72,0.88]$. This implies that if this candidate was a true planet orbiting KOI-5A, it would have a non-zero eccentricity. This is suspicious, in particular given the short orbital period of the candidate, and a possible explanation is that the candidate does not transit KOI-5A but rather KOI-5B instead. Because KOI-5B is much fainter, the candidate would consequentially be larger and might not be planetary in nature.

The second candidate's posterior distribution is given in Figure~\subref{fig:koi5c} and is consistent with a circular orbit around KOI-5A ($e \in [0,0.4]$, relative density $\in [0.45,1.11]$). This could imply that this is a genuine planet orbiting KOI-5A. However, due to the large error bar caused by the small size of the planet, it is difficult to exclude KOI-5B as a host for this candidate without knowing more about this companion star.

\subsubsection{KOI-270} 

KOI-270 contains two transiting planet candidates which transit every 12 and 33 days, thus far unconfirmed. KOI-270 has a stellar companion, separated by only $0.05$ arcsec and with the same magnitude in both $J$ and $Ks$ band \citep{adams2012}. Therefore KOI-270 appears to consist of two very similar stars and we dilute the light curve by a factor two to account for this. We find no evidence for TTVs but note that only limited short-cadence data is available.

After accounting for the flux dilution, the planetary radii are 2.1 and 2.8 $R_\oplus$. Both planets are consistent with circularity ($[0,0.31]$ and $[0,0.28]$, see Figures~\subref{fig:koi270b} and \subref{fig:koi270c}), which means their transits match the asteroseismic stellar density. The relative density intervals are $[0.94,1.29]$ and $[0.80,1.11]$ respectively. Both candidates are likely true planets and KOI-270A is a plausible planet host star. However, with KOI-270B presumably very similar to KOI-270A, we cannot rule out the planets orbit this star instead. In this case the transits would still be caused by genuine planets with similar properties, so we find that KOI-270's two candidates are indeed planets orbiting either KOI-270A or KOI-270B, and the planets are further referred to as Kepler-449b and Kepler-449c.

\subsubsection{KOI-279} 

KOI-279 contains three planetary candidates which transit every 7.5, 15 and 28 days, previously unconfirmed as planets. For the outer planet, a long period TTV signal was clearly measured (see Figure~\ref{fig:oc_partone}) and included, while for the inner two planets no sinusoidal TTVs were included although an increased scatter in the transit times of the middle planet was seen.

The reason for the lack of confirmation for this system is the presence of a second star (at 0.9 arcsec) to which we refer as KOI-279B which is significantly fainter and contributes 6\% flux\footnote{Based on WIYN Speckle images and Keck spectra; Mark Everett and David R. Ciardi, from https://cfop.ipac.caltech.edu}. After removing this flux contamination assuming the candidates orbit KOI-279A and including the TTV signal for the outer planet candidate we proceed to measure the orbital eccentricity. The posterior distributions are reported in Figures \subref{fig:koi279b}, \subref{fig:koi279c} and \subref{fig:koi279d}. 

We find the outer two planet candidates' orbits to be tightly constrained to be circular or close to circular, while the inner planet similarly appears close to circular but is less tightly constrained due to its small size. This implies that the stellar density derived from the candidates' transits is consistent with the asteroseismic stellar density \citep{huber2013}, with relative densities $[0.88,1.08]$, $[0.90,1.13]$ and $[0.56,1.51]$ respectively. The range of periods and the TTV signal is further evidence that the planets orbit the same star. We find that the three candidates are indeed planets orbiting KOI-279(A), and they are subsequently named Kepler-450b, Kepler-450c and Kepler-450d.

\subsubsection{Kepler-92 (KOI-285)} 

Kepler-92 contains three planets, of which the inner two (13 and 26 day periods) were validated based on their TTV signal \citep{xie2014}. The eccentricity of the planets could not be determined due to a mass-eccentricity degeneracy \citep{xie2014}. Due to a limited amount of short cadence data, we pick up only a hint of the TTVs and we choose not to include them. 

The planets are consistent with circularity ($[0,0.27]$ and $[0,0.25]$ at 68\% confidence, respectively). Our eccentricity posteriors for the planets are shown in Figure~\subref{fig:kepler92b} (Kepler-92b), Figure~\subref{fig:kepler92c} (Kepler-92c).

There's a third planetary candidate observed transiting every 49 days, which has not yet been validated or confirmed as true planet orbiting Kepler-92. We model the transit under the assumption that it does. We find a modal eccentricity value of 0.07 (and a 68\% confidence interval of $[0.03,0.41]$, see Figure~\subref{fig:kepler92d}). Adaptive optics observations have revealed two other stars at $1.4$ and $2.3$ arcsec, the brightest is estimated to be 5.6 magnitudes fainter in the \textit{Kepler} bandpass \citep{adams2012} so that their flux contributions are negliglible. Given the planet candidate's period and similar size to the two confirmed planets, as well as their agreement with the stellar density for (close to) circular orbits, all planets are likely to orbit the same star (Kepler-92), and KOI-285.03 is subsequently named Kepler-92d.

\section{Discussion}
\label{sec:discussion}

We discuss two important implications of our eccentricity distribution here. In Section~\ref{sec:habitability} we discuss the influence of orbital eccentricity on habitability. In Section~\ref{sec:occurrence_rates} the consequences of the orbital eccentricity distribution on exoplanet occurrence rates is discussed.
 
\subsection{Habitability}
\label{sec:habitability}

Earth's orbit is almost circular with a current eccentricity ($e$) of 0.017. The influence of the orbital eccentricity on habitability has been investigated using planet climate models \citep{williams2002,dressing2010}. Our results allow one of the first looks at the orbital eccentricities of small and potentially rocky planets and indicate that low eccentricities are the rule. In fact we can not find a clear candidate for a planet on an elliptic orbit among the 74 planets in our sample. The few planets with densities away from unity Figure~\ref{fig:relativedensities} also have the largest uncertainties (See Appendix~\ref{sec:individual_posterior_discussion} for a discussion of individual systems and Table~\ref{tab:parameter_table} for an overview).

If this extends to planets on longer orbital periods or to planets orbiting lower mass stars (the planets in our sample are all outside the habitable zone) then this influences habitability in two ways. Planets on circular orbits have more stable climates than planets on eccentric orbits which can have large seasonal variations, even though large oceans might temper the climate impact of moderate eccentricities \citep{williams2002}. Secondly the location of the habitable zone itself depends on the orbital eccentricity. For moderately eccentric orbits the outer edge of the habitable zone is increased \citep{spiegel2010,dressing2010,kopparapu2013}, i.e.\ moderately eccentric planets could be habitable further away from the host star than planets on circular planets. However our results suggest that this might not occur.

\subsection{Occurrence rates}
\label{sec:occurrence_rates}

The eccentricity distribution is a key parameter needed to reliably estimate planetary occurrence rates inferred from transit surveys. This is because the transit probability depends on eccentricity \citep{barnes2007}. Planets on orbits with $e = 0.5$ are 33\% more likely to transit, and in the extreme case of HD\,80606b ($e = 0.92$) \citep{naef2001} the transit probability increased by 640\%. A recent estimate based on the eccentricity distribution derived from RV observations shows that the overall transit probability changes by 10\% \citep{kipping2014_occurencerate}. This can significantly change the planet occurrence estimate, e.g.\ the number of planets smaller than $4~R_\oplus$ around cool stars is estimated to 3\% precision before the effect of eccentricity is taken into account \citep{dressing2013}. Our analysis shows that neglecting eccentricity is a valid assumption when considering transiting multiple planet systems.

Beyond the influence on the global occurrence rate the eccentricity distribution also influences the relative occurrence between different types of planets. Because single more massive planets show a wider range of eccentricities than multi-planet systems with smaller planets, the occurrence of larger planets is overestimated compared to smaller planets. These effects are important when comparing occurrence rates of different types of planets but have so far not been taken into account \citep{petigura2013,foremanmackey2014}.

\section{Conclusions}
\label{sec:conclusions}

We have measured the eccentricity distribution of 74 planets orbiting 28 stars, making use of photometry alone. For this we made use of the influence of eccentricity on the duration of planetary transits. Several complications are avoided by carefully selecting this sample. Planetary false positives and third light blending are sidestepped in our selection of (primarily) confirmed multi-transiting planet systems around bright host stars. Issues due to inaccurate stellar parameters are overcome owing to the power of asteroseismology to determine stellar densities and other stellar parameters. The use of short cadence data, newly derived orbital periods and a careful analysis of possible TTVs prevent a bias towards high impact parameters.

We find that most of the systems we considered are likely to reside on orbits which are close to circular. The eccentricity is well-described by a Rayleigh distribution with $\sigma = 0.049 \pm 0.013$. This is distinctly different from RV measurements \citep{wright2009,latham2011,mayor2011}, possibly due to the smaller planets in our sample. It is similar to low eccentricities reported for TTV systems \citep{hadden2014} and to the eccentricities found in the solar system.

Our findings have important consequences:

\begin{itemize}
 \item Constraining orbital eccentricities is an important step towards understanding planetary formation. Several mechanisms for eccentricity excitation and damping have previously been suggested based on evidence of eccentric orbits from RV observations. If planet-planet scattering \citep{ford2008_planetplanet} is important, it appears to result in low eccentricity in systems with multiple planets, at least for those systems with low mutual inclinations. This could be related to the small planet size, the planetary multiplicity or the orbital distance, or a combination of these.
 
 \item While no Earth twins are present in our sample, our findings cover planets with small radii and a wide range of orbital periods. It seems plausible that low eccentricity orbits would also be common in solar system analogues, influencing habitability and the location of the habitable zone.
 
 \item Orbital eccentricities influence planet occurrence rates derived from transit surveys because eccentric planets are more likely to transit. Our findings indicate that the transit probability of multi-planet systems is different from that of systems with single, massive planets.
 
 \item We have compared the individual eccentricity estimates with accurately determined stellar parameters, such as the stellar temperature \citep{huber2013,silvaaguirre2015} and age \citep{silvaaguirre2015}, and found no trend. It would be interesting to compare the eccentricity measurements with measurements of stellar inclination, which might be possible using asteroseismology \citep[e.g.][]{chaplin2013,vaneylen2014,lund2014} for some stars in our sample.
 
 \item With circular orbits common in systems with multiple transiting planets, the stellar density can be reliably estimated from transit observations of such systems. This can be used to characterize the host stars of such systems and to rule out planetary false positives. We use this to validate planets in two systems with planetary candidates (KOI-270, now Kepler-449, and KOI-279, now Kepler-450), as well as one planet in a system with previously known planets (KOI-285.03, now Kepler-92d). 
 
 \item We anticipate that the methods used here will be useful in the context of the future photometry missions \textit{TESS} \citep{ricker2014} and \textit{PLATO} \citep{rauer2014}, both of which will allow for asteroseismic studies of a large number of targets. Transit durations will be useful to confirm the validity of transit signals in compact multi-planet systems, in particular for the smallest and most interesting candidates that are hardest to confirm using other methods. For systems where independent stellar density measurements exist the method will also provide further information on orbital eccentricities.
 
\end{itemize}

\acknowledgements

We are grateful to Victor Silva Aguirre for making available an early version of the asteroseismic parameters to us. We thank Mikkel N. Lund for suggestions in the early stage of this work, and Josh Winn for fruitful discussions as well as suggestions on the manuscript. We are grateful for the many valuable suggestions by the anonymous referee, which have significantly improved this manuscript. We thank David Kipping, Dan Fabrycky and Daniel Huber for insightful comments and suggestions. Part of this manuscript was written at MIT and I appreciate the hospitality of the researchers and staff at the Institute for Astrophysics and Space Research. This research made use of the Grendel HPC-cluster for computations. Funding for the Stellar Astrophysics Centre is provided by The Danish National Research Foundation (Grant agreement no.: DNRF106). The research is supported by the ASTERISK project (ASTERoseismic Investigations with SONG and Kepler) funded by the European Research Council (Grant agreement no.: 267864). We acknowledge ASK for covering travels in relation to this publication.

\bibliographystyle{bibstyle}
\bibliography{vaneylen_references}


\appendix

\begin{figure*}[ht]
\centering
\resizebox{\hsize}{!}{\includegraphics{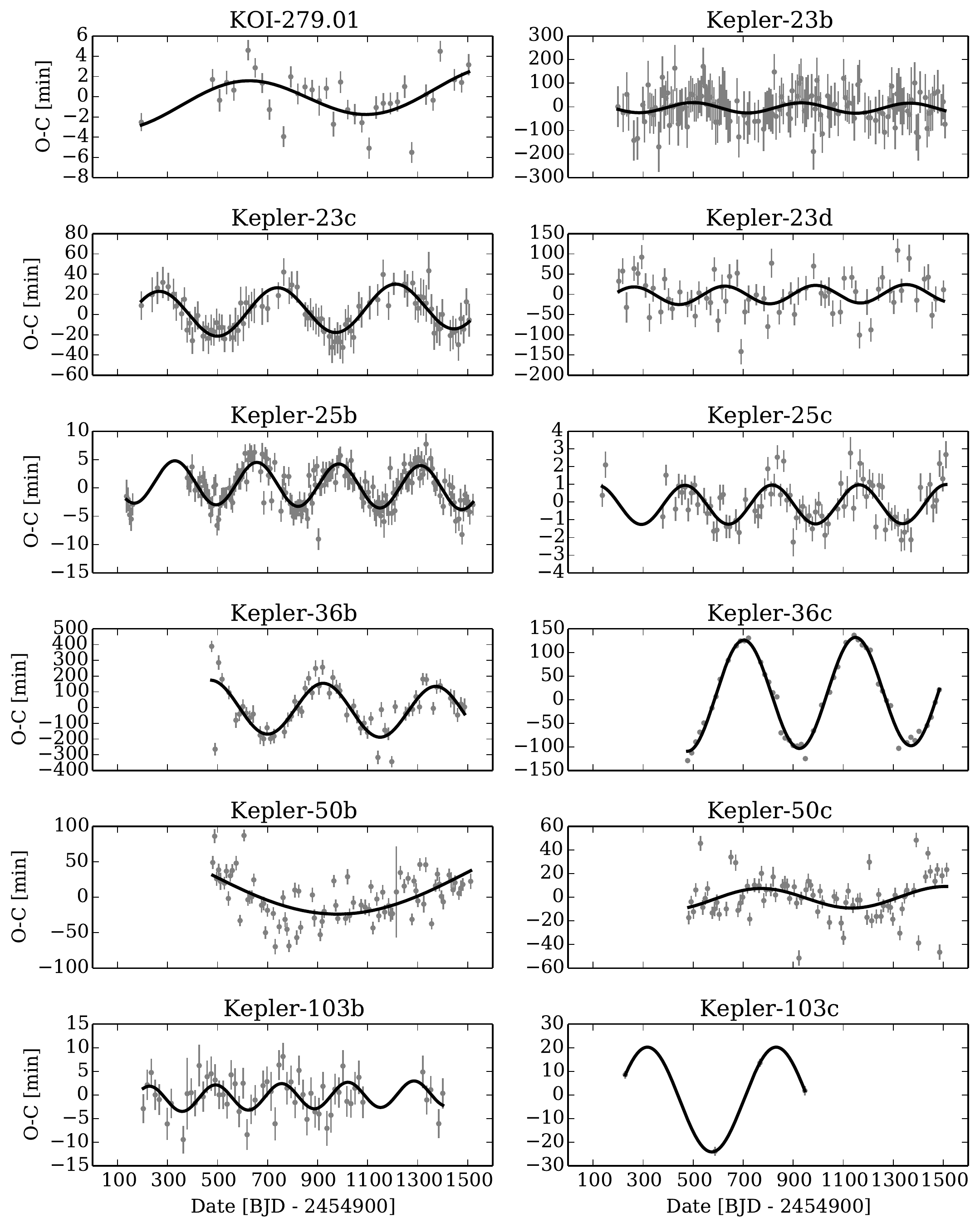}}
\caption{The observed minus calculated transit times are shown for systems with detected TTVs. A sinusoidal fit to the O-C times is shown.\label{fig:oc_partone}
}
\end{figure*}

\begin{figure*}[ht]
 \ContinuedFloat
\centering
\resizebox{\hsize}{!}{\includegraphics{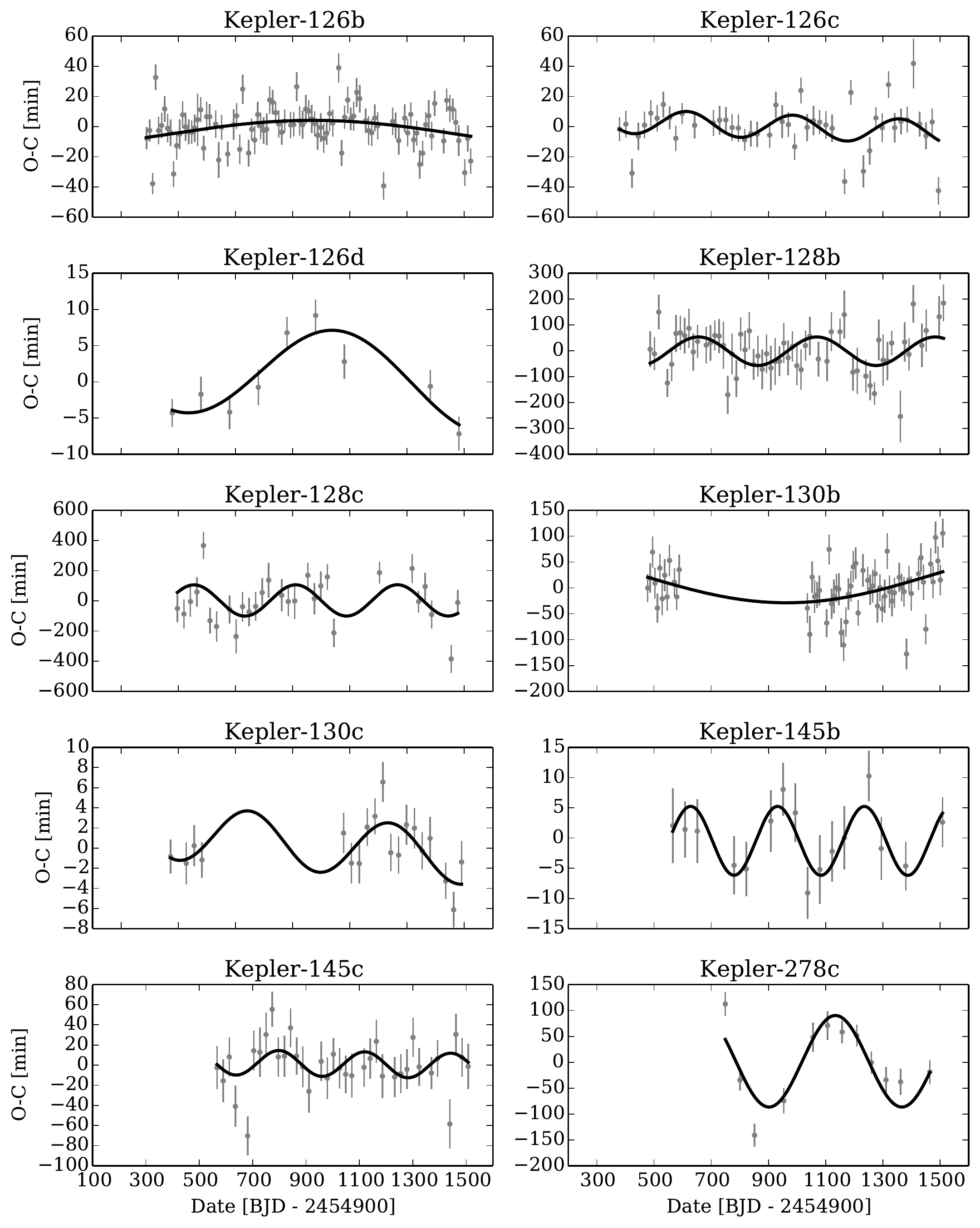}}
\caption{(Continued) The observed minus calculated transit times are shown for systems with detected TTVs. A sinusoidal fit to the O-C times is shown.\label{fig:oc_parttwo}
}
\end{figure*}

\section{Individual planet systems}
\label{sec:individual_posterior_discussion}

Here we discuss the eccentricity posterior measurements for each star-planet system in our sample in detail. Our posterior distributions follow the convention of the illustration in Figure~\ref{fig:illustration_b_ecc} to separate eccentricity measurements with angles in [0,180]$^\circ$ from those with $\omega \in [180,360]^\circ$, where the former are encoded with a minus sign, and we show the correlation with $b$ for reasons discussed in Section~\ref{sec:methods}. All final values are summarized in Table~\ref{tab:parameter_table}.

In what follows, when best values are reported, they are the modal value of the distribution. When confidence intervals are reported, they represent the 68\% highest probability density confidence interval unless stated otherwise.

We note that for individual systems, an unknown angle of periastron $\omega$ influences the uncertainty of the measurement of $e$ as discussed in Section~\ref{sec:results_omega} and consequentially the uncertainties on measurements of individual planets are larger than when looking at the ensemble of planets as a whole (see Section~\ref{sec:multis_loweccentricity}). \\

\textbf{Kepler-10 (KOI-72)} contains two planets. Kepler-10b \citep{batalha2011} is \textit{Kepler}'s first rocky planet and has a short 0.88 day period. Kepler-10c is a Super-Earth in a 45 day orbit \citep{fressin2011}. A detailed asteroseismic analysis also revealed that it is one of the oldest exoplanet systems (10.41 $\pm$ 1.36 Gyr) \citep{fogtmannschulz2014}. 

We find no evidence of TTVs and present our eccentricity distributions in Figure~\subref{fig:kepler10b} and \subref{fig:kepler10c}. Due to the small size of the planets, the eccentricity distribution is degenerate with impact parameter. However, low eccentricities are clearly favored for both planets. For Kepler-10b, a circular orbit is expected because of tidal circularisation; we find $e \in [0,0.19]$. For Kepler-10c, the mode of the eccentricity is 0.05, the 68\% confidence interval is $[0,0.22]$. Despite Kepler-10c's small size, the planet was detected using RV measurements due to its high density  \citep{dumusque2014}, and the RV observations favor a low eccentricity ($e \in [0,0.14]$). Kepler-10 is the only system in our sample for which RV eccentricity measurements are available.\\

\textbf{Kepler-23 (KOI-168)} contains three planets which were confirmed making use of their timing variations \citep{ford2012}. With about three times more data available now, we reanalyze the transit times and fit a sinusoidal TTV model to the measurements. A TTV signal is visible for all three planets , which orbit in 7, 10 and 15 days around the host star (see Figure~\ref{fig:oc_partone}). The observed TTV period of 472 days for Kepler-23c matches the predicted 470 days for a 3:2 period ratio with Kepler-23b \citep{ford2012}. 

After removing the TTV signal, we model the planetary transits. The planets are small (1.7, 3.1 and 2.2 $R_\oplus$) and consequentially, a degeneracy between eccentricity and impact parameter is observed. Nevertheless, the eccentricities are likely low, with modal values of 0.06, 0.02 and 0.08, respectively. The $1-\sigma$ confidence intervals are consistent with circularity, i.e. $[0,0.31]$ (Figure~\subref{fig:kepler23b}), $[0,0.39]$ (Figure~\subref{fig:kepler23c}) and $[0,0.31]$ (Figure~\subref{fig:kepler23d}). The TTVs were fitted using an assumption about circularity but the observed TTV amplitude was larger than expected and could be caused by (moderately) eccentric orbits \citep{ford2012}.\\

\textbf{Kepler-25 (KOI-244)} contains two planets in a near 2:1 resonance, discovered due to their anti-correlated TTVs \citep{steffen2012}. A third, non-transiting planet was discovered with RV observations \citep{marcy2014}. The latter is a large planet (minimum mass 90 $\pm$ 14 $M_\oplus$) in a long 123 day orbit, best-fitted with an eccentricity of $0.18 \pm 0.10$ \citep{marcy2014}. The RV observations point to a low density for the transiting planets but do not have the sensitivity to measure eccentricities \citep{marcy2014}. Due to the fast stellar rotation Kepler-25 has been a target for Rossiter-McLaughlin (RM) observations despite the small transit depth, and  the star was found to be closely aligned ($2 \pm 5$ deg) with the plane of the transiting planets \citep{albrecht2013}. However rotational splittings of the asteroseismic signal of the star find $i_\textrm{star} = 65.4^{10.6}_{-6.4}$, which indicates a slight misalignment \citep{benomar2014}.

After removing the small TTVs (3.8 and 1.1 minute amplitudes, respectively; see Figure~\ref{fig:oc_partone}), we find both planets' eccentricity to be tightly constrained. Both orbits are consistent with circularity, and respectively have $e \in [0,0.06]$ and $e \in [0,0.13]$ to 68\% confidence. The posteriors are shown in Figures~\subref{fig:kepler25b} and \subref{fig:kepler25c}. From TTVs a low eccentricity for the planet pair is also measured \citep{wu2013}.\\

\textbf{Kepler-37 (KOI-245)} contains three small planets \citep{barclay2014}. The innermost one is the smallest known exoplanet, similar in size to the moon. We refine its radius to $0.354 \pm 0.014~R_\oplus$. We find a circular orbit is likely, with a model eccentricity of 0.08 and a 68\% confidence interval of $[0,0.29]$ (Figure~\subref{fig:kepler37b}). The initial analysis \citep{barclay2014} yielded measurements of $e \cos \omega$ and $e \sin \omega$ which were consistent with circularity but were less constraining.

Kepler-37c's radius is only $0.7~R_\oplus$ and we find similar eccentricity constraints as for Kepler-37b (Figure~\subref{fig:kepler37c}). The outer transiting planet possibly has a small but non-zero eccentricity ($e \in [0.05,0.22]$, Figure~\subref{fig:kepler37d}), although the orbit is also consistent with circularity within $2\sigma$. 

RV follow-up observations did not detect any of the planet signals and yield only loose upper limits for the planetary mass; no additional non-transiting planets were discovered \citep{marcy2014}.\\

\textbf{Kepler-65 (KOI-85)} contains three small planets with short-periods (2, 6 and 8 day periods) which were previously validated \citep{chaplin2013}. A TTV signal in Kepler-65d was detected but it was noted that uncertainties in the transit times might be underestimated \citep{chaplin2013}. We find no evidence for TTVs in any of the planets. The rotational splitting in the asteroseismic signal was also analyzed and the host star was found to be aligned with the orbital plane of the planets \citep{chaplin2013}.

We find the eccentricity of all three planets to be consistent with circularity. The $68\%$ confidence intervals are $[0,0.18]$, $[0,0.12]$ and $[0,0.27]$ for Kepler-65b, Kepler-65c and Kepler-65d respectively. The full distributions are shown in Figures~\subref{fig:kepler65b}, \subref{fig:kepler65c} and \subref{fig:kepler65d}.\\

\textbf{Kepler-68 (KOI-246)} contains two transiting planets, Kepler-68b and Kepler-68c \citep{gilliland2013} on 5 and 10 day orbits. An additional large non-transiting planet (Kepler-68d) in a 625 day orbit with an eccentricity of $0.10 \pm 0.04$ was discovered \citep{marcy2014}. The inner transiting planet has a planet mass of 5.97 $M_\oplus$ \citep{marcy2014}. 

The transit duration was previously compared to a stellar density estimate and both planets were consistent with circularity, although the outer (transiting) planet could have an eccentricity of up to 0.2 \citep{gilliland2013}. We find the inner planet to have a tightly constrained orbit ($e \in [0,0.13]$), consistent with circularity. Due to the small size ($< 1~R_\oplus$) of the outer transiting planet, its eccentricity is largely unconstrained and correlated with its impact parameter. The eccentricity distributions are shown in Figures~\subref{fig:kepler68b} and \subref{fig:kepler68c}. We find no evidence of TTVs.\\

\textbf{Kepler-100 (KOI-41)} has three planets which were validated based on RV measurements \citep{marcy2014} which showed no companion stars. None of the planets were detected in RV, but upper limits on the planetary mass could be placed.

We find a hint of a TTV signal for the inner two planets (Kepler-100b and Kepler-100c) but do not include it in our analysis. Their orbital periods are 6.8 and 12.8 days. The inner planet is the smallest ($1.3~R_\oplus$) and a moderate eccentricity constraint is placed ($e \in [0,0.4]$, Figure~\subref{fig:kepler100b}). Kepler-100c is $2.2~R_\oplus$ and has its orbital eccentricity within $[0.01,0.17]$ (Figure~\subref{fig:kepler100c}).  

Kepler-100d ($1.9~R_\oplus$, $P = 35$ days) peaks at a significant eccentricity (0.38). However, care must be taken when interpreting this value, because of the large degeneracy with impact parameter (see Figure~\subref{fig:kepler100d}). Depending on the impact parameters, different eccentricities are possible, although very large eccentricities ($> 0.65$) are outside the $2\sigma$ confidence interval.\\

\textbf{Kepler-103 (KOI-108)} contains two planets which were validated based on RV measurements \citep{marcy2014}, although only upper limits on the masses could be placed and their eccentricities could not be determined. The inner planet orbits the star in 16 days and has a $3.5~R_\oplus$ radius. The outer planet is bigger ($5.3~R_\oplus$) and has a 180 day period. Consequentially, only 4 transits were observed. Nevertheless, a clear TTV signal is measured for both planets (see Figure~\ref{fig:oc_parttwo}). The TTVs have periods of 264 and 514 days and amplitudes of 2.7 and 22.2 minutes, respectively. To our knowledge these TTVs were previously undetected, although it was noted that this interesting system warrants a detailed TTV search \citep{marcy2014}. 

The eccentricity posteriors are shown in Figures~\subref{fig:kepler103b} and \subref{fig:kepler103c} and the distributions peak at eccentricities 0.025 and 0.027 respectively, while $68\%$ confidence intervals are $[0,0.21]$ and $[0,0.20]$.\\

\textbf{Kepler-107 (KOI-117)} contains four planets which were validated as part of a large multi-transiting planet validation effort \citep{rowe2014}, based on a statistical framework \citep{lissauer2014}. The planets orbit on short periods of 3, 5, 8 and 15 days and are all small (1-3 $R_\oplus$). We find no evidence for TTVs. 

Despite their small sizes, we find good constraints on the eccentricity; to 68\% confidence, they are: $[0,0.22]$ (Kepler-107b, Figure~\subref{fig:kepler107b}), $[0,0.28]$ (Kepler-107c, Figure~\subref{fig:kepler107c}), $[0,0.39]$ (Kepler-107d, Figure~\subref{fig:kepler107d}) and $[0,0.19]$ (Kepler-107e, Figure~\subref{fig:kepler107e}). \\

\textbf{Kepler-108 (KOI-119)} contains two transiting planets that were validated \citep{rowe2014} based on a statistical framework \citep{lissauer2014}. The planets orbit on relatively long periods of 50 and 190 days. Only seven transits were observed in short cadence for Kepler-109b, and only two for Kepler-109c. The constraints on their eccentricity are shown in Figure~\subref{fig:kepler108b} and \subref{fig:kepler108c}. We find that the inner, giant planet, is almost certainly slightly eccentric ($e \in [0.1,0.41]$). The outer planet is consistent with circularity ($e \in [0,0.23]$). \\

\textbf{Kepler-109 (KOI-123)} contains two transiting Super-earth planets. Kepler-109b and Kepler-109c orbit on periods of 6.5 and 21 days and were validated statistically \citep{rowe2014,lissauer2014}. RV constraints rule out a rocky composition for the planets \citep{marcy2014}.

The posterior eccentricity distributions are given in Figures~\subref{fig:kepler109b} and \subref{fig:kepler109c}. The eccentricity of Kepler-109b shows a degeneracy with impact parameter, but is nevertheless constrained to $[0.01,0.31]$ with $68\%$ confidence. Kepler-109c has a modal eccentricity of 0.025 and has a $68\%$ confidence interval $[0,0.22]$.\\

\textbf{Kepler-126 (KOI-260)} contains three transiting planets which were validated statistically \citep{rowe2014,lissauer2014}. We find evidence of long period TTVs in all three planets which to our knowledge have not been previously reported. Our best sinusoidal fits have periods of 2052, 372 and 1052 days, respectively (see Figure~\ref{fig:oc_parttwo}). 

The inner two planets are small ($1.5~R_\oplus$) and orbit in 10 and 21 days. Their eccentricities are constrained to $[0,0.16]$ and $[0,0.36]$. The outer planet has a period of 100 days and a radius of 2.5~$R_\oplus$. Its eccentricity is tightly constrained to $[0,0.11]$. The eccentricity distributions are shown in Figures~\subref{fig:kepler126b}, \subref{fig:kepler126c} and \subref{fig:kepler126d}.\\

\textbf{Kepler-127 (KOI-271)} contains three planets on 14, 29 and 49 day orbits (1.4, 2.4 and 2.7 $R_\oplus$) which were validated statistically \citep{rowe2014,lissauer2014}. We find marginal evidence for TTVs but do not include them in our analysis. The small size of the inner planet causes a large degeneracy between $e$ and $b$ causing it to be essentially unconstrained (see Figure~\subref{fig:kepler127b}). For the other two planets, we find modal values of 0.03 and 0.1 and $1-\sigma$ confidence intervals $[0,0.17]$ (Kepler-127c, Figure~\subref{fig:kepler127c}) and $[0,0.31]$ (Kepler-127d, Figure~\subref{fig:kepler127d}).\\

\textbf{Kepler-129 (KOI-275)} contains two planets which have 16 and 82 day periods. They were validated statistically \citep{rowe2014,lissauer2014}. Only limited amount of short cadence observations are available, and respectively only 12 and 2 transits are available. No TTV evidence was found. Both eccentricity posterior distributions (Figure~\subref{fig:kepler129b} and \subref{fig:kepler129c}) point towards circular orbits: $e \in [0,0.25]$ and $[0,0.35]$. \\

\textbf{Kepler-130 (KOI-282)} contains three transiting planets on orbits of 8, 27 and 87 days, which were validated statistically \citep{rowe2014,lissauer2014}. Carefully measuring their transit times we detect TTVs in the inner two planets. The best sinusoidal fit to the transit times is shown in Figure~\ref{fig:oc_parttwo} and the TTVs have periods of 2000 and 500 days (see Table~\ref{tab:ttv_table}). To our knowledge these TTVs were not previously reported. 

After removing the TTV signal we model the transits. The inner two planets (Figures~\subref{fig:kepler130b} and \subref{fig:kepler130c}) orbit in circular or low-eccentricity orbits ($e \in [0,0.24]$ and $[0,0.28]$, respectively). For the outer planet, the impact parameter and the eccentricity are unconstrained and correlated (see Figure~\subref{fig:kepler130d} due to the small transit depth and we caution against blindly using the modal value: within $2\sigma$, all eccentricities between 0 and 0.89 are allowed. \\

\textbf{Kepler-145 (KOI-370)} contains two transiting planets validated statistically \citep{rowe2014,lissauer2014}. The planets were independently confirmed \citep{xie2014} based on a mutual TTV signal. The TTV signal in the inner planet is only marginally significant, but has a similar period to that of the outer one, and we choose to include it (see Figure~\ref{fig:oc_parttwo}).

The planets orbit on 23 and 43 day periods. The inner planet's transits are too shallow for any meaningful constraints on eccentricity, which is heavily correlated with impact parameter (see Figure~\subref{fig:kepler145b}). The outer planet favors circular orbits or small eccentricties, as shown in Figure~\subref{fig:kepler145c}, with a 68\% confidence interval of $[0,0.22]$.\\

\textbf{Kepler-197 (KOI-623)} contains four transiting planets with periods 5, 10, 15 and 25 days, which were validated statistically \citep{rowe2014,lissauer2014}. All planets are small (1-1.2~$R_\oplus$). We find no evidence of TTVs for any of the planets. 

We find low eccentricities or circular orbits for the three inner planets: $[0,0.27]$ (Kepler-197b, Figure~\subref{fig:kepler197b}), $[0,0.22]$ (Kepler-197c, Figure~\subref{fig:kepler197c}) and $[0,0.24]$ (Kepler-197d, Figure~\subref{fig:kepler197d}). The outer planet shows a small but non-zero eccentricity with a mode of 0.27 and a 68\% confidence interval $[0.21,0.63]$. Given the small transit depth (and large eccentricity error bar), some caution is required, as unseen TTVs or a misidentified period could cause this measurement; however, we find no evidence of this to be the case. The posterior distribution is shown in Figure~\ref{fig:kepler197e}.\\

\textbf{Kepler-278 (KOI-1221)} contains two transiting planets with periods of 30 and 51 days, validated statistically \citep{rowe2014,lissauer2014}. We include a TTV signal detected in the outer planet (see Figure~\ref{fig:oc_parttwo}), but we note that this is a giant star ($2.9~R_\odot$) \citep{huber2013} and the light curve shows significant variability, most likely due to stellar spots. 

Consequentially, it is difficult to measure the planetary transits for this star. The inner planet is most likely close to circular, with a modal value of 0.03 and a 68\% confidence interval at $[0,0.36]$ (see Figure~\subref{fig:kepler278b}). The outer planet could be eccentric, but we caution against overinterpreting this result due to the large degeneracy with impact parameter (Figure~\subref{fig:kepler278c}) and the poor quality of the transit light curves.\\

\textbf{Kepler-338 (KOI-1930)} contains four planets, with orbital periods of 9, 13, 24 and 44 days, and was validated statistically \citep{rowe2014,lissauer2014}. We found no convincing TTV signal. The inner planet (which somewhat confusingly is called Kepler-338e) is $1.6~R_\oplus$ and its eccentricity is constrained to $[0,0.27]$ (Figure~\subref{fig:kepler338e}). The other three planets are all about $2.5~R_\oplus$ and have similar eccentricity constraints: $[0,0.31]$ (Kepler-338b, Figure~\subref{fig:kepler338b}), $[0,0.26]$ (Kepler-338c, Figure~\subref{fig:kepler338c}) and $[0,0.24]$ (Kepler-338d, Figure~\subref{fig:kepler338d}).\\

\textbf{Kepler-444  (KOI-3158)} contains five transiting planets which all orbit the host star in a period less than 10 days. This highly interesting system was characterised and validated very recently  \citep{campante2015}. All five planets are small with radii between 0.38 and 0.68 $R_\oplus$. 

We find no evidence of TTVs and find a clear degeneracy between $b$ and $e$ due to the small transit depths. The four inner planets all have 68\% confidence intervals consistent with zero eccentricity: $[0,0.29]$ (Kepler-444b, Figure~\subref{fig:kepler444b}), $[0,0.29]$ (Kepler-444c, Figure~\subref{fig:kepler444c}), $[0,0.28]$ (Kepler-444c, Figure~\subref{fig:kepler444c}), $[0,0.36]$ (Kepler-444d, Figure~\subref{fig:kepler444d}) and $[0,0.29]$ (Kepler-444e, Figure~\subref{fig:kepler444e}). The outer planet has a modal value of 0.58, however, we caution against overinterpreting this due to the large degeneracy with the impact parameter and the large error bar.

\section{Systems which were excluded}
\label{sec:excluded_systems}
Several systems were part of our initial sample but were excluded from the final sample because the eccentricities could not reliably be modeled. They are presented here. In all cases, the presence of TTVs which could not be adequately removed using a sinusoidal model is the cause of their exclusion.\\

\textbf{Kepler-36 (KOI-277)} consists of two planets in very close orbits with periods of 13.8 and 16.2 days \citep{carter2012}. Their densities are very different, with the inner planet rocky while the outer planet has a lower density \citep{carter2012}. With more data available we reanalyze the transit times and their large TTV signal and find amplitudes of around three and two hours respectively (see Figure~\ref{fig:oc_partone}). After removing the TTVs, significant residuals in the timing variations remain present, particularly for Kepler-36c. They indicate that a sinusoidal model may not be adequate to fully remove the large TTV signal, which is perhaps unsurprising given the close proximity of the two planets. An alternative to the sinusoidal model is to directly use the measured times of individual transits; however, we find that this typically leads to `overfitting', smearing out the folded transits by including noise on individual timing measurements.

It is possible the eccentricity can be determined from the transits if a full dynamical model is employed, predicting the times of transits. This is outside the scope of this work.\\

\textbf{Kepler-50 (KOI-262)} is a two-planet system with neighbouring orbits on a near 6:5 resonance (7.8 and 9.3 day periods). It was validated owing to the planets' mutual TTV signal \citep{steffen2013}. The system was later analyzed and the host star was found to be well-aligned with the orbital plane of the planets \citep{chaplin2013}.

The TTV signal is shown in Figure~\ref{fig:oc_partone} but shows significant residuals, indicating a sinusoidal model might not be adequate. We note that this case is similar to Kepler-36 with two planets in high order resonance orbits. A full dynamical model seems to be required to adequately model the TTVs and the eccentricity, but this is outside the scope of this work and we further exclude Kepler-50 from our sample.\\

\textbf{Kepler-56 (KOI-1241)} contains two transiting planets with mutual TTVs \citep{steffen2013}. The planet's host star was found to be misaligned compared to the planetary orbital plane \citep{huber2013}, which triggered further analysis \citep{li2014}. The limited amount of data, the data quality and the small size of the planets make it difficult to measure the TTV signal and we do not include this planet in our further analysis.\\

\textbf{Kepler-128 (KOI-274)} consists of two small planets which orbit close to a 2:3 resonance (periods of 15 and 22 days). A TTV signal was previously detected \citep{xie2014} and we show our best sinusoidal fit in Figure~\ref{fig:oc_parttwo}. Due to the small size of the planets their individual times are measured poorly, and it is difficult to measure the TTV signal correctly.

\clearpage

\begin{figure*}[t]\begin{widepage}

  \subfloat[KOI-5b\label{fig:koi5b}]{\includegraphics[width=0.33\linewidth]{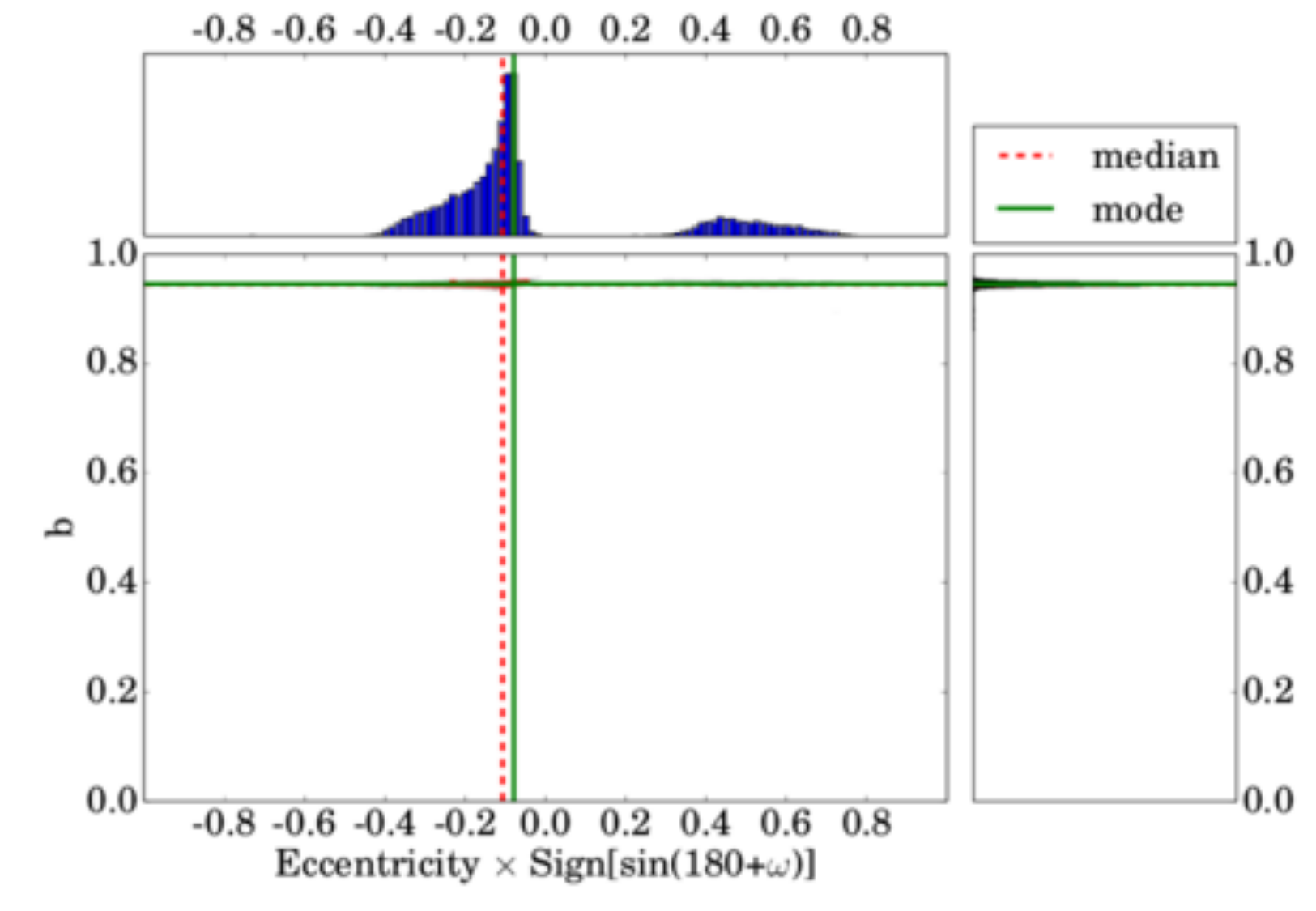}}  
  \subfloat[KOI-5c\label{fig:koi5c}]{\includegraphics[width=0.33\linewidth]{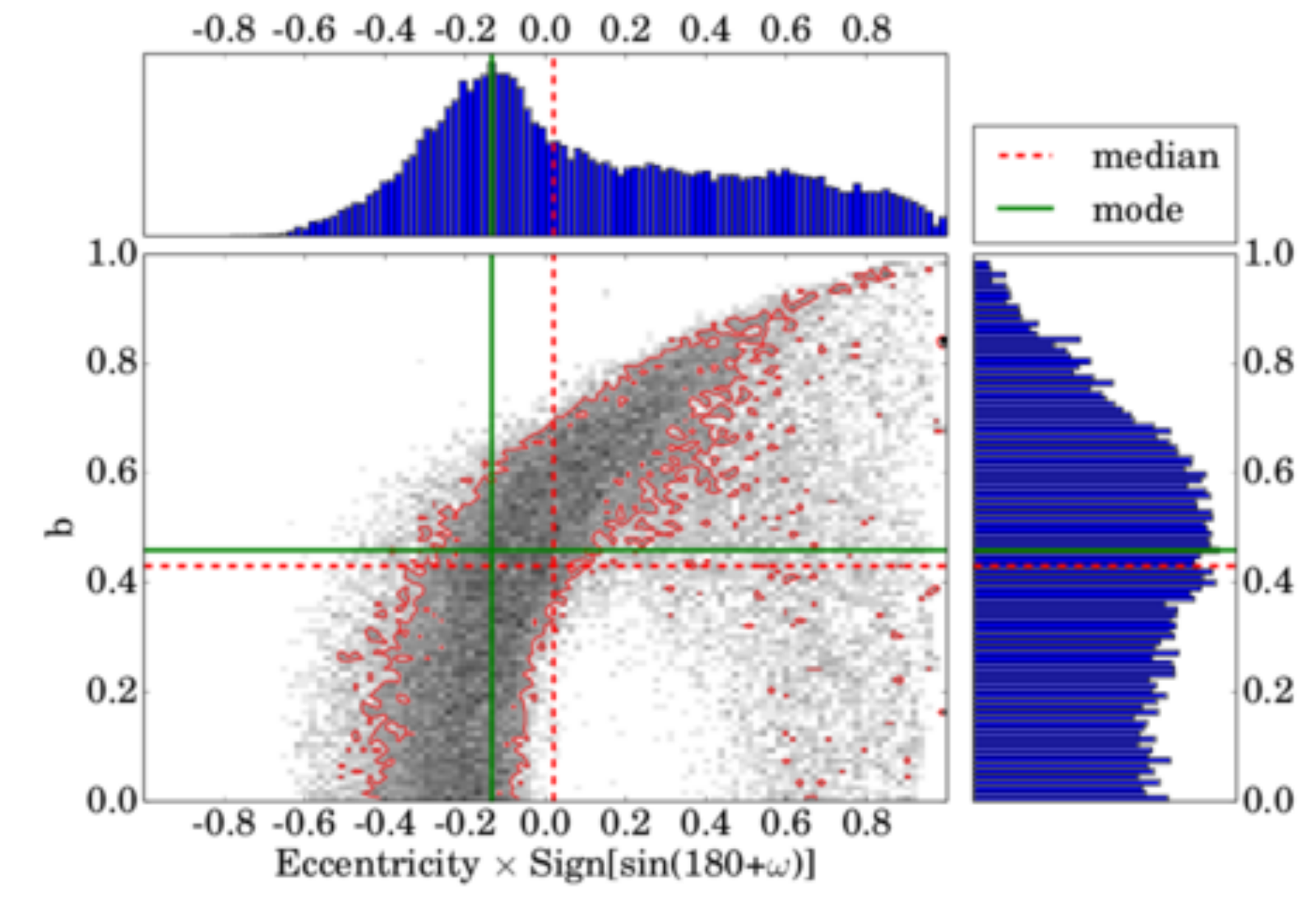}}
  \subfloat[KOI-270b\label{fig:koi270b}]{\includegraphics[width=0.33\linewidth]{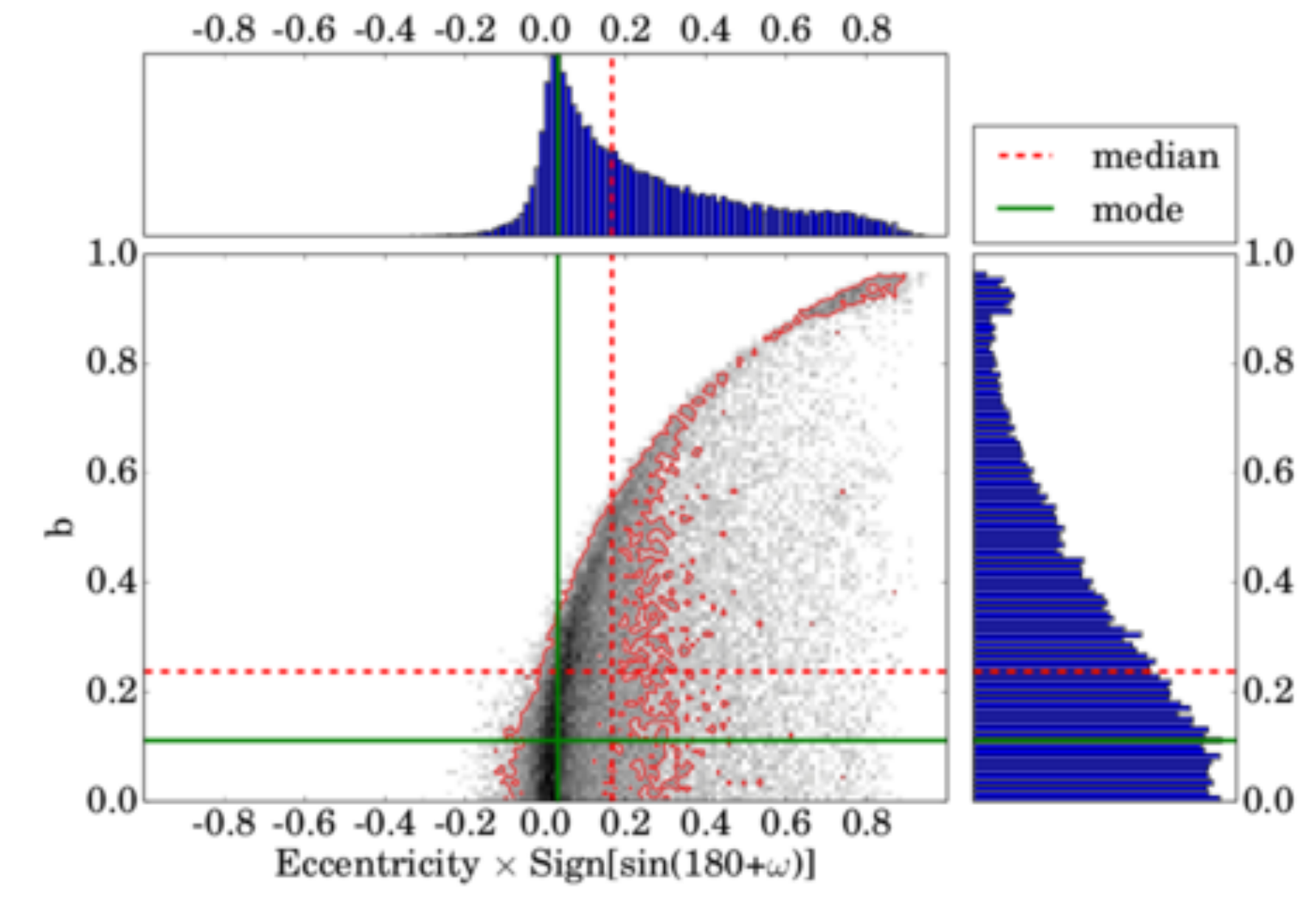}}
  
  \subfloat[KOI-270c\label{fig:koi270c}]{\includegraphics[width=0.33\linewidth]{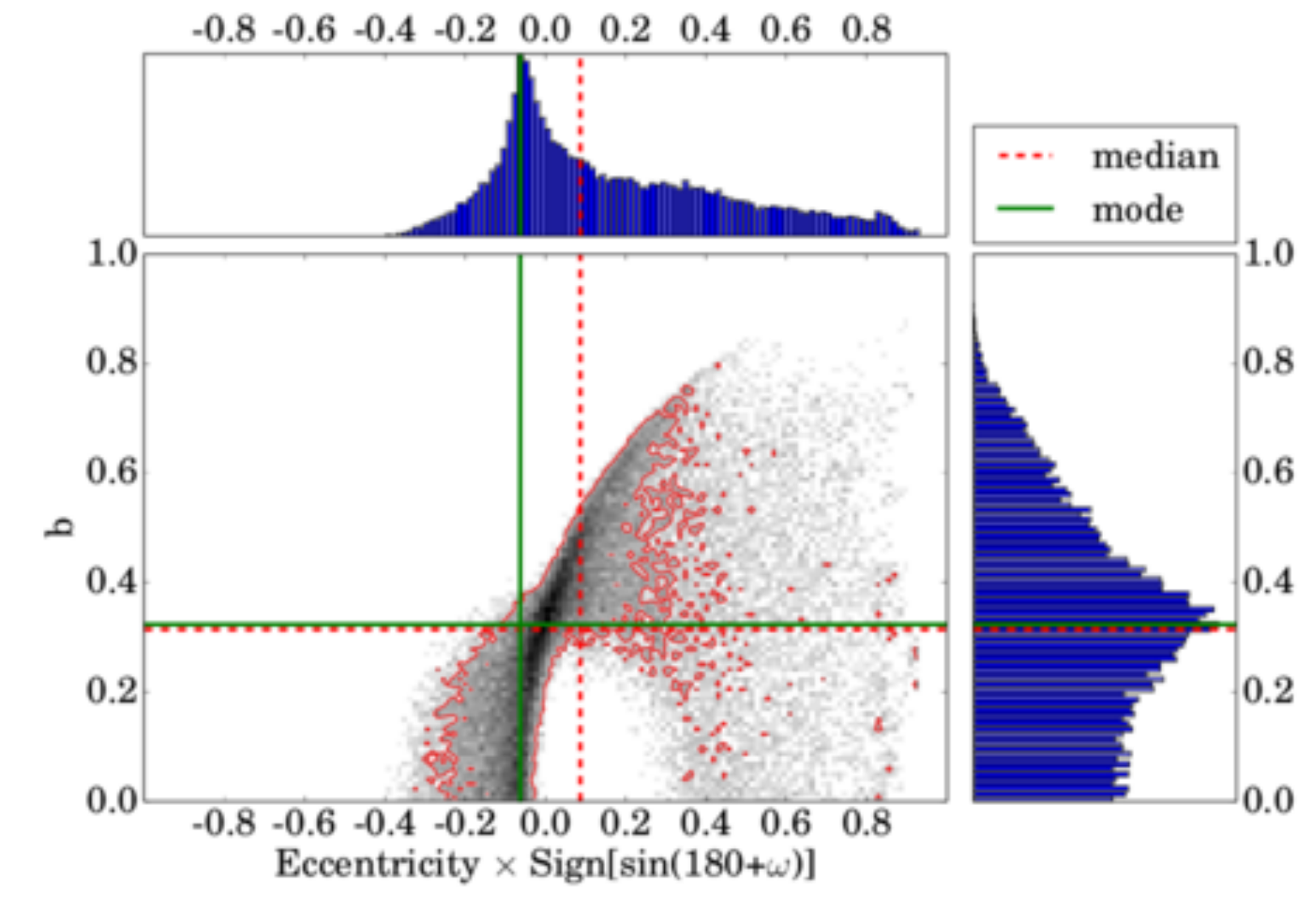}}
  \subfloat[KOI-279b\label{fig:koi279b}]{\includegraphics[width=0.33\linewidth]{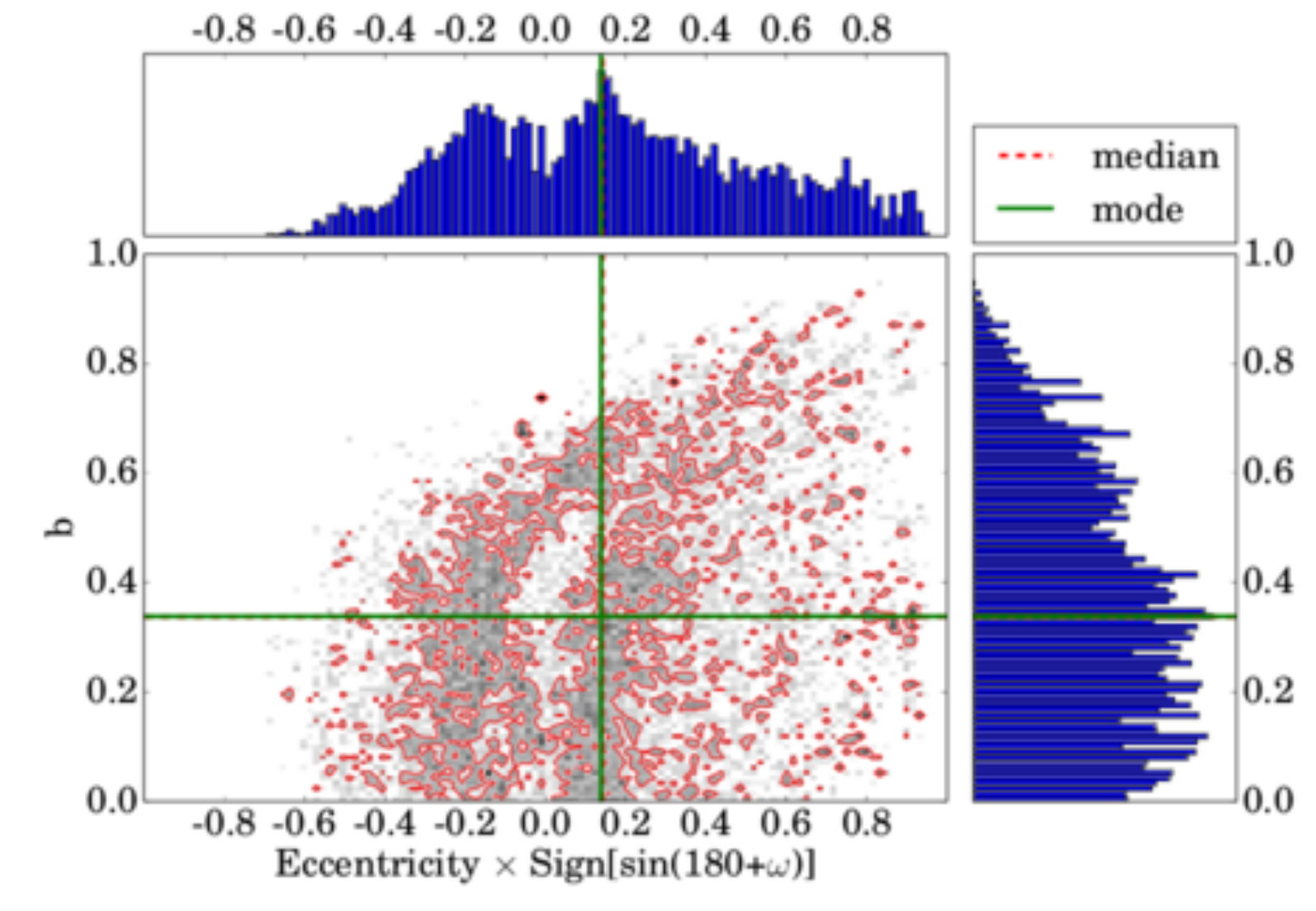}}
  \subfloat[KOI-279c\label{fig:koi279c}]{\includegraphics[width=0.33\linewidth]{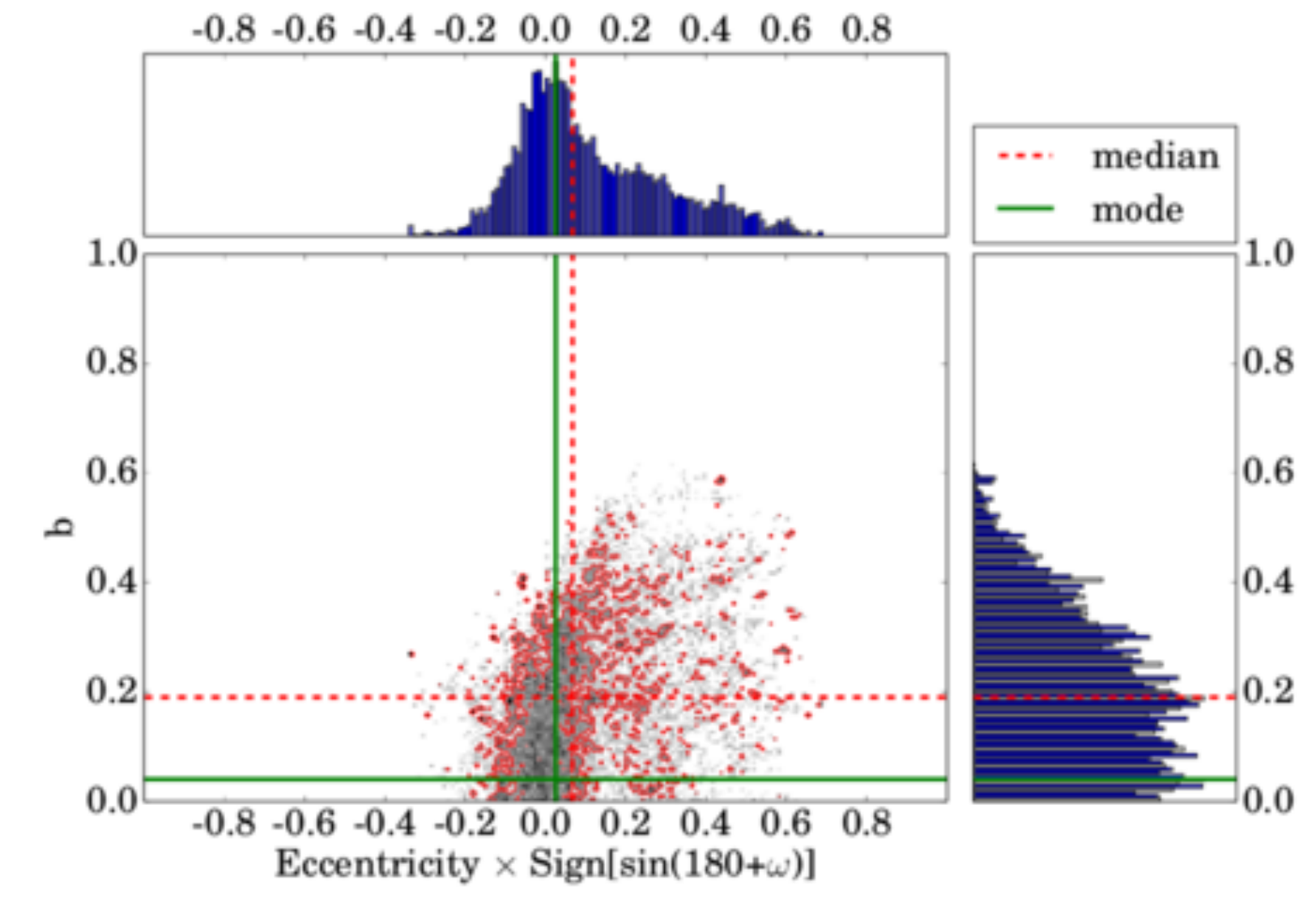}}
  
  \subfloat[KOI-279d\label{fig:koi279d}]{\includegraphics[width=0.33\linewidth]{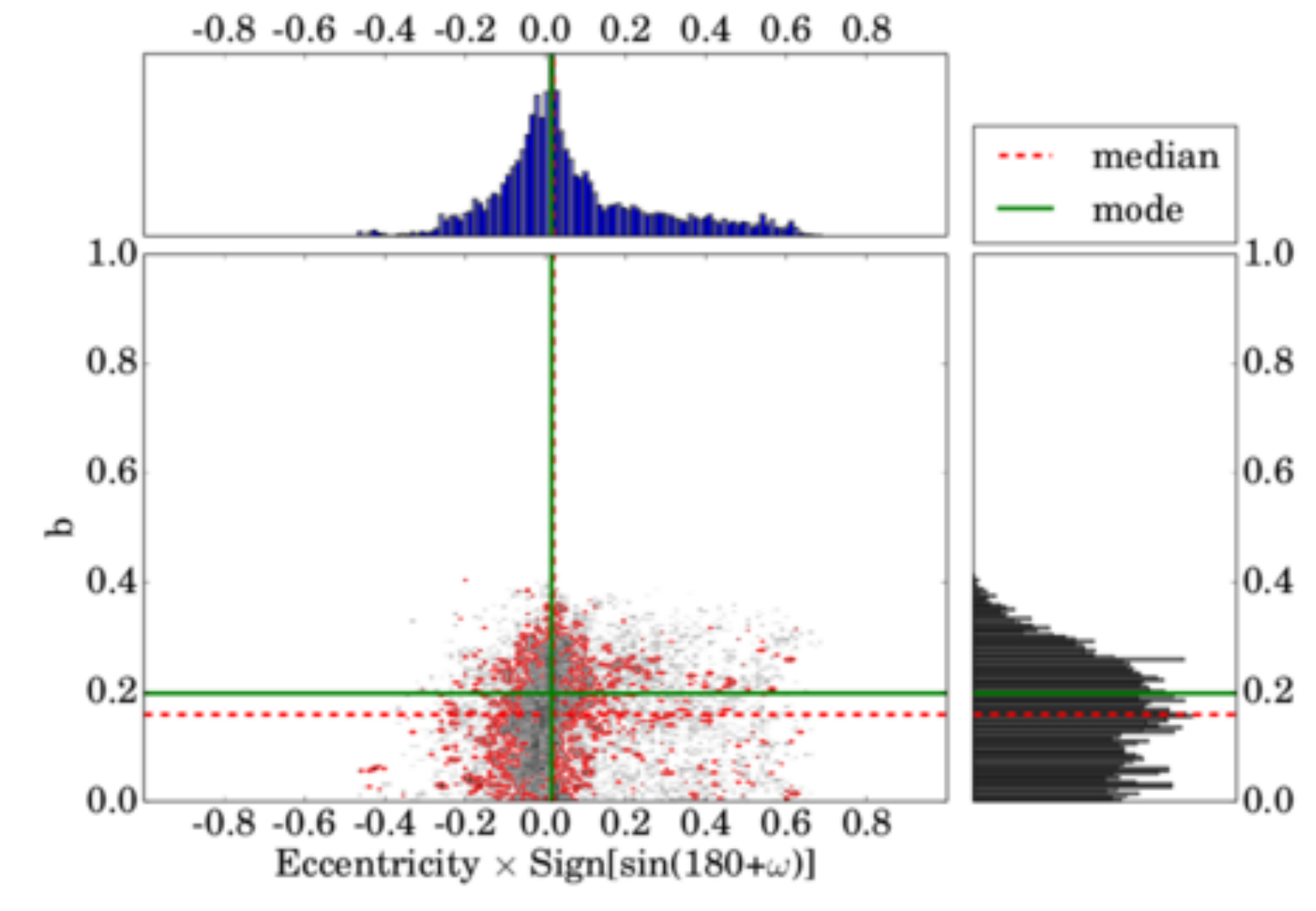}}
  \subfloat[Kepler-92b\label{fig:kepler92b}]{\includegraphics[width=0.33\linewidth]{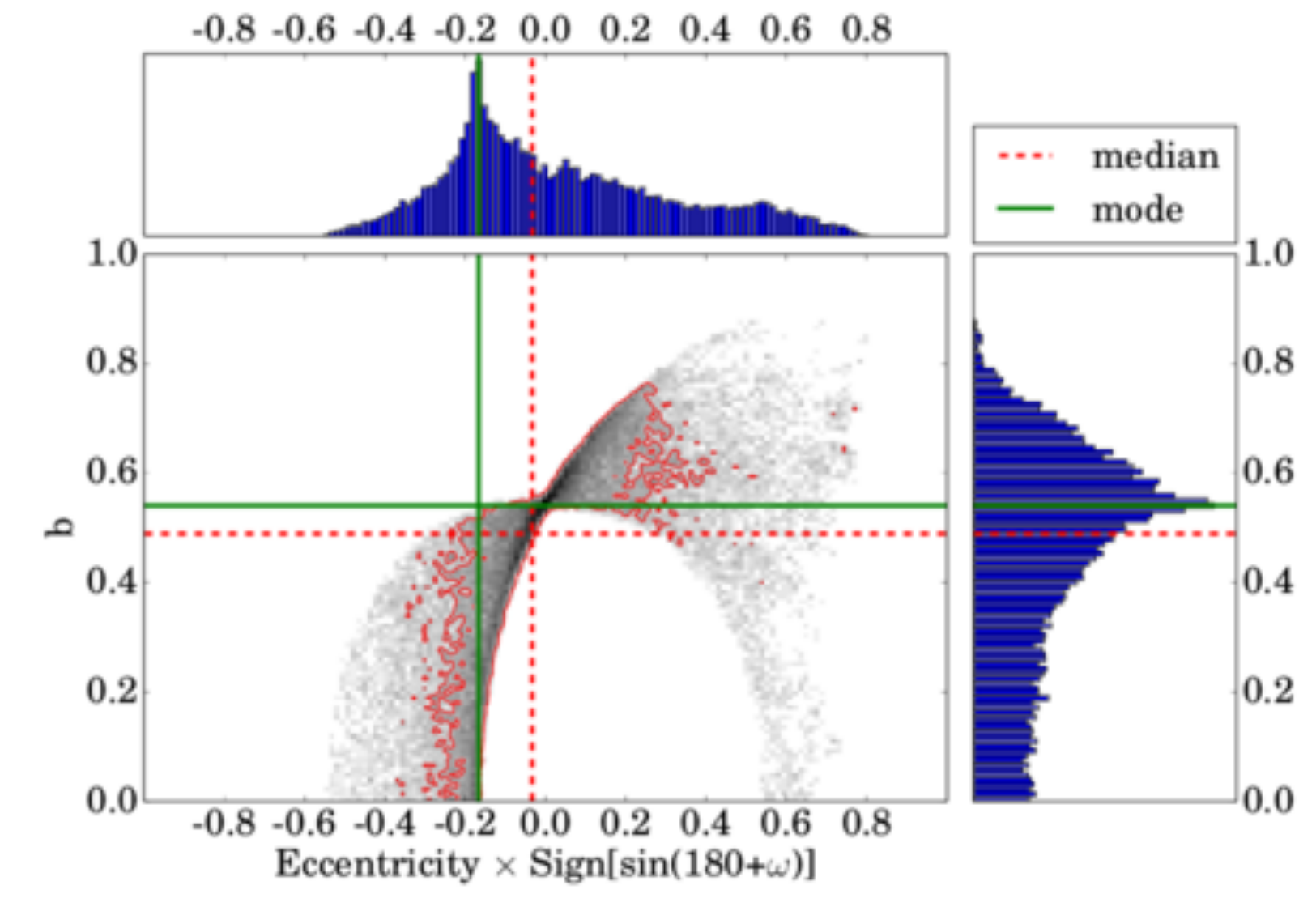}}  
  \subfloat[Kepler-92c\label{fig:kepler92c}]{\includegraphics[width=0.33\linewidth]{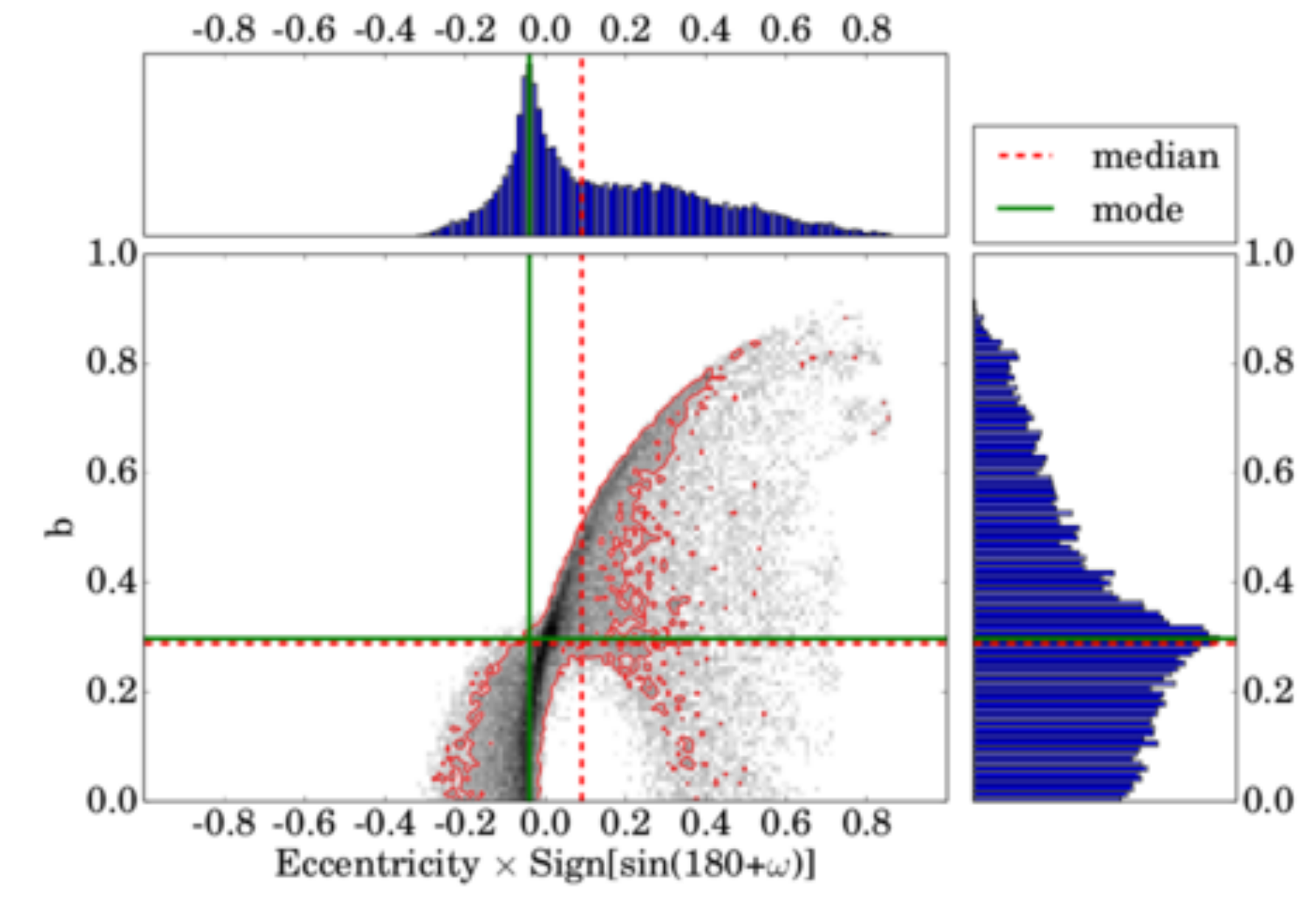}} 
  
  \subfloat[Kepler-92d\label{fig:kepler92d}]{\includegraphics[width=0.33\linewidth]{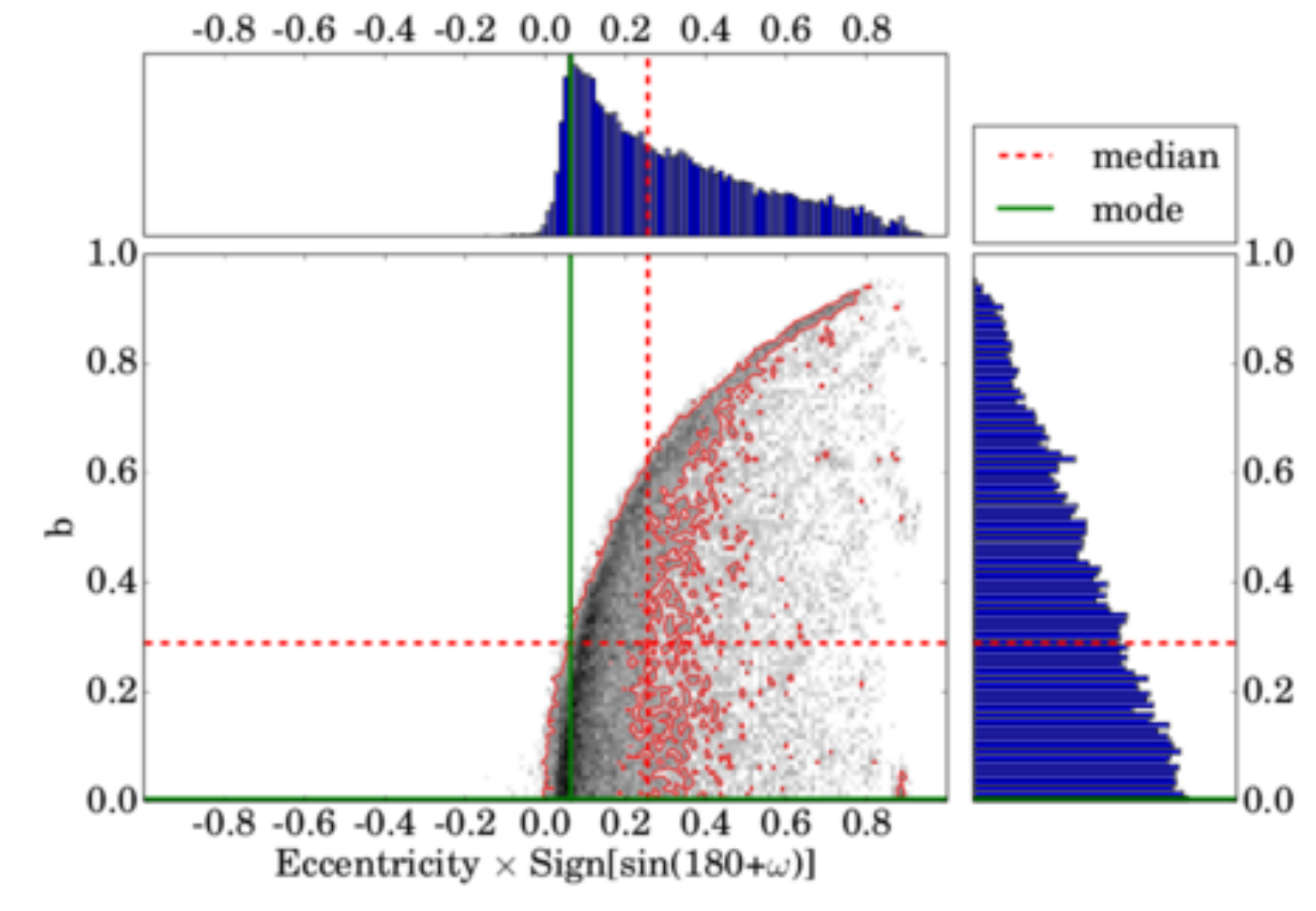}}  
 \subfloat[Kepler-10b\label{fig:kepler10b}]{\includegraphics[width=0.33\linewidth]{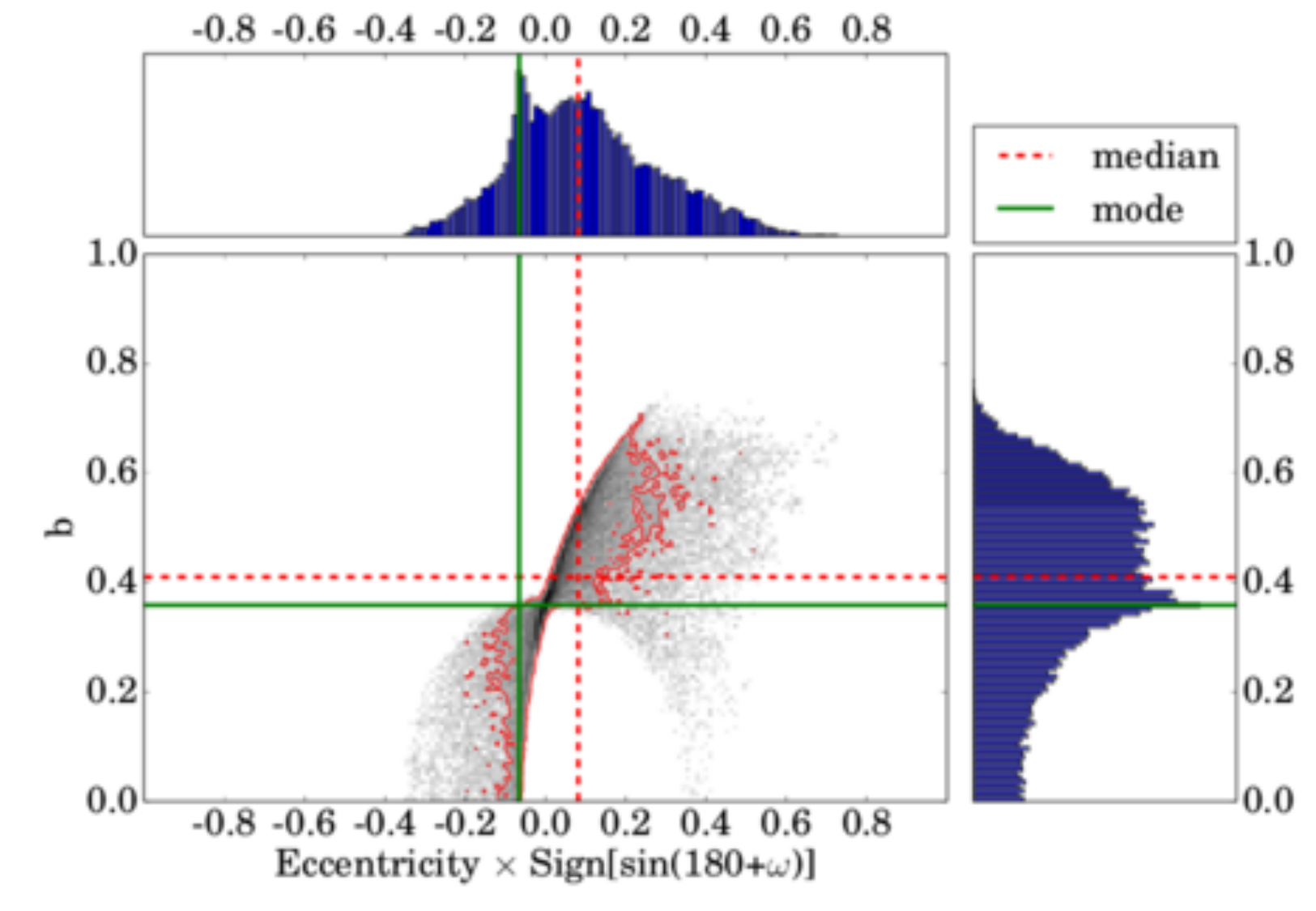}}
 \subfloat[Kepler-10c\label{fig:kepler10c}]{\includegraphics[width=0.33\linewidth]{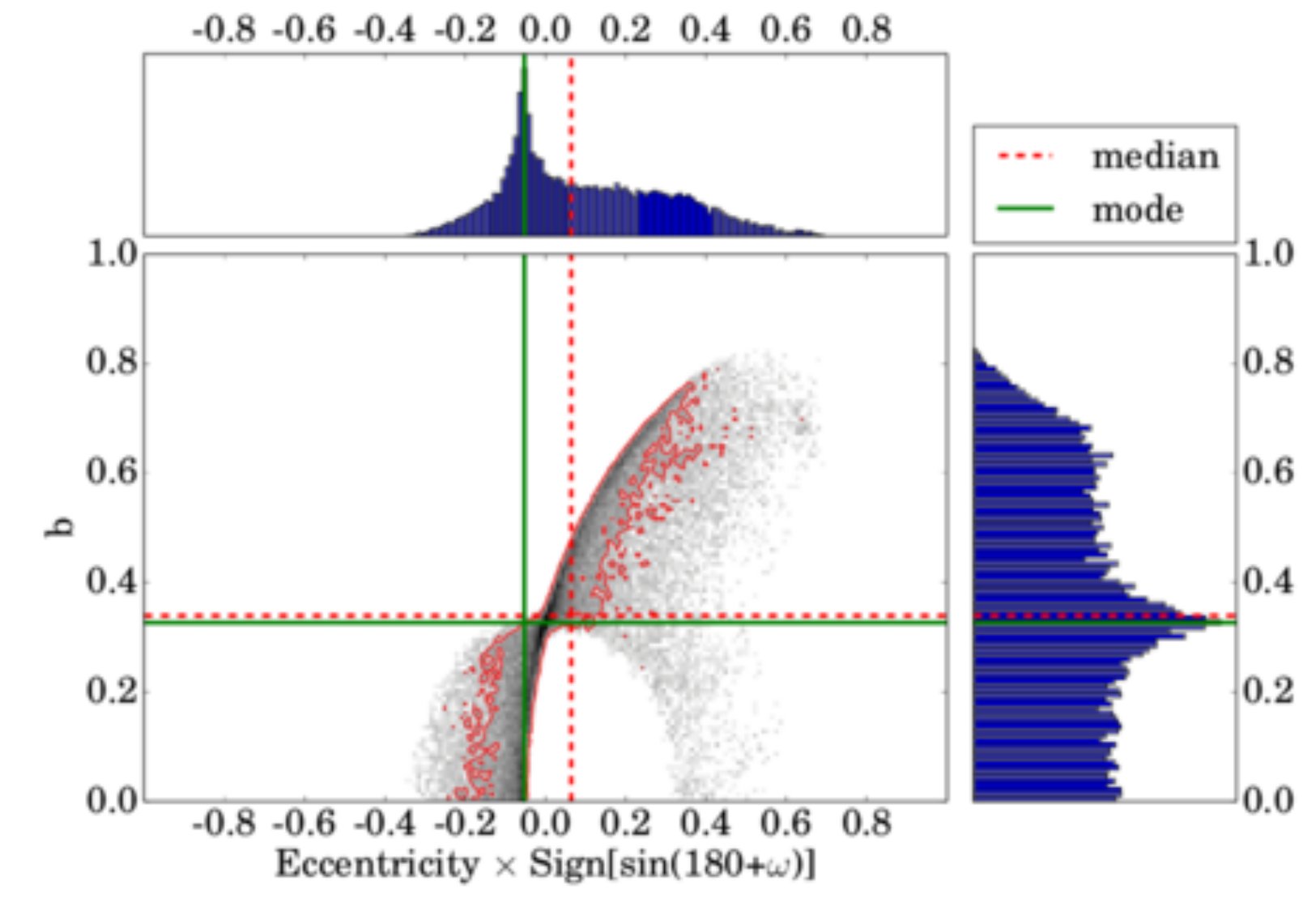}}
 
 \end{widepage}
 \caption{Posterior distributions of individual planets\label{fig:individual_posterior}}
\end{figure*}

\begin{figure*}[t]
 \ContinuedFloat
    \begin{widepage}
    
     \subfloat[Kepler-23b\label{fig:kepler23b}]{\includegraphics[width=0.33\linewidth]{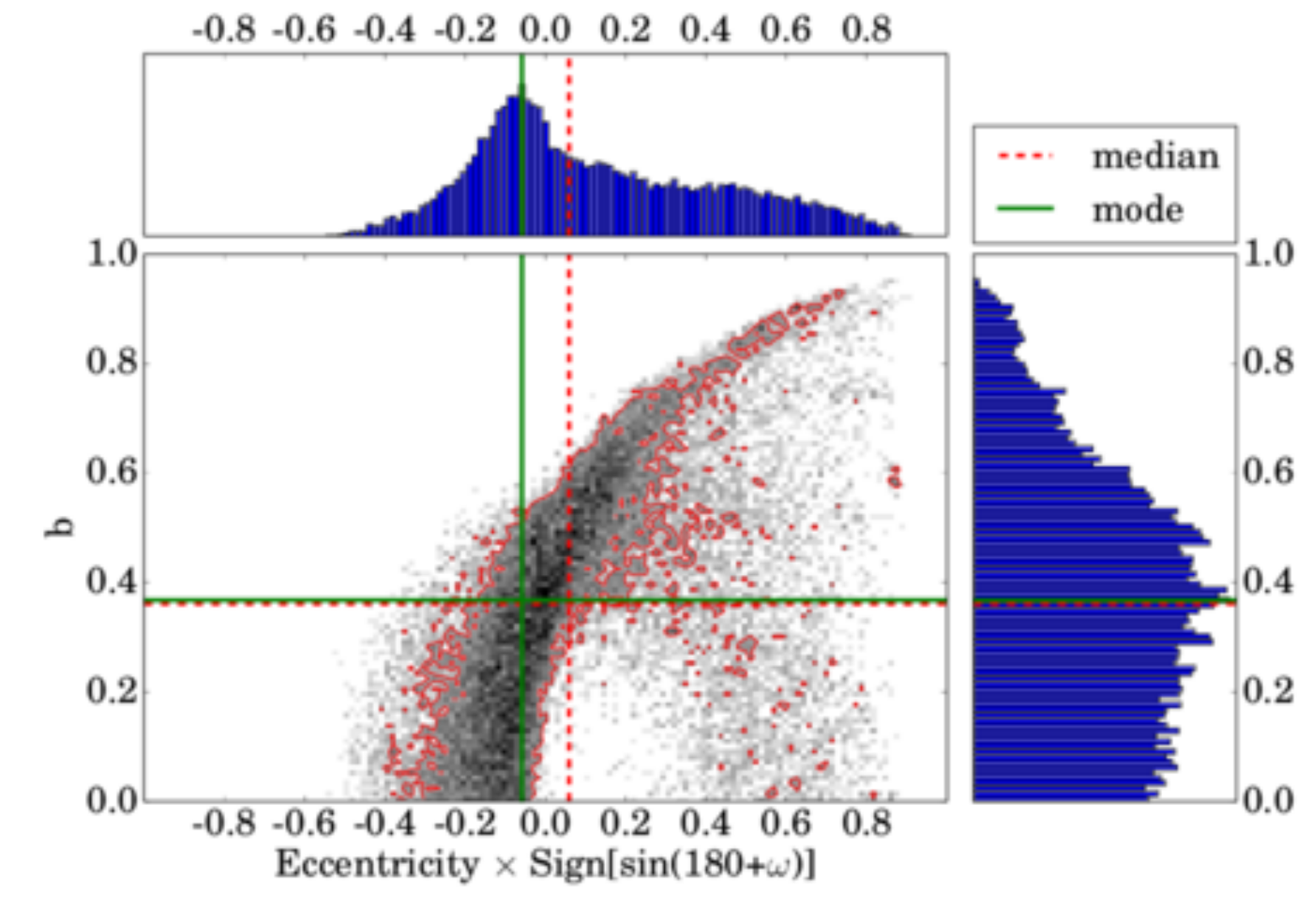}}
 \subfloat[Kepler-23c\label{fig:kepler23c}]{\includegraphics[width=0.33\linewidth]{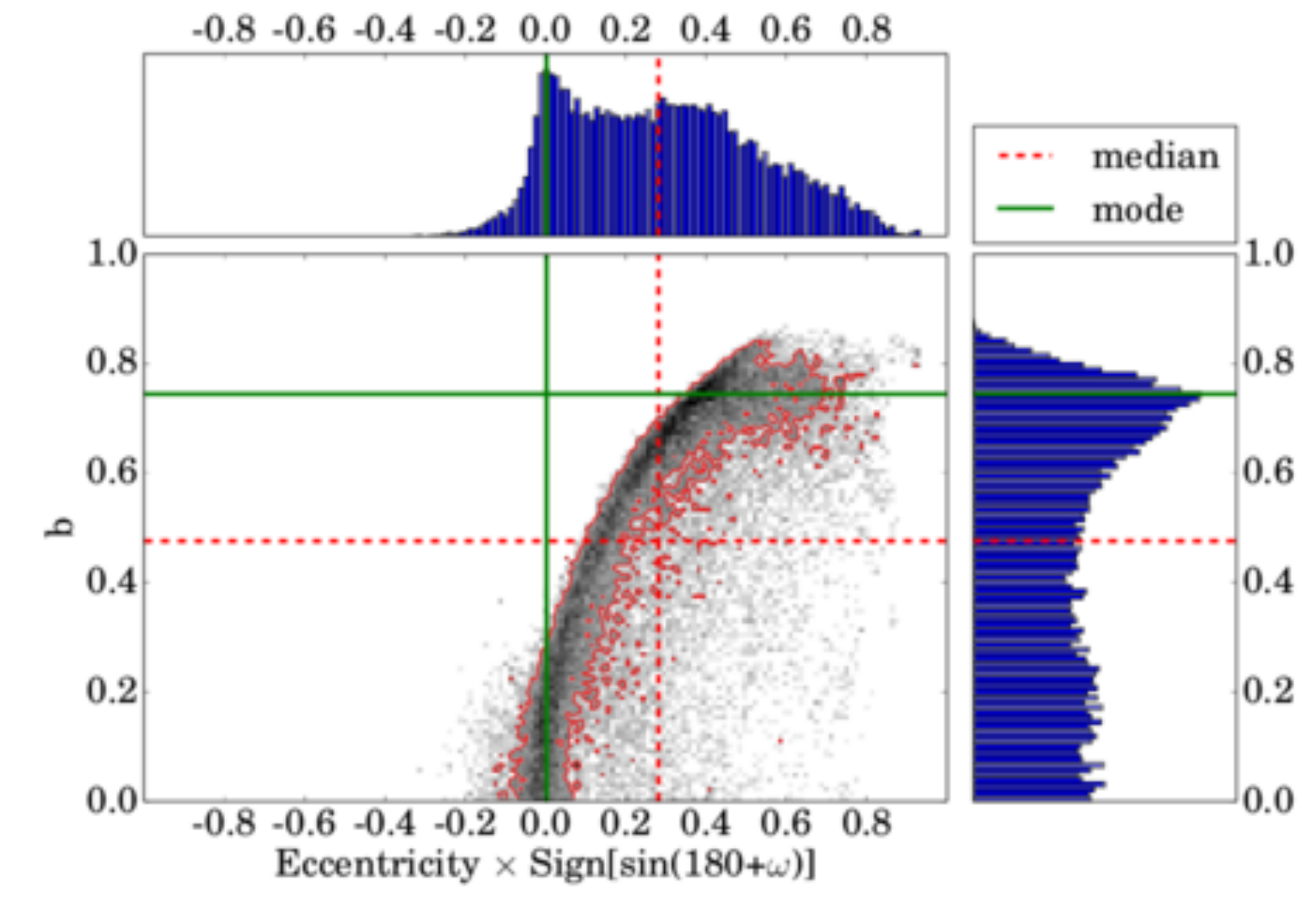}}
 \subfloat[Kepler-23d\label{fig:kepler23d}]{\includegraphics[width=0.33\linewidth]{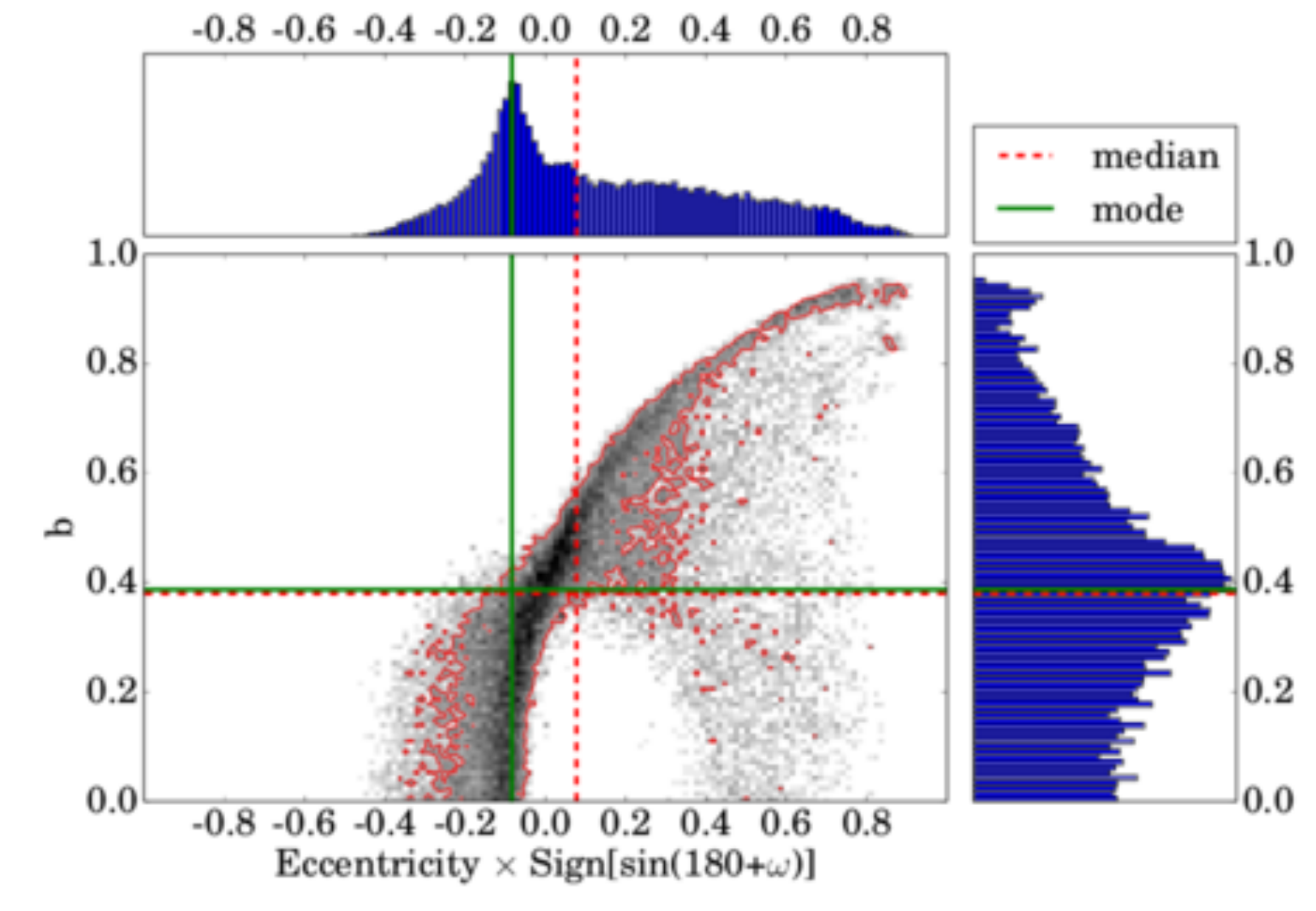}}
 
 \subfloat[Kepler-25b\label{fig:kepler25b}]{\includegraphics[width=0.33\linewidth]{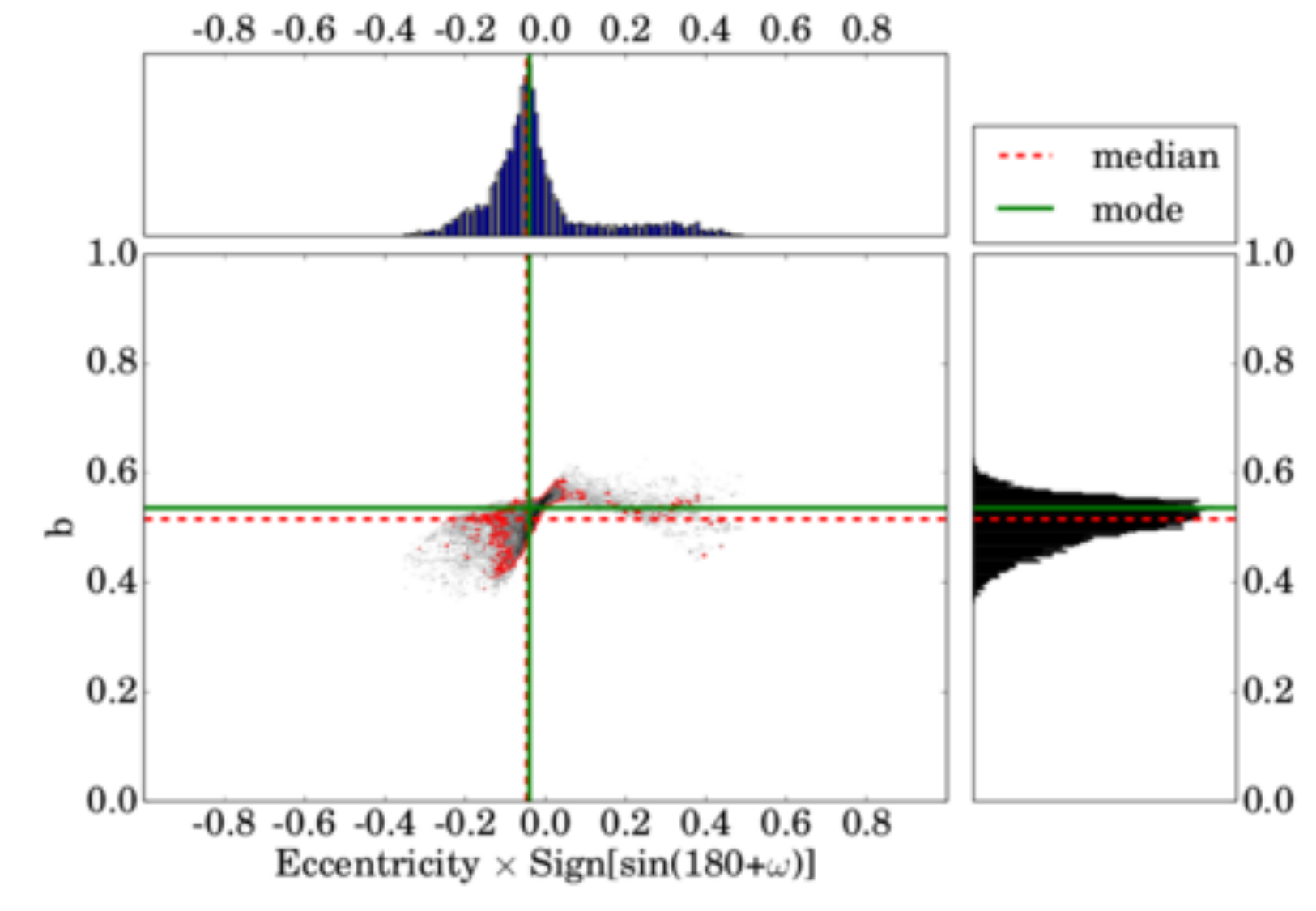}}
 \subfloat[Kepler-25c\label{fig:kepler25c}]{\includegraphics[width=0.33\linewidth]{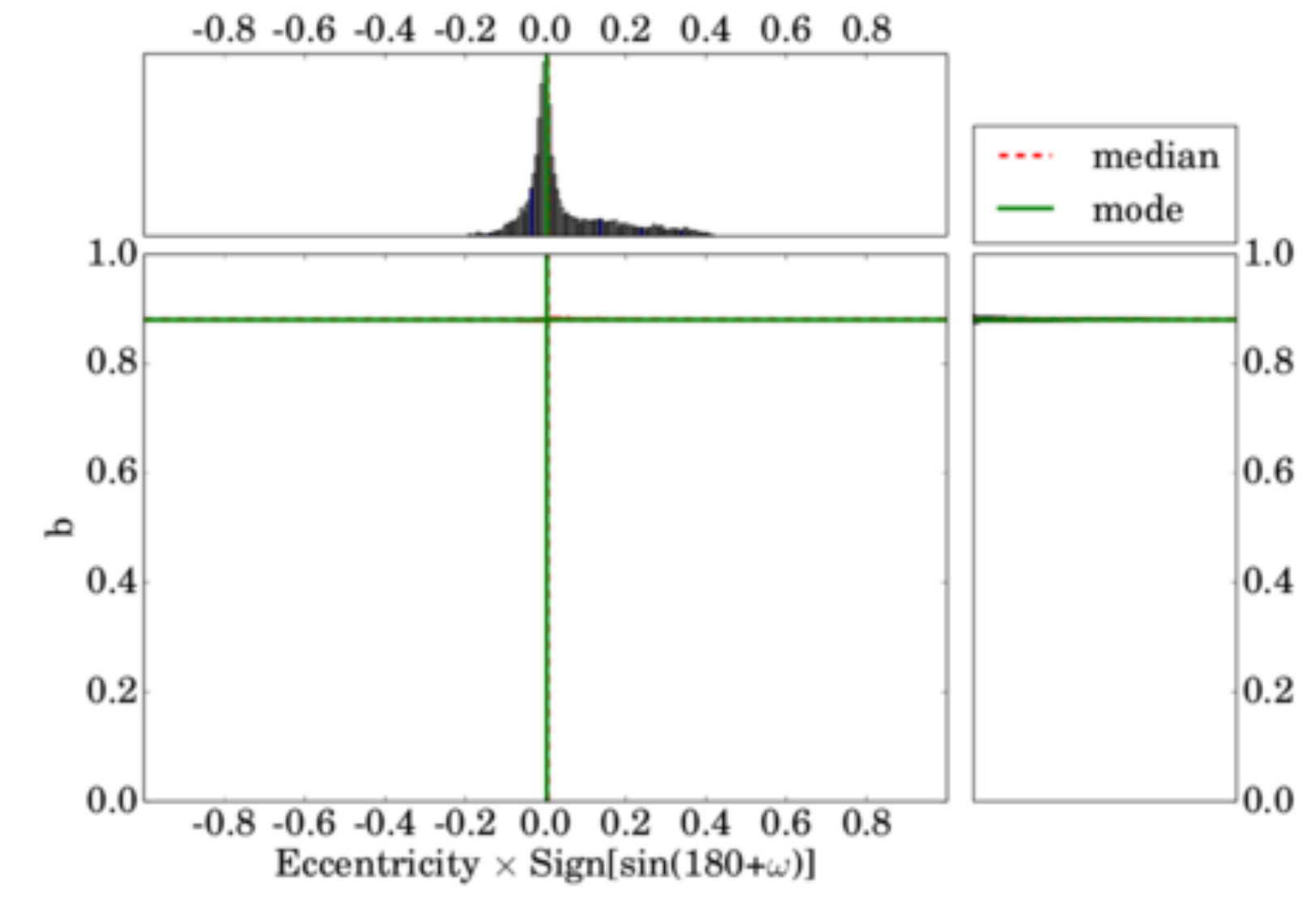}}
  \subfloat[Kepler-37b\label{fig:kepler37b}]{\includegraphics[width=0.33\linewidth]{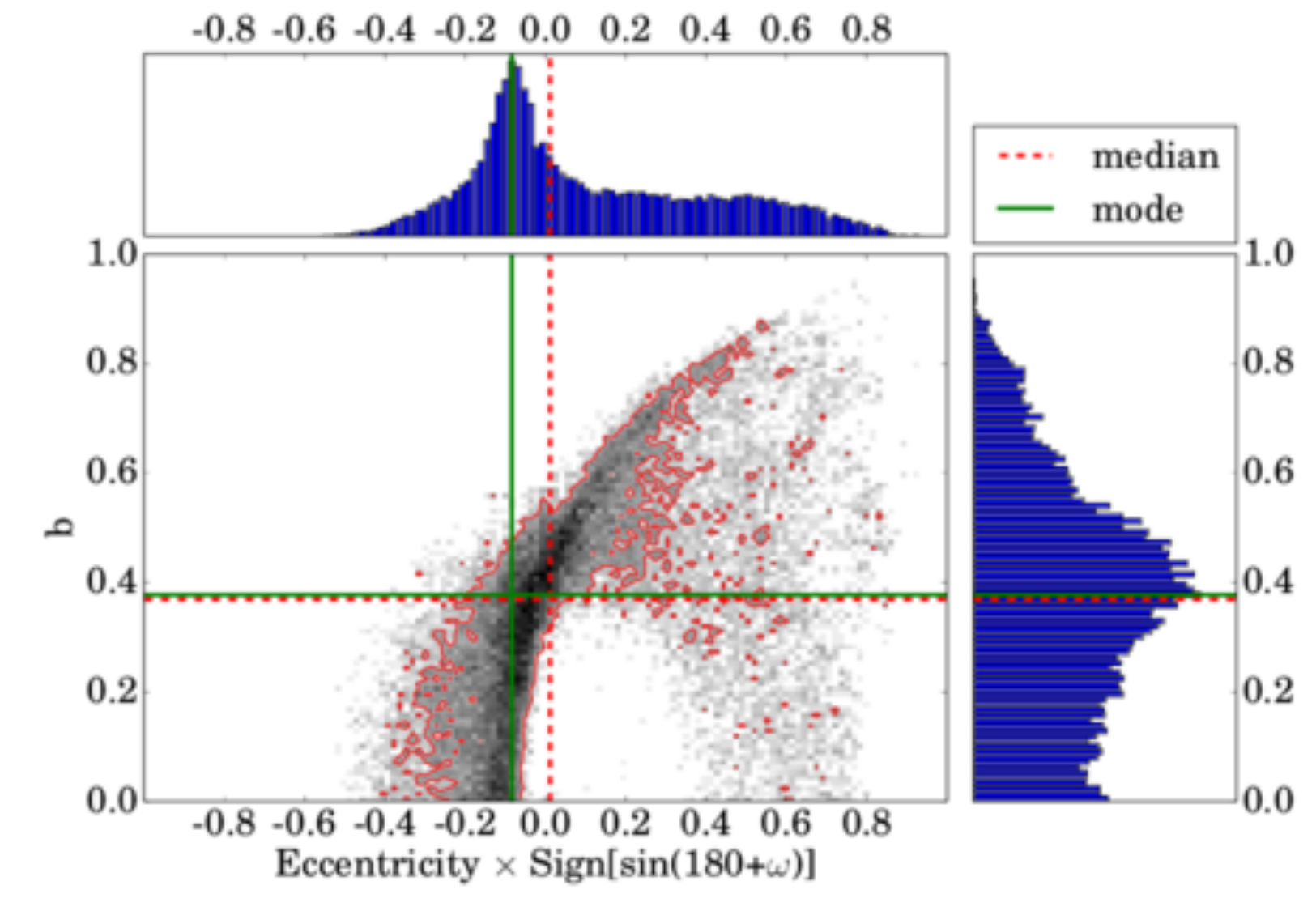}}
  
 \subfloat[Kepler-37c\label{fig:kepler37c}]{\includegraphics[width=0.33\linewidth]{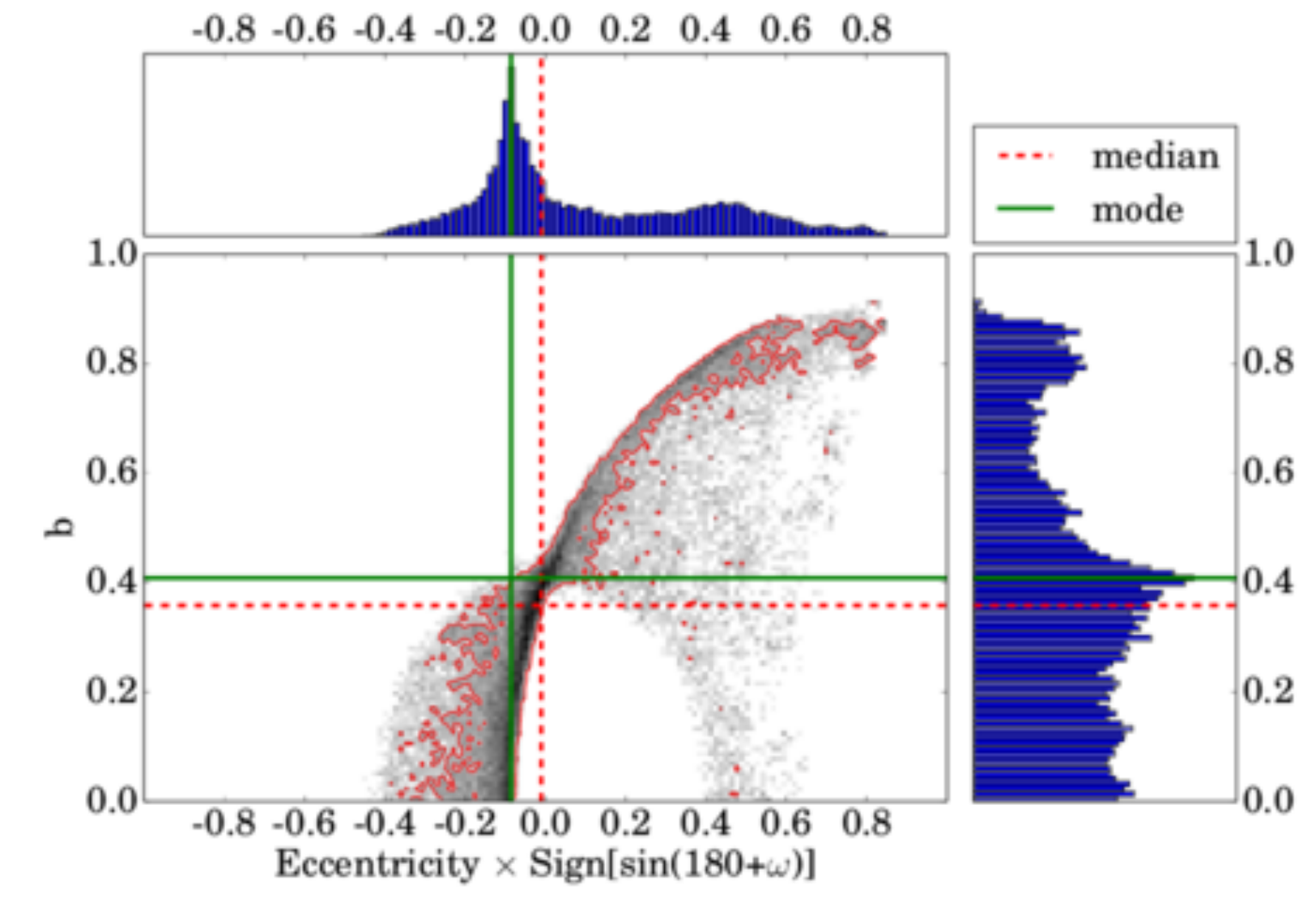}}
 \subfloat[Kepler-37d\label{fig:kepler37d}]{\includegraphics[width=0.33\linewidth]{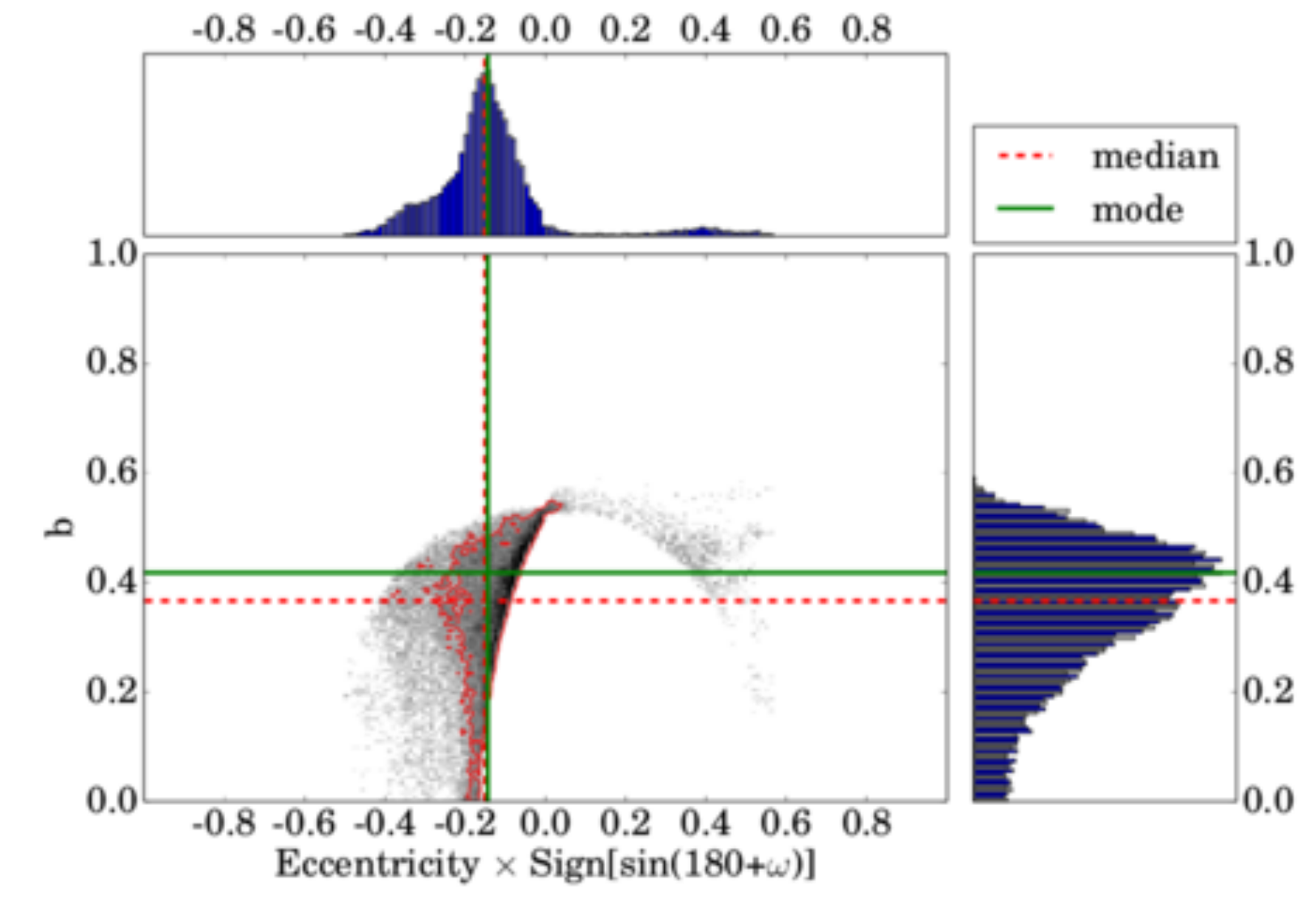}} 
  \subfloat[Kepler-65b\label{fig:kepler65b}]{\includegraphics[width=0.33\linewidth]{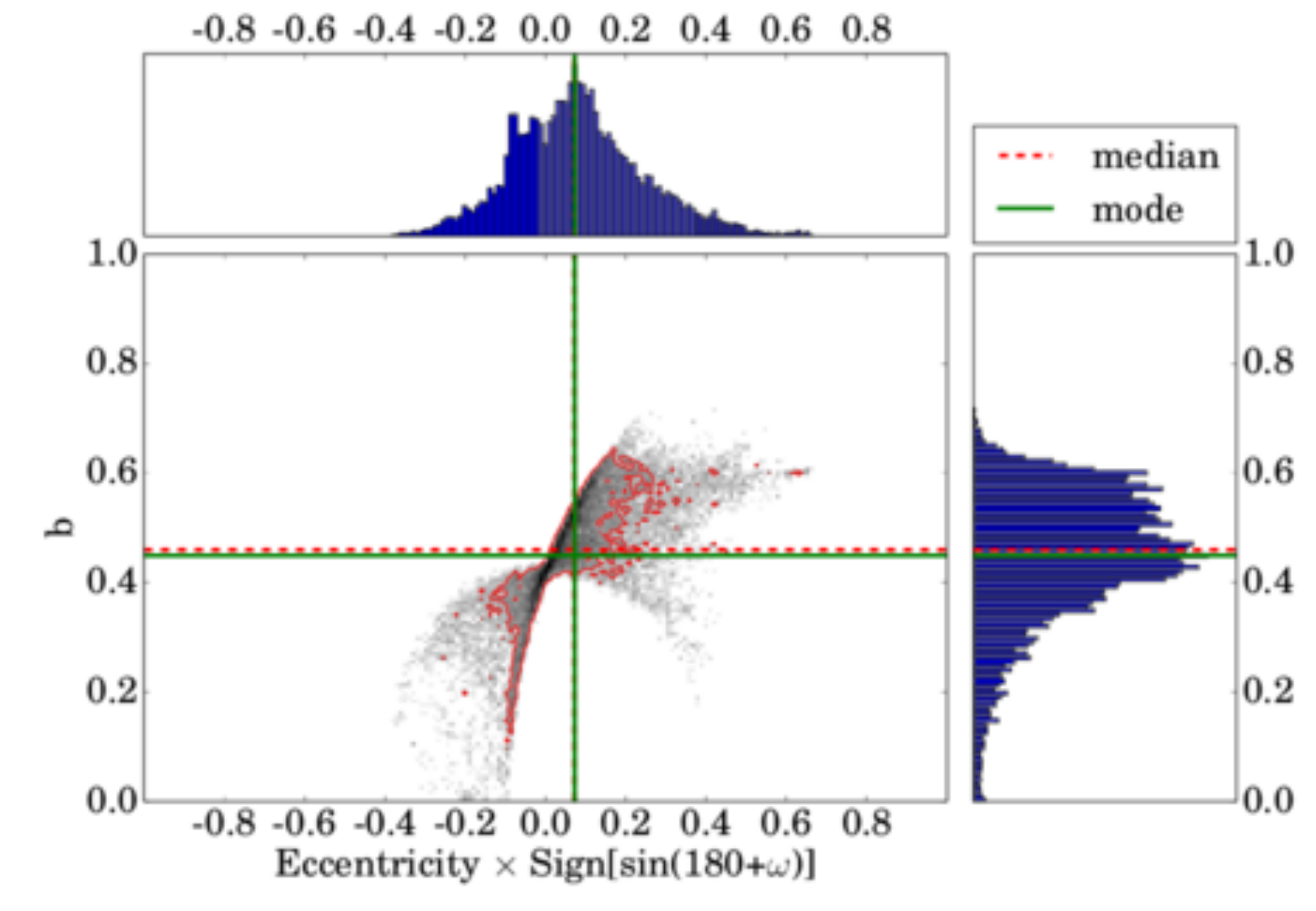}}
  
  \subfloat[Kepler-65c\label{fig:kepler65c}]{\includegraphics[width=0.33\linewidth]{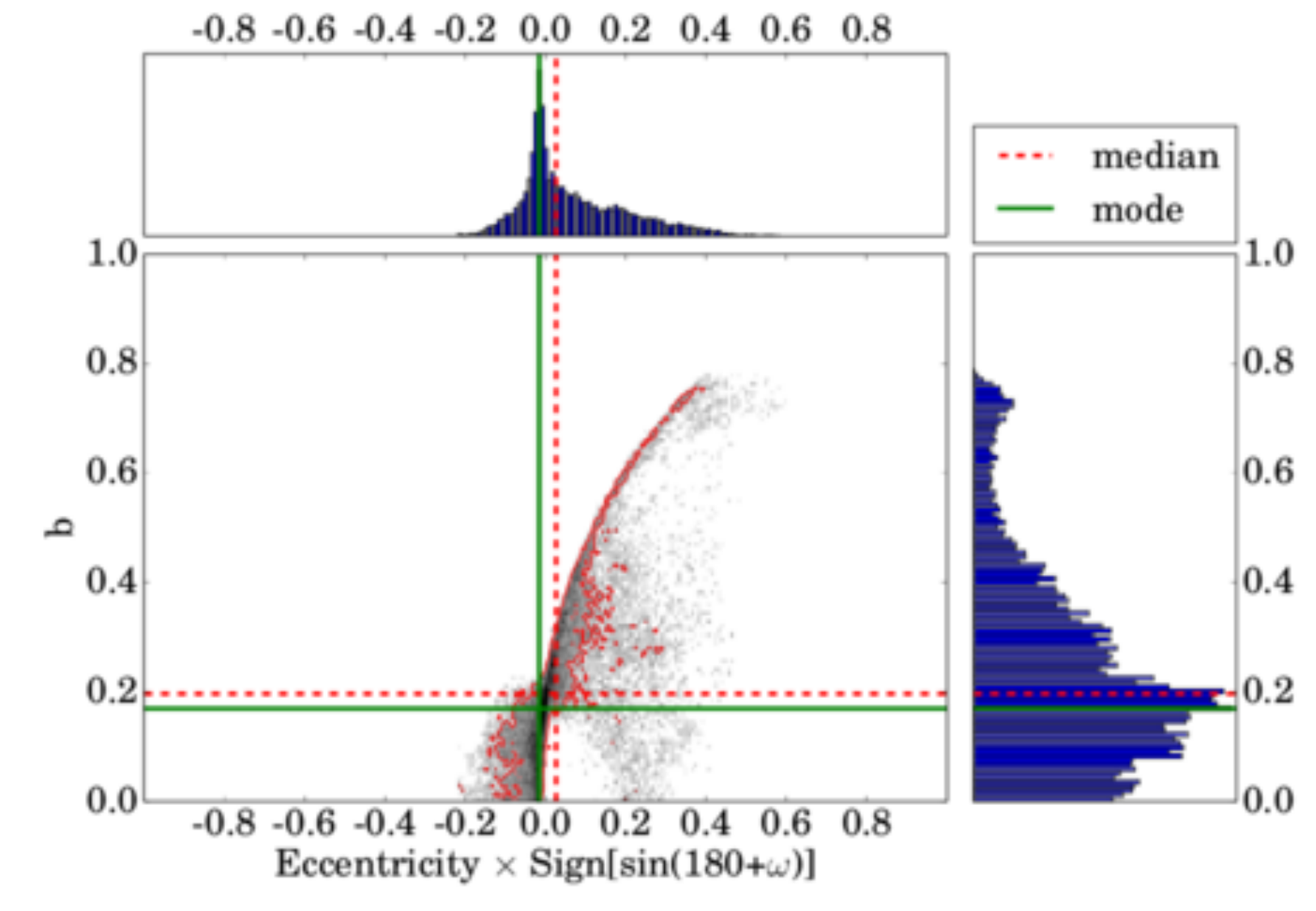}} 
  \subfloat[Kepler-65d\label{fig:kepler65d}]{\includegraphics[width=0.33\linewidth]{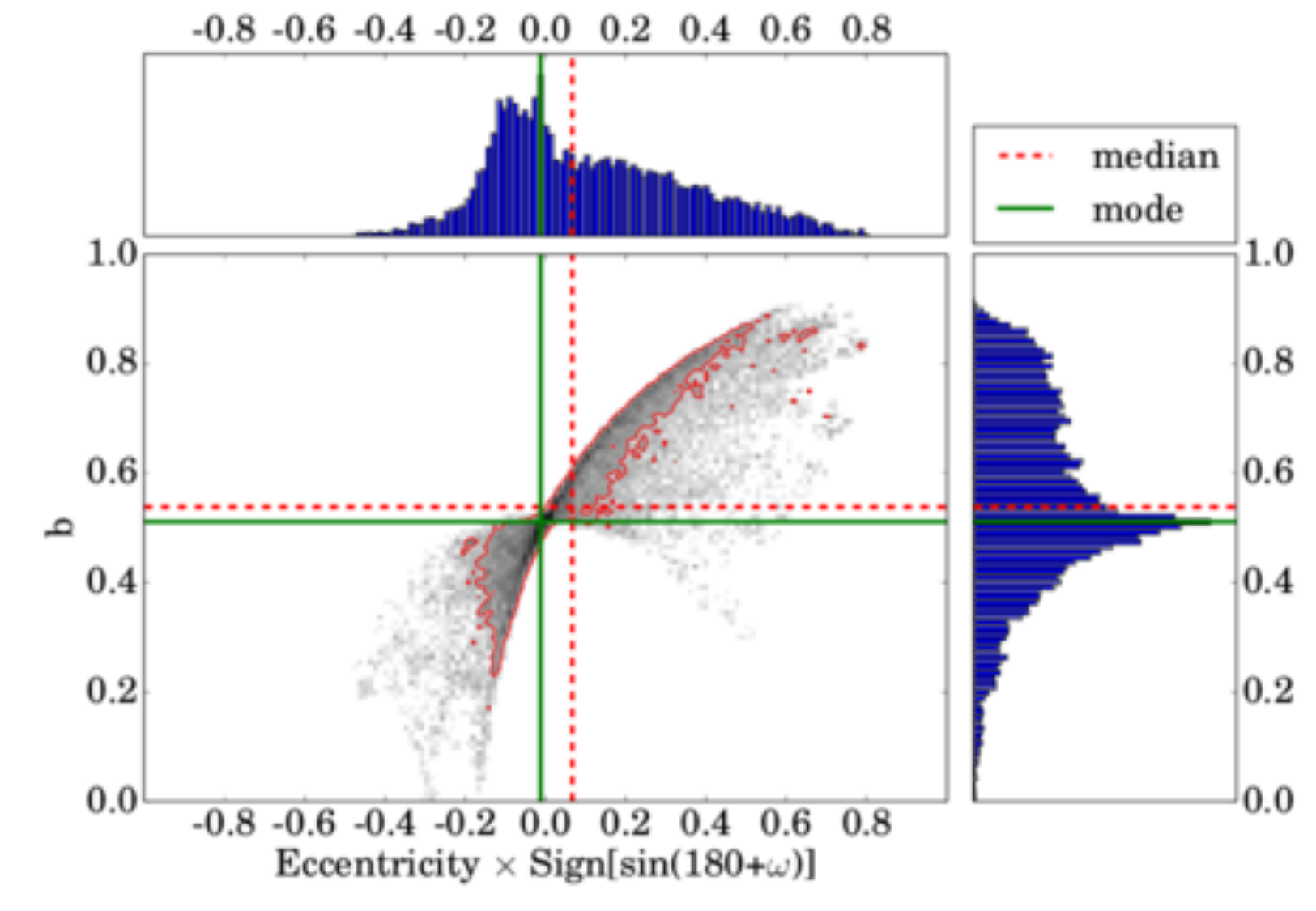}}                  
  \subfloat[Kepler-68b\label{fig:kepler68b}]{\includegraphics[width=0.33\linewidth]{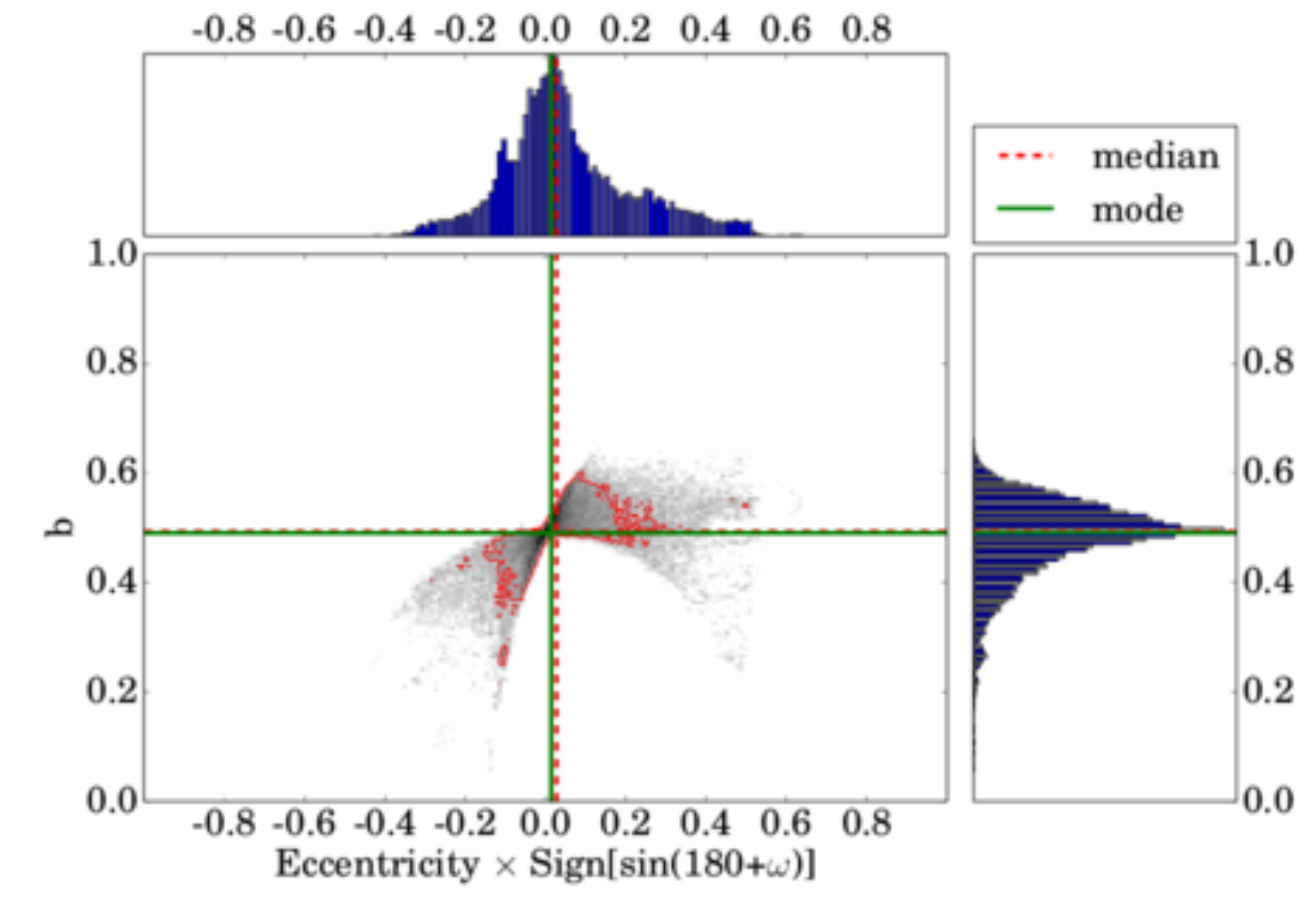}}
  
  \end{widepage}
 \caption{(continued) Posterior distributions of individual planets}
  \end{figure*}

\begin{figure*}[t]
 \ContinuedFloat
    \begin{widepage}
  \subfloat[Kepler-68c\label{fig:kepler68c}]{\includegraphics[width=0.33\linewidth]{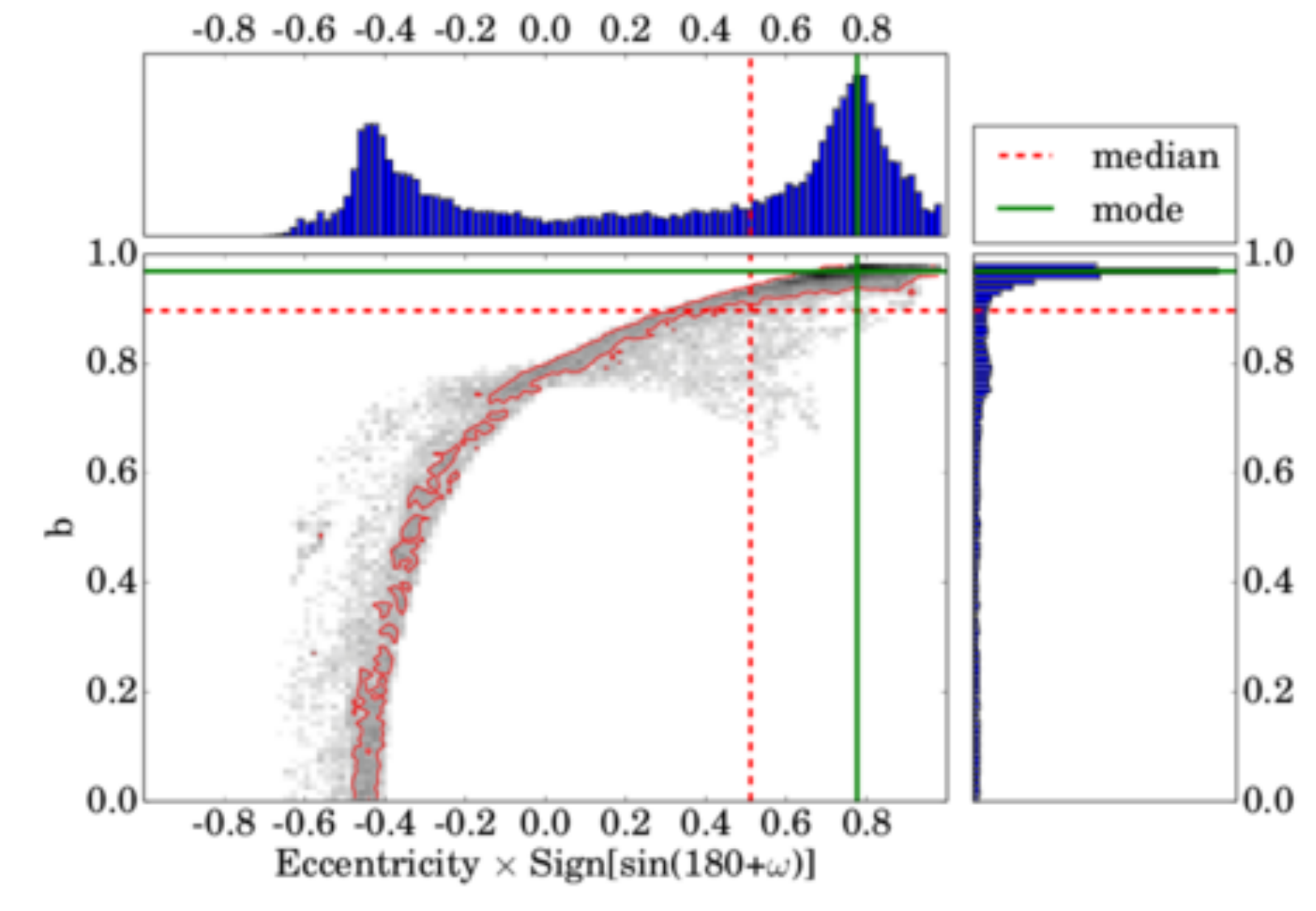}} 
  \subfloat[Kepler-100b\label{fig:kepler100b}]{\includegraphics[width=0.33\linewidth]{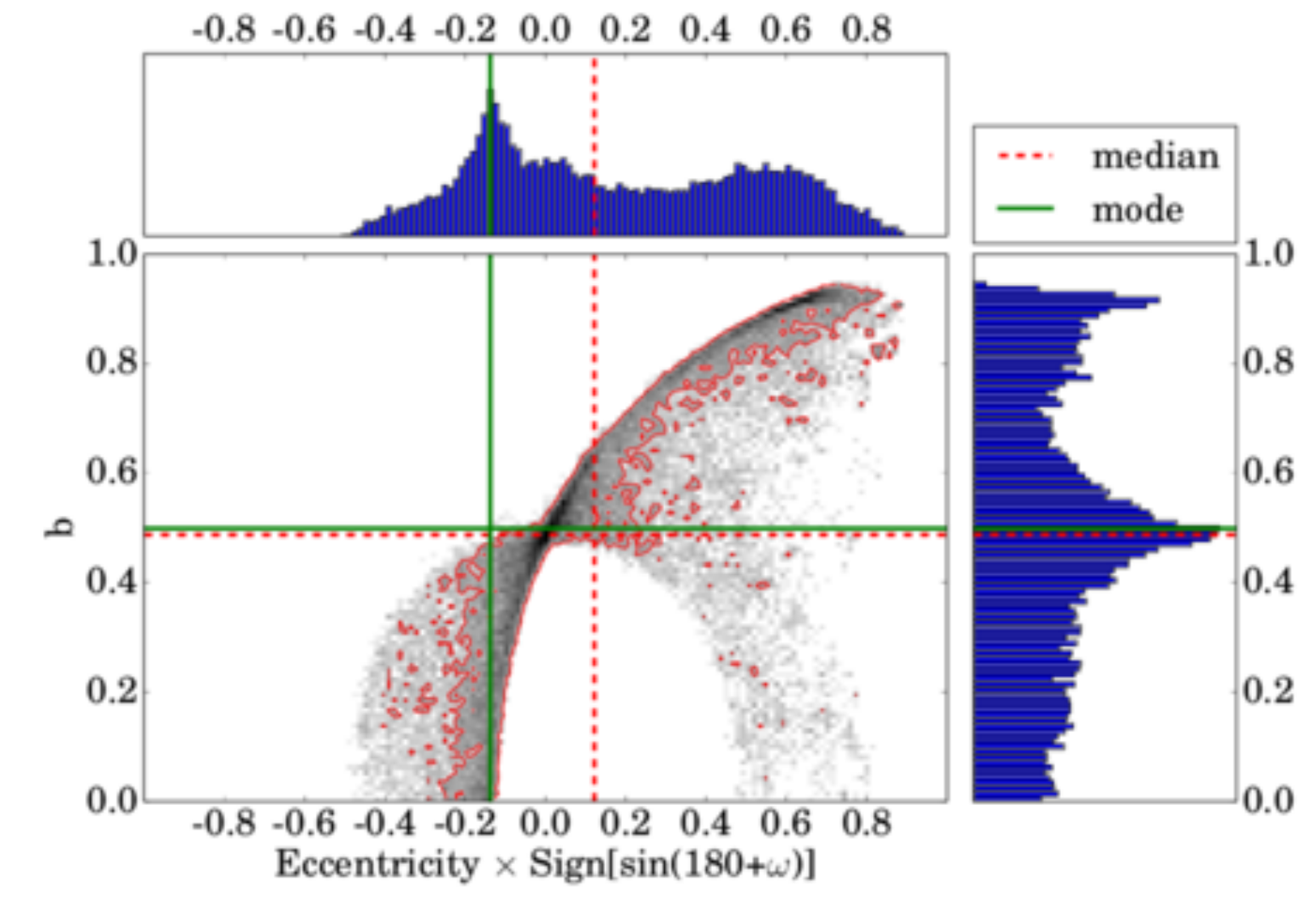}}  
  \subfloat[Kepler-100c\label{fig:kepler100c}]{\includegraphics[width=0.33\linewidth]{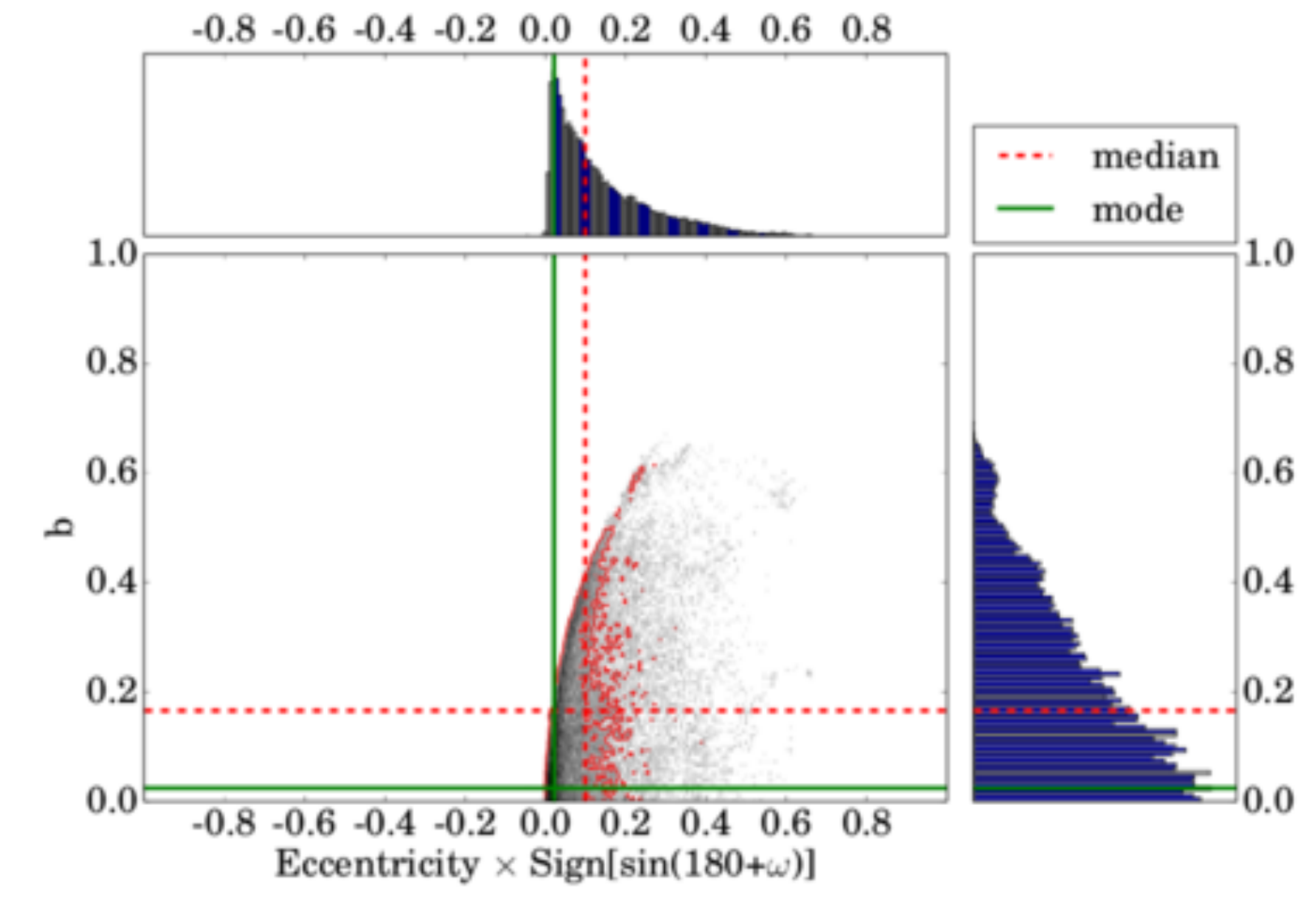}}

  \subfloat[Kepler-100d\label{fig:kepler100d}]{\includegraphics[width=0.33\linewidth]{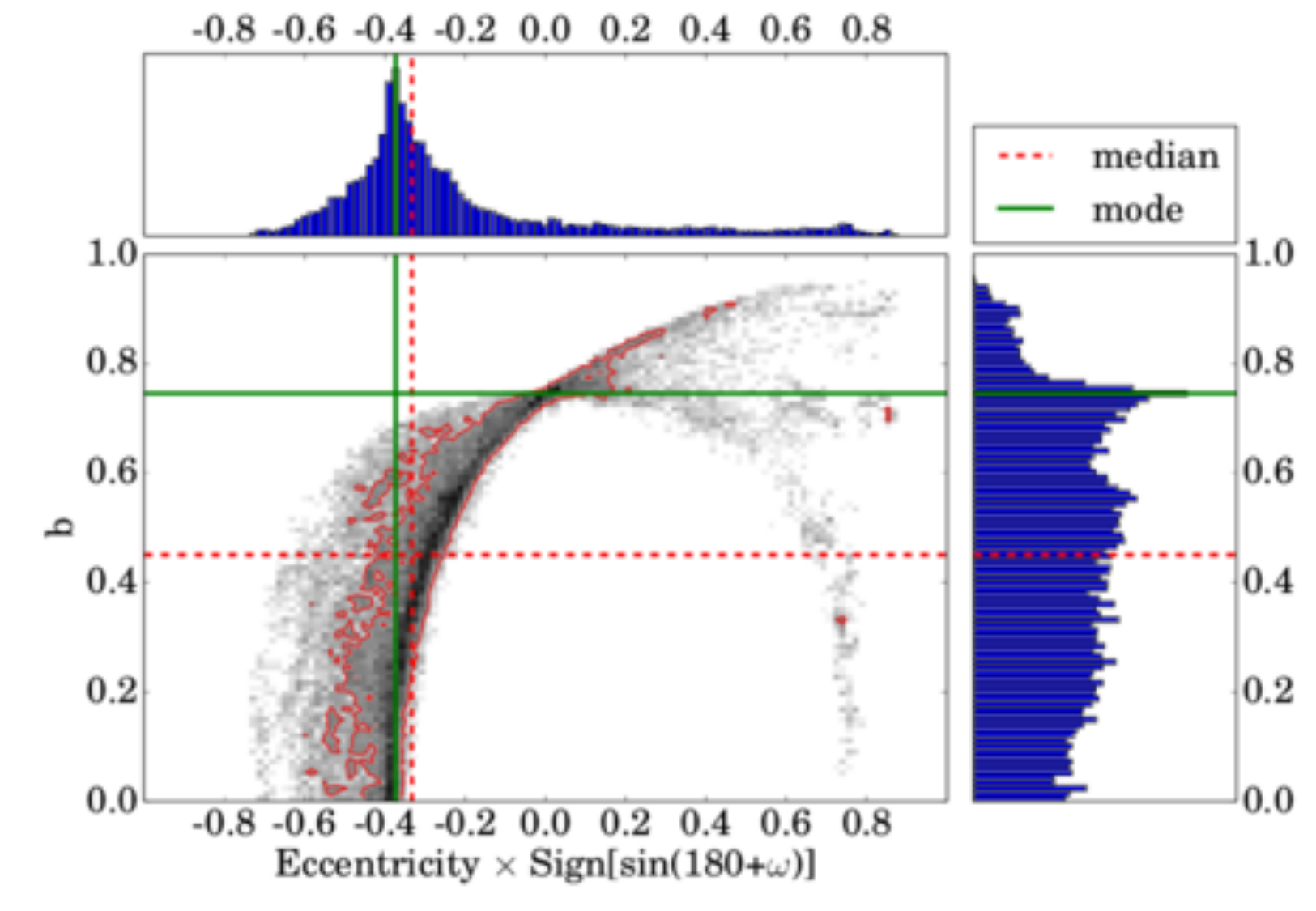}}  
  \subfloat[Kepler-103b\label{fig:kepler103b}]{\includegraphics[width=0.33\linewidth]{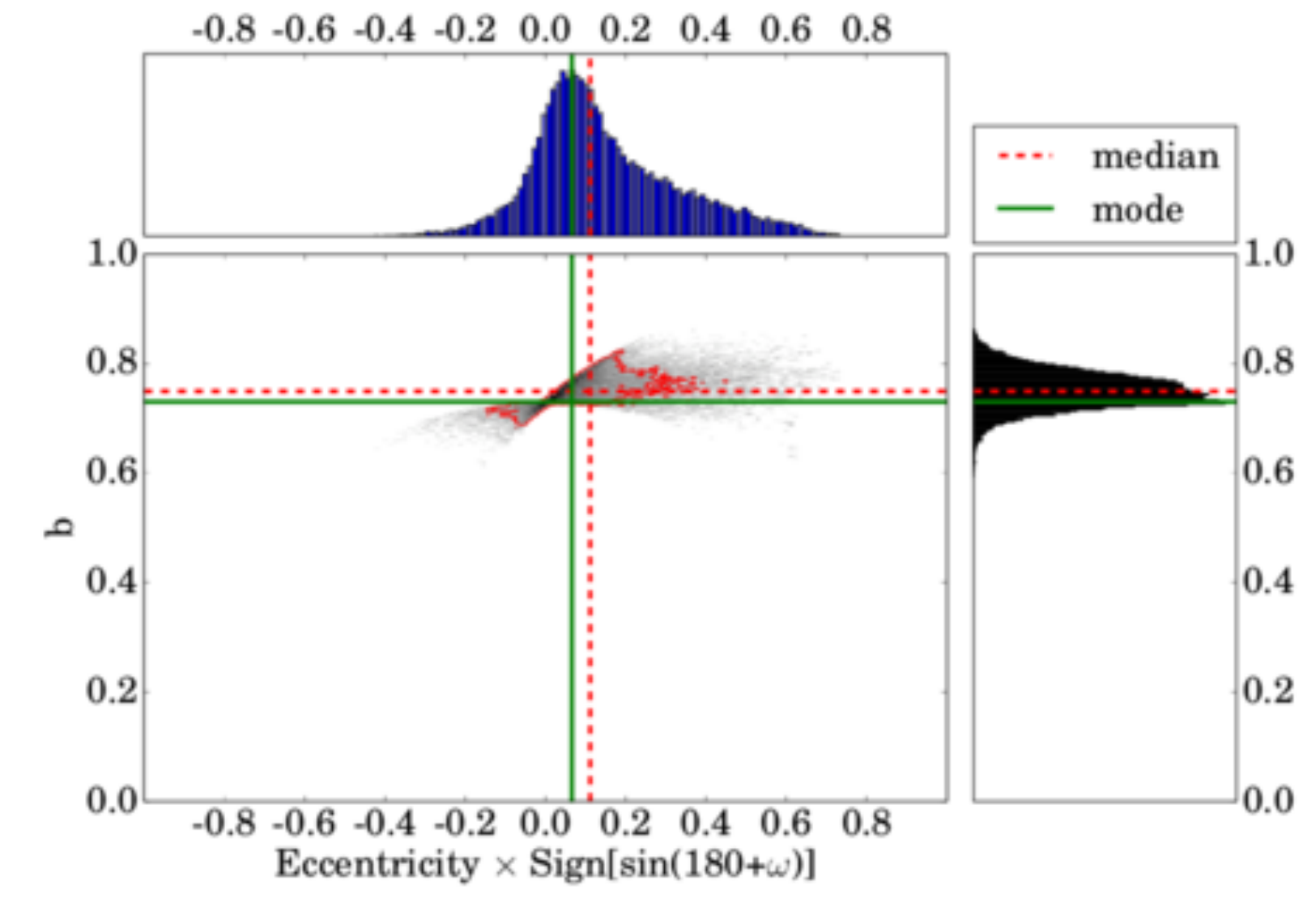}}
  \subfloat[Kepler-103c\label{fig:kepler103c}]{\includegraphics[width=0.33\linewidth]{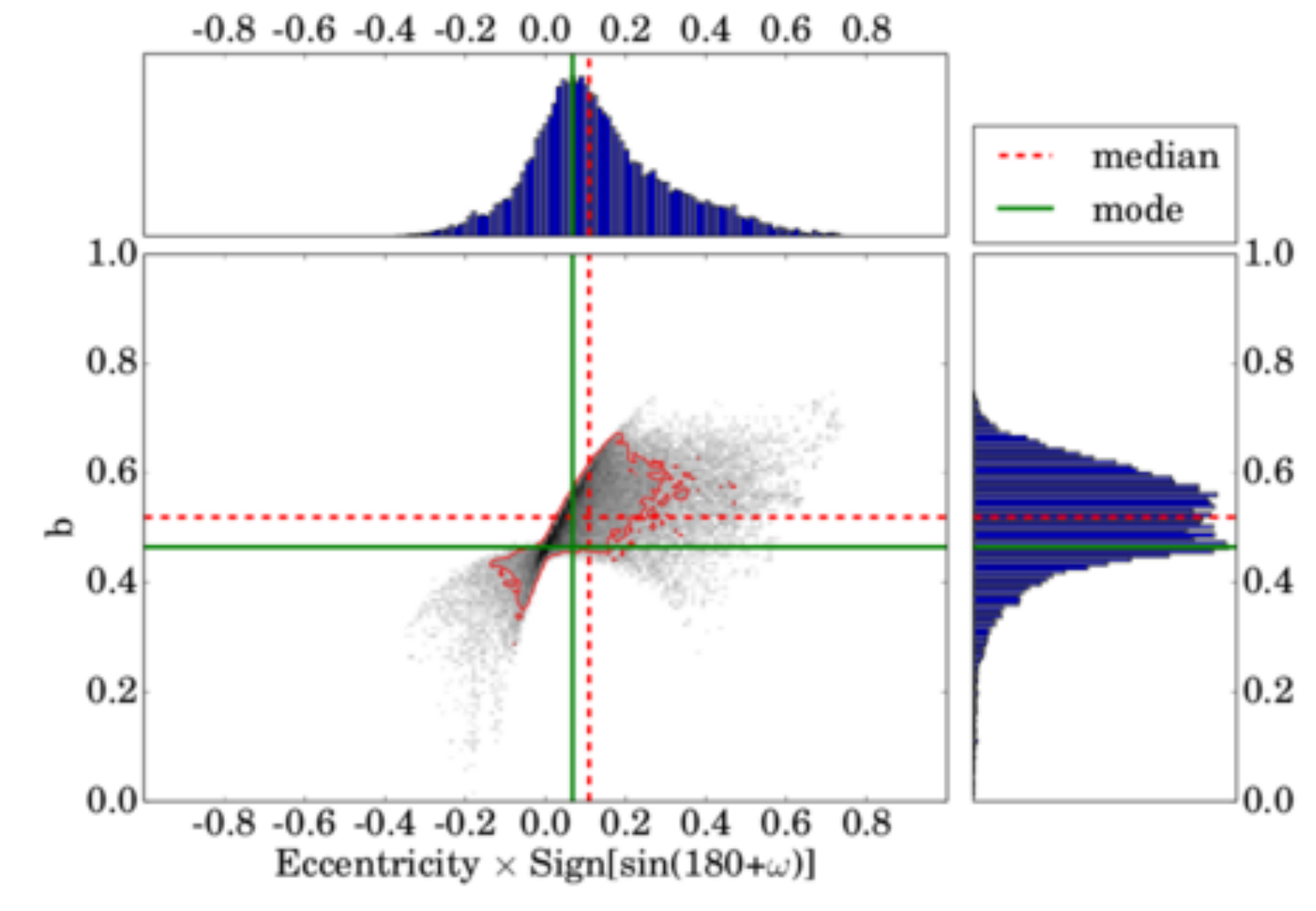}} 
  
  \subfloat[Kepler-107b\label{fig:kepler107b}]{\includegraphics[width=0.33\linewidth]{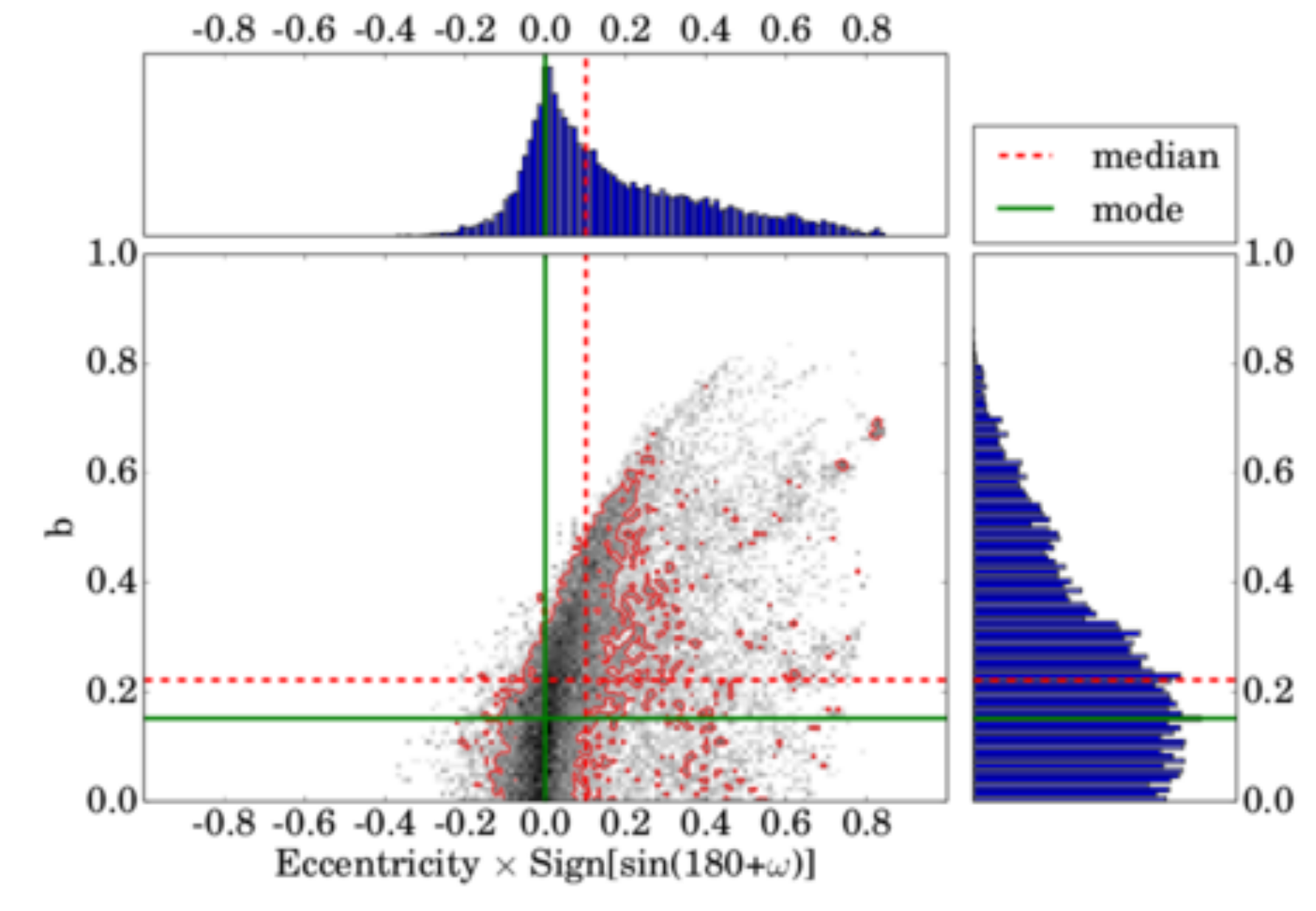}}
  \subfloat[Kepler-107c\label{fig:kepler107c}]{\includegraphics[width=0.33\linewidth]{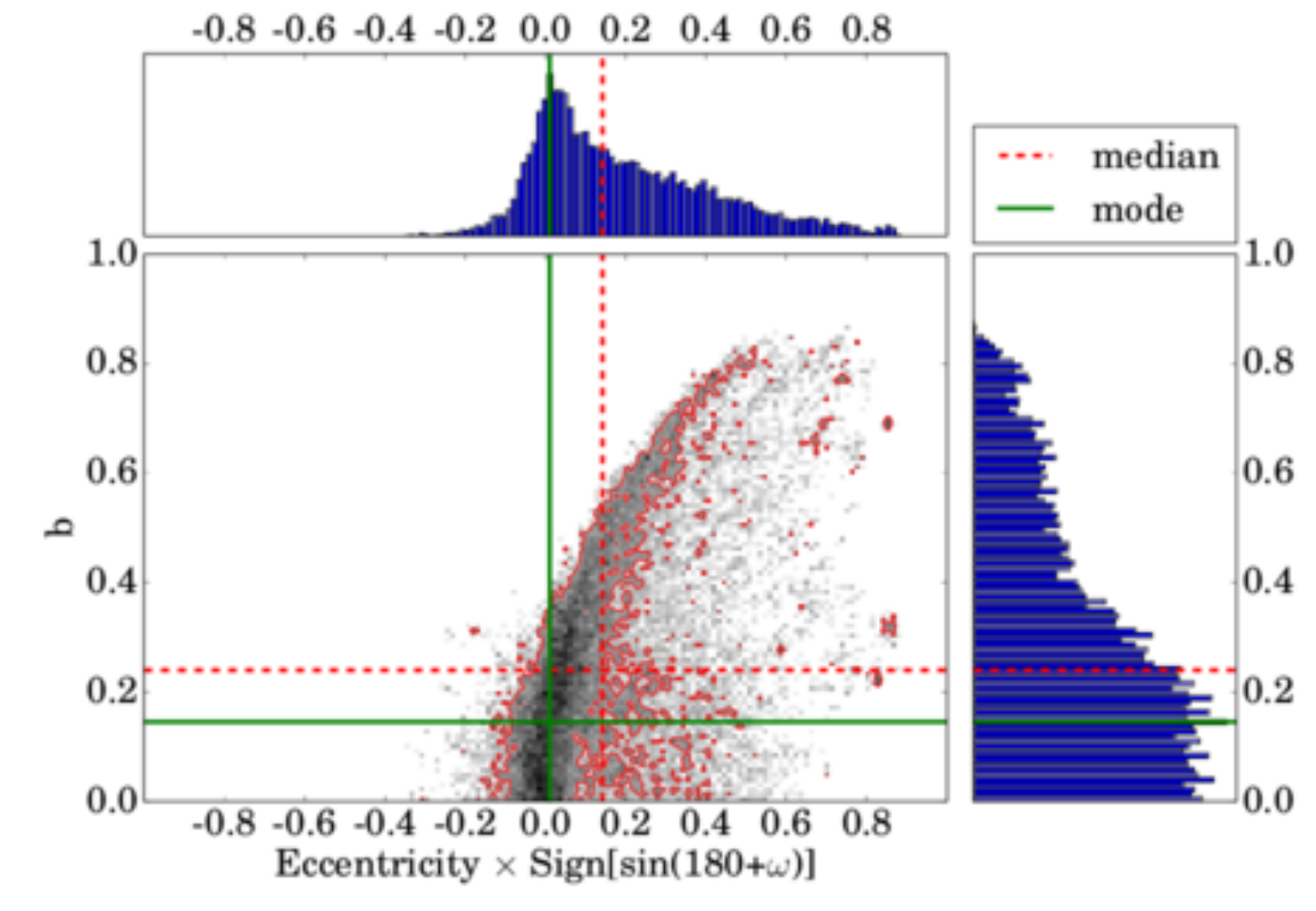}}
  \subfloat[Kepler-107d\label{fig:kepler107d}]{\includegraphics[width=0.33\linewidth]{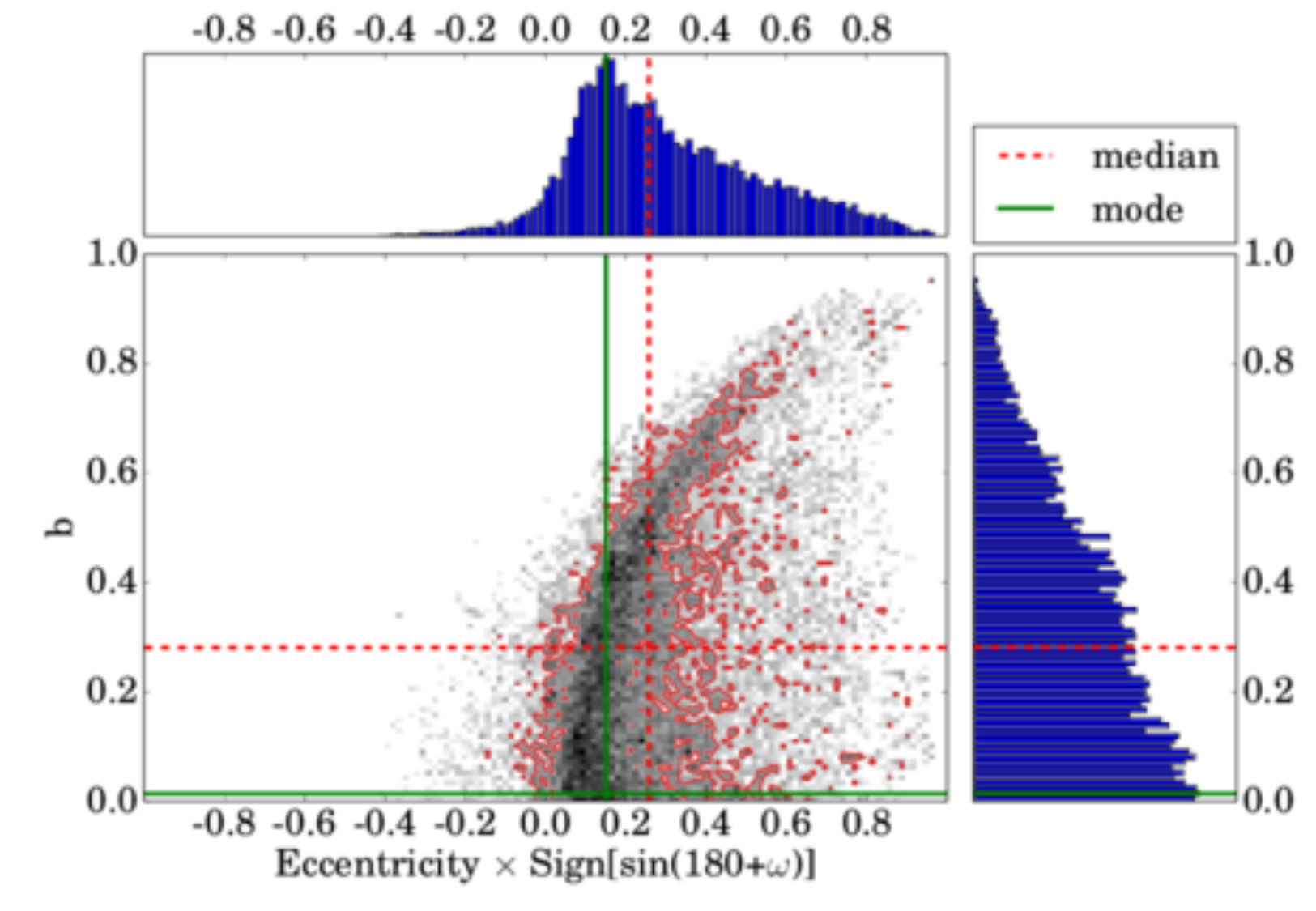}}
  
  \subfloat[Kepler-107e\label{fig:kepler107e}]{\includegraphics[width=0.33\linewidth]{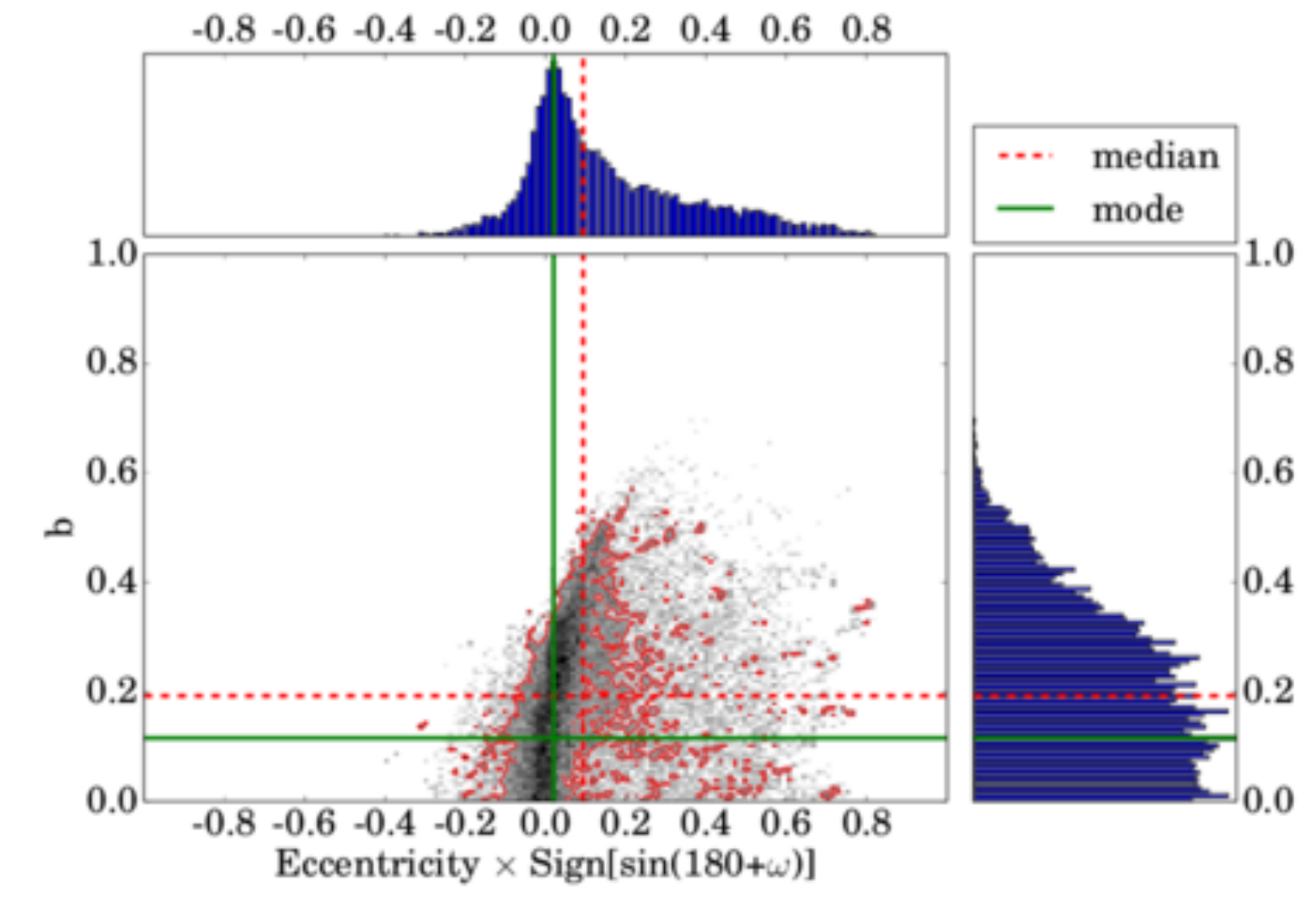}}  
  \subfloat[Kepler-108b\label{fig:kepler108b}]{\includegraphics[width=0.33\linewidth]{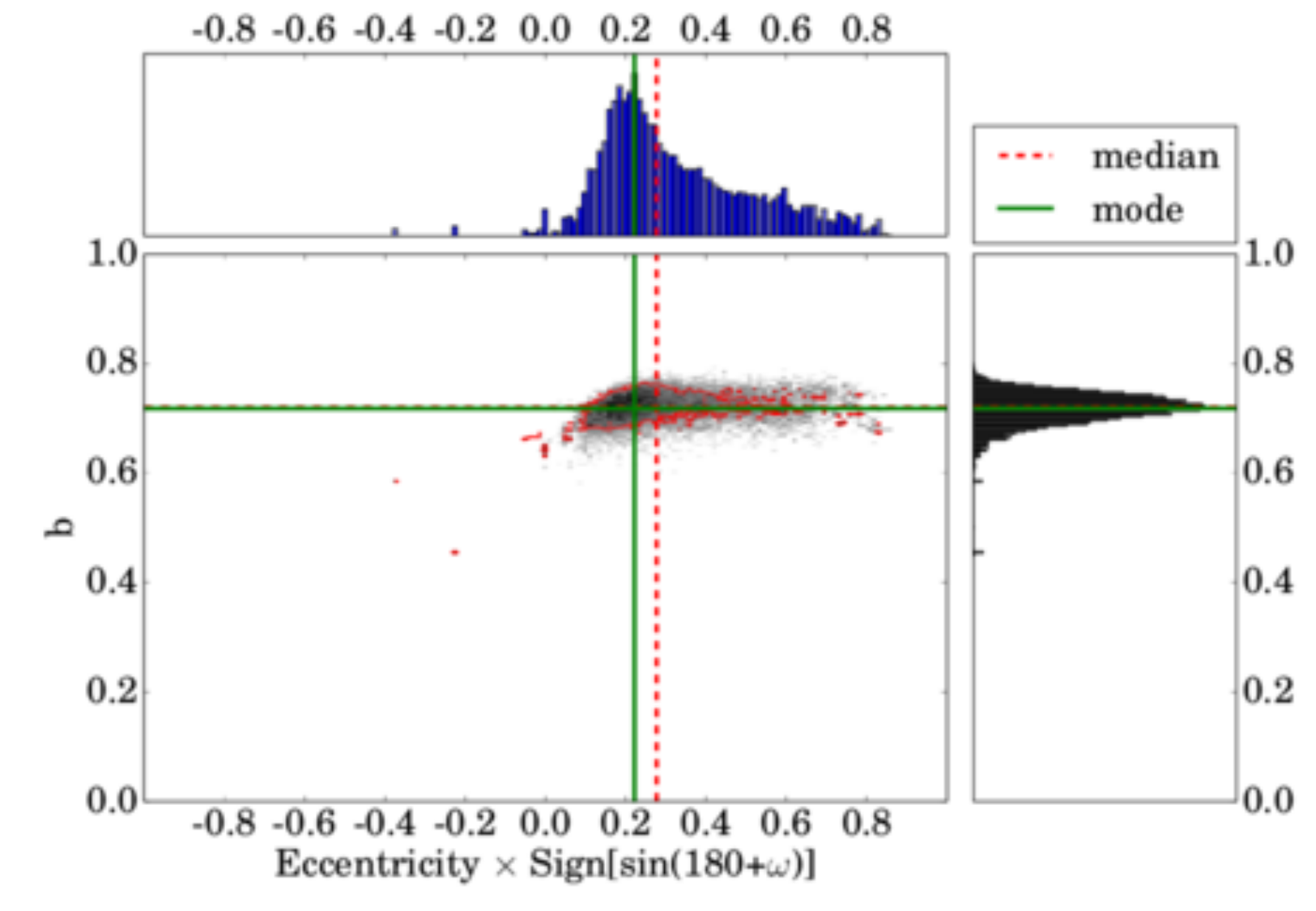}}
  \subfloat[Kepler-108c\label{fig:kepler108c}]{\includegraphics[width=0.33\linewidth]{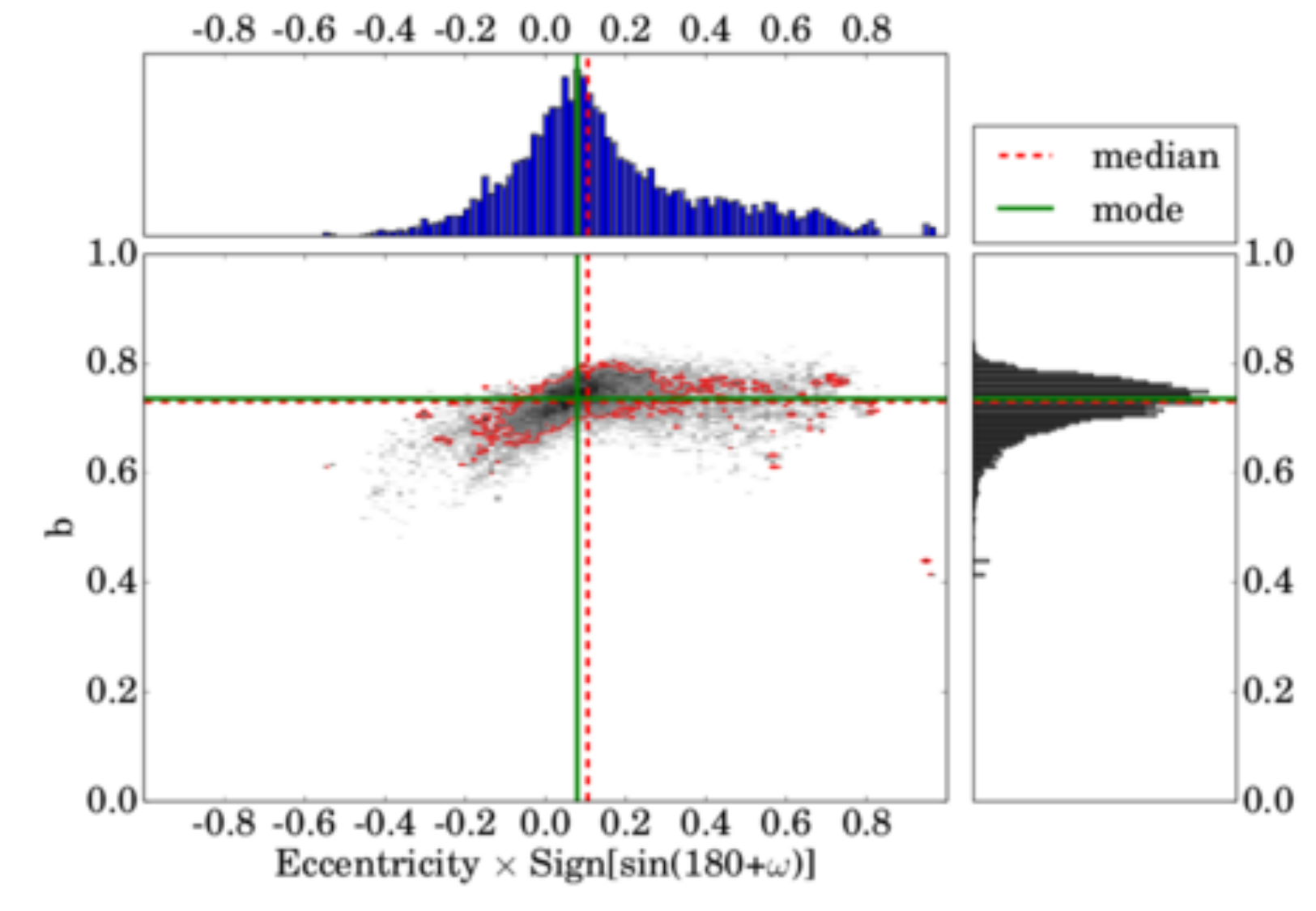}}
\end{widepage}
 \caption{(continued) Posterior distributions of individual planets}
\end{figure*}

\begin{figure*}[t]
 \ContinuedFloat
\begin{widepage}
  
  \subfloat[Kepler-109b\label{fig:kepler109b}]{\includegraphics[width=0.33\linewidth]{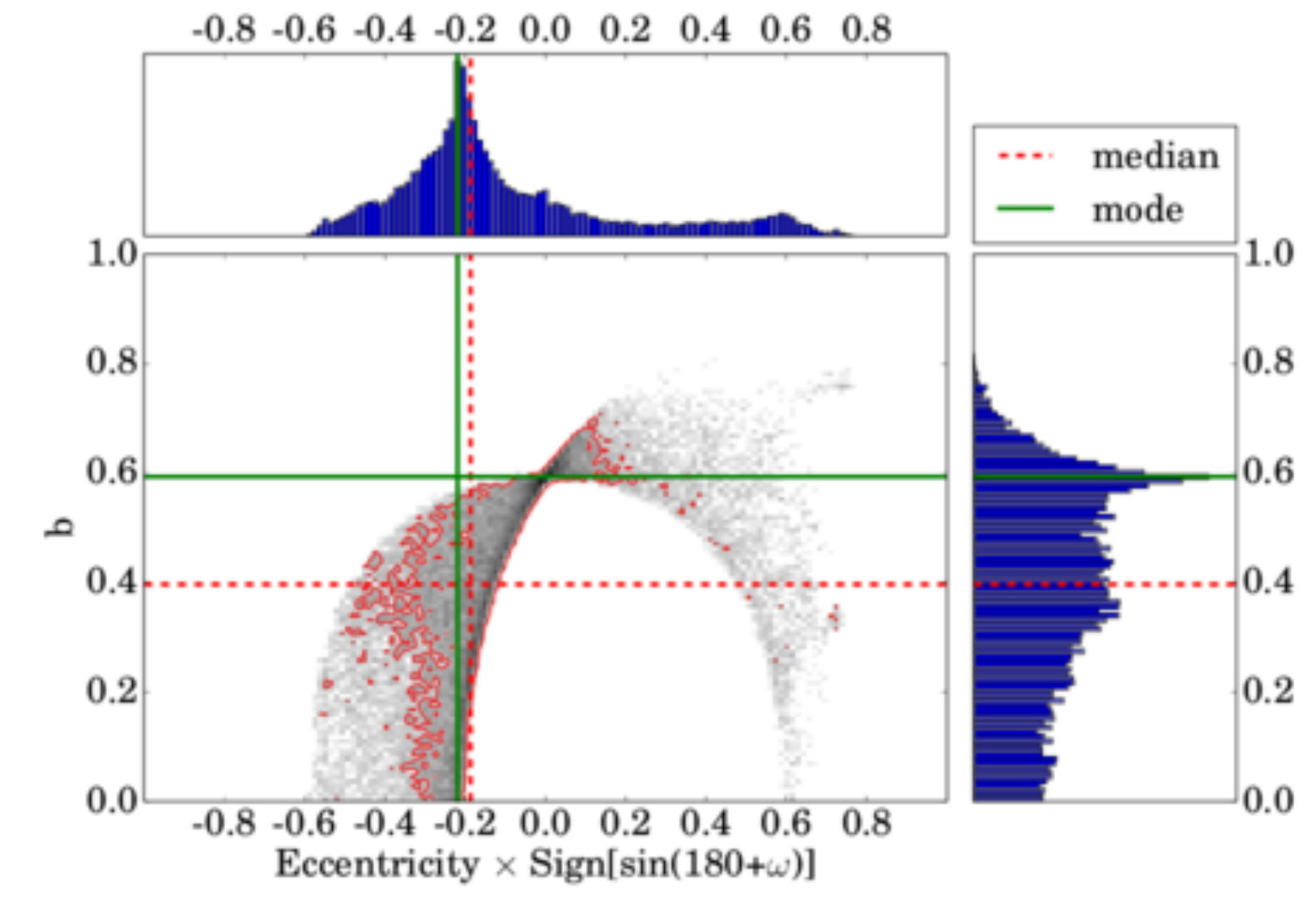}}  
  \subfloat[Kepler-109c\label{fig:kepler109c}]{\includegraphics[width=0.33\linewidth]{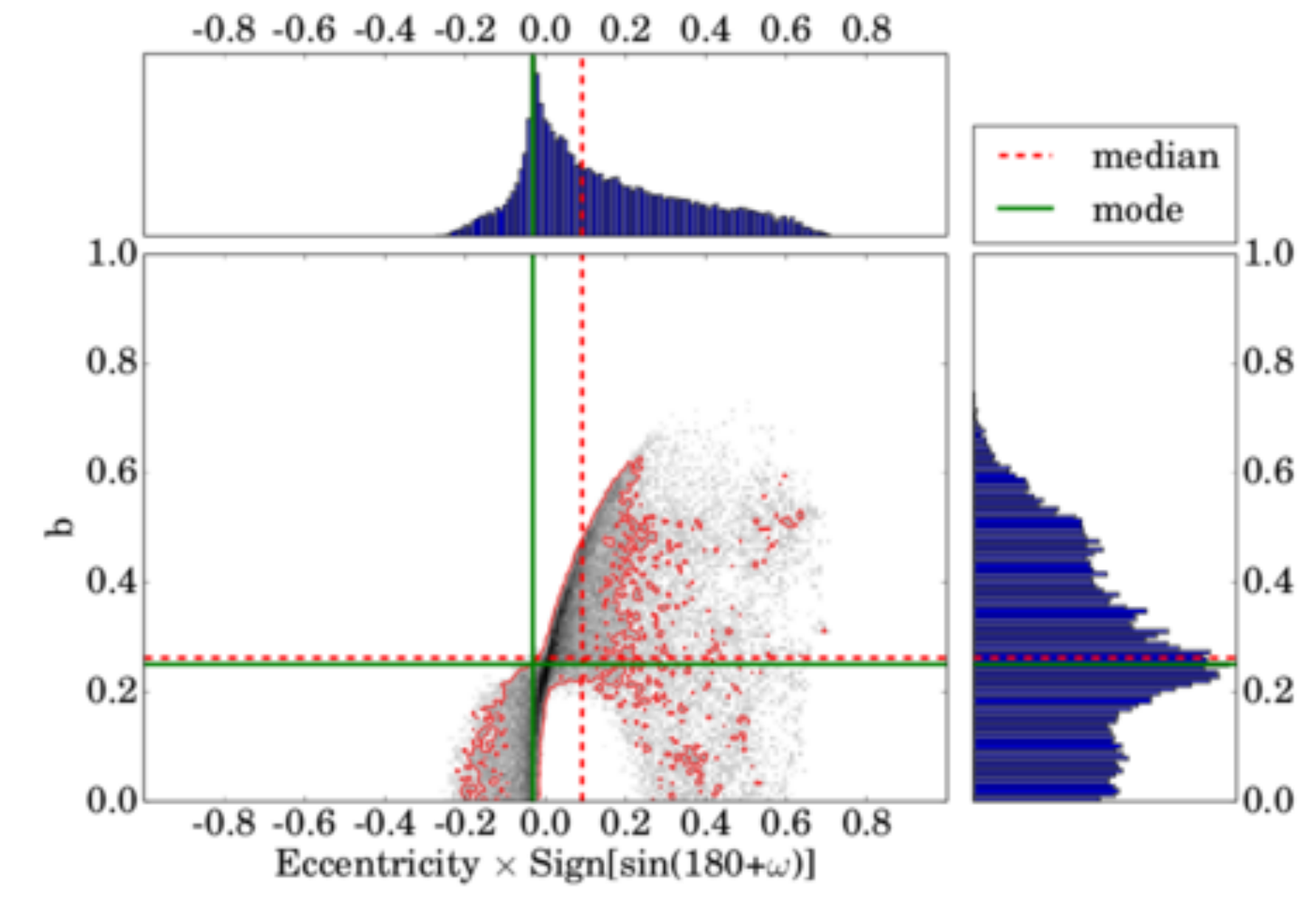}} 
  \subfloat[Kepler-126b\label{fig:kepler126b}]{\includegraphics[width=0.33\linewidth]{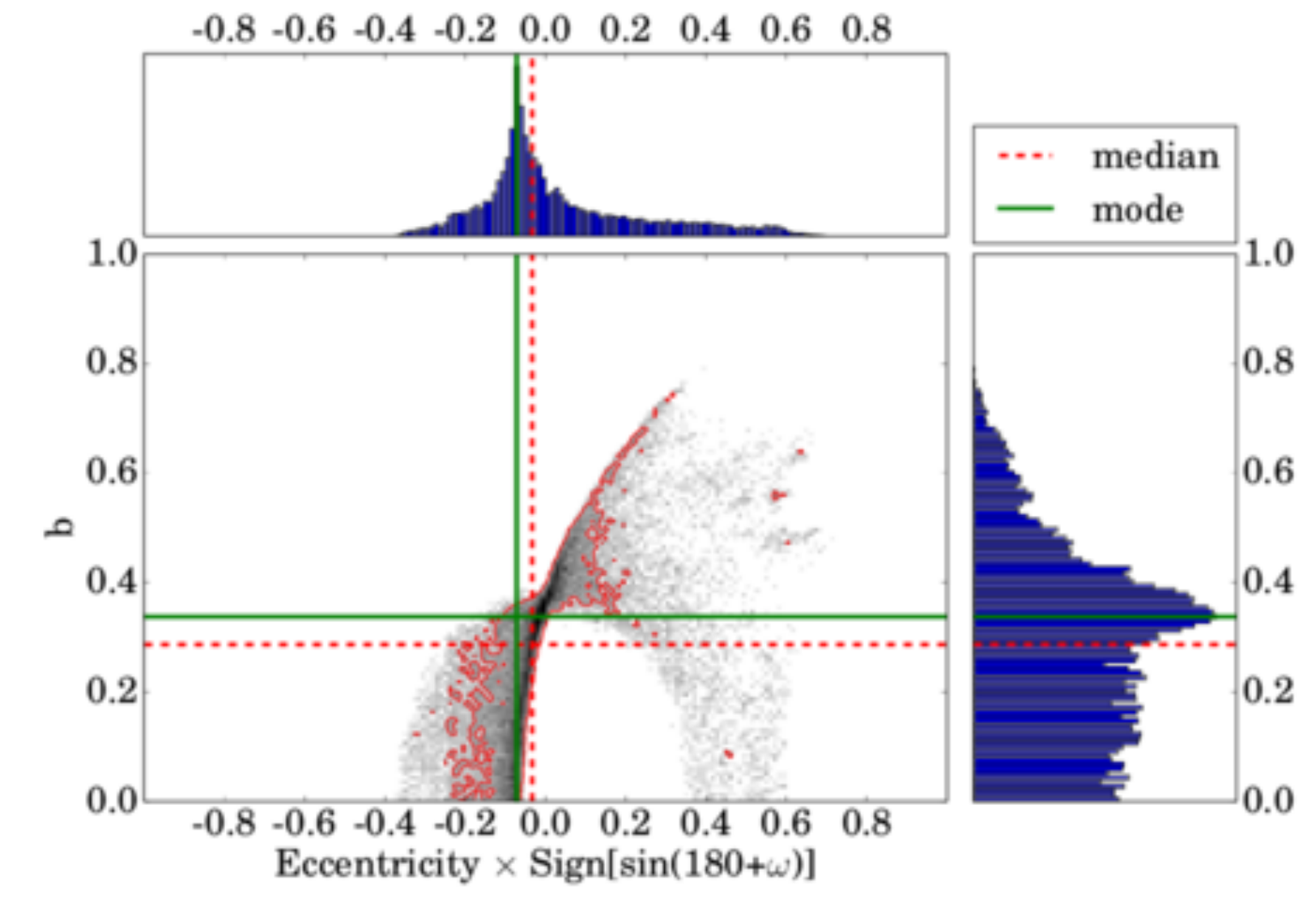}}
  
  \subfloat[Kepler-126c\label{fig:kepler126c}]{\includegraphics[width=0.33\linewidth]{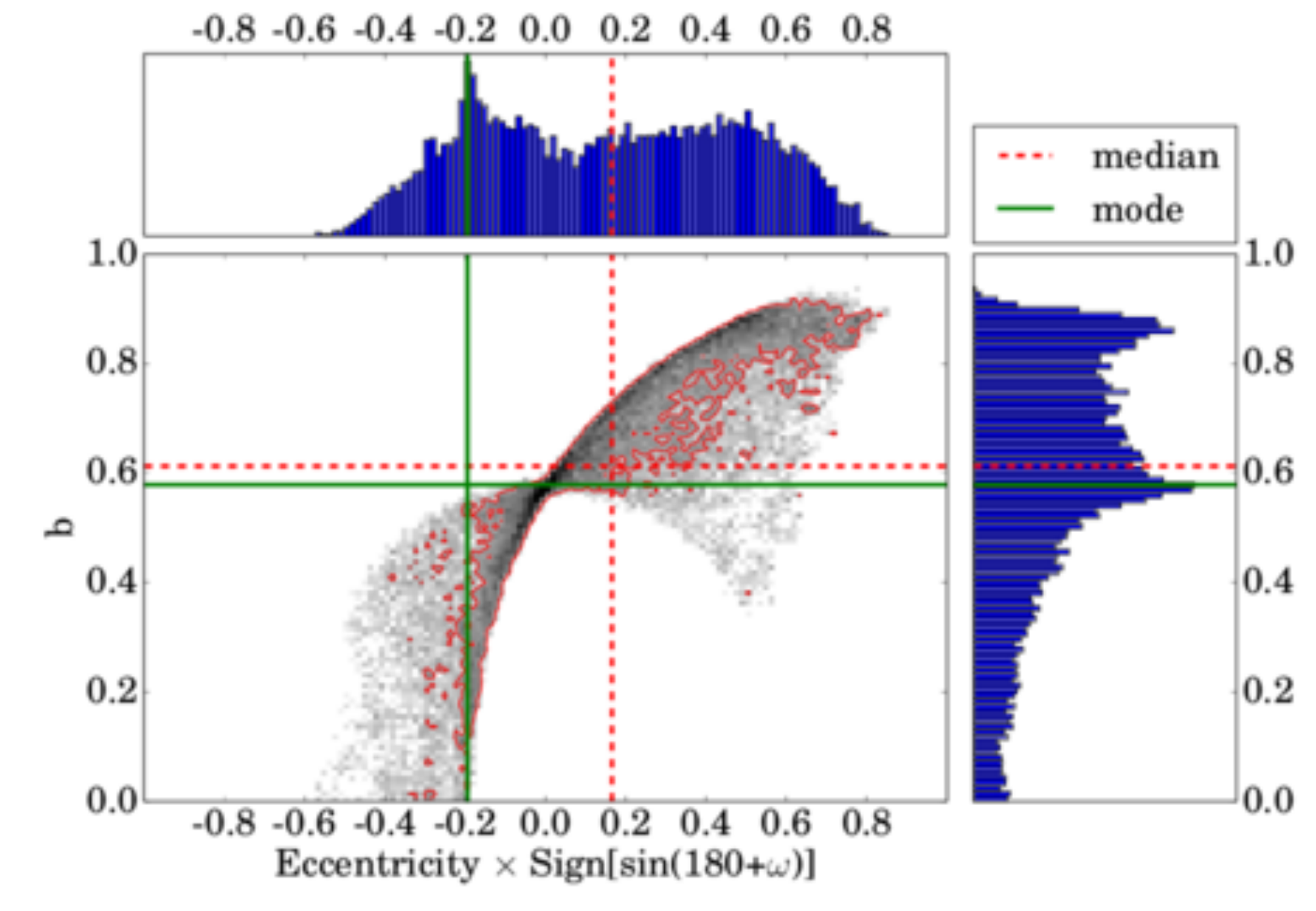}}
  \subfloat[Kepler-126d\label{fig:kepler126d}]{\includegraphics[width=0.33\linewidth]{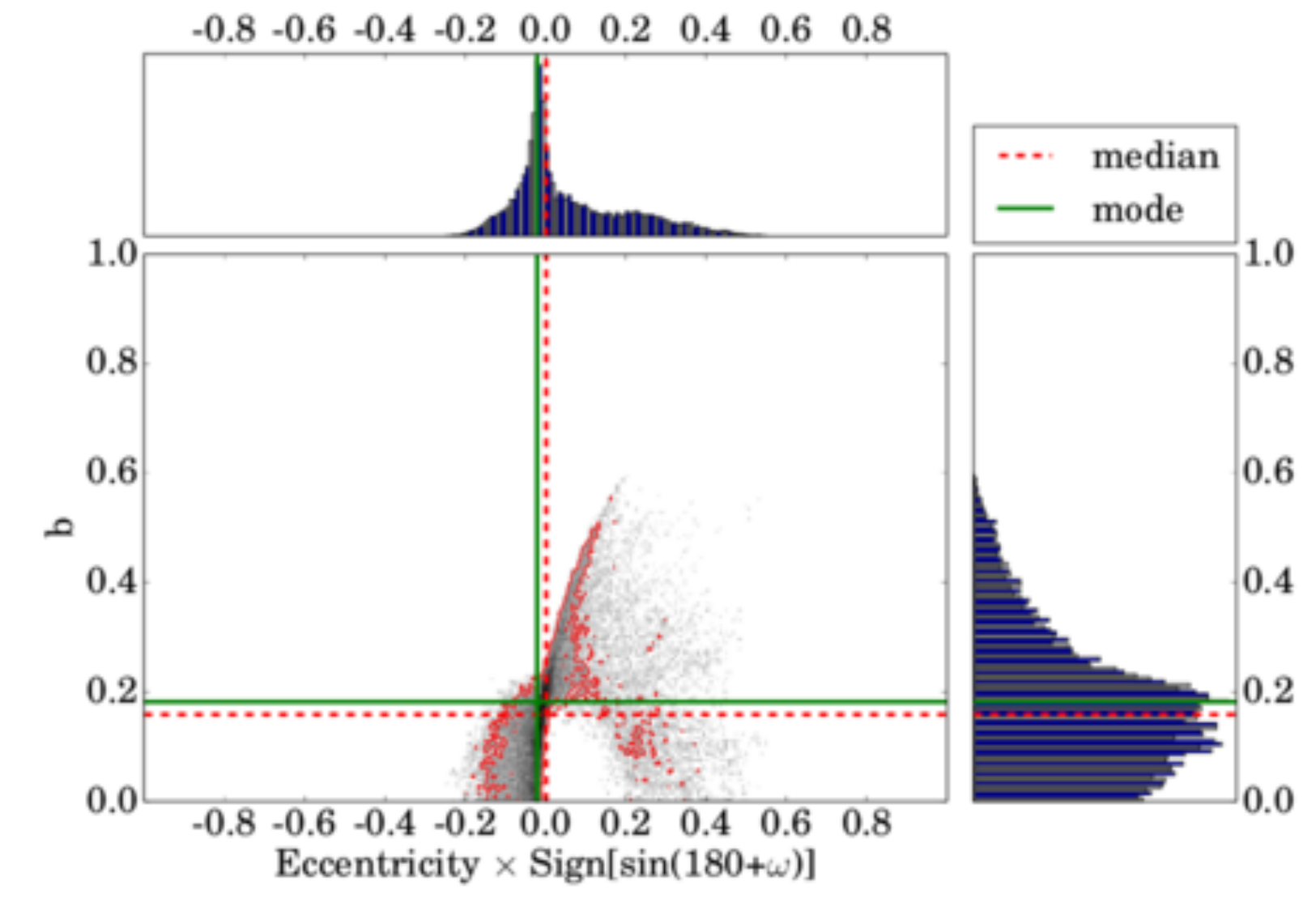}}
  \subfloat[Kepler-127b\label{fig:kepler127b}]{\includegraphics[width=0.33\linewidth]{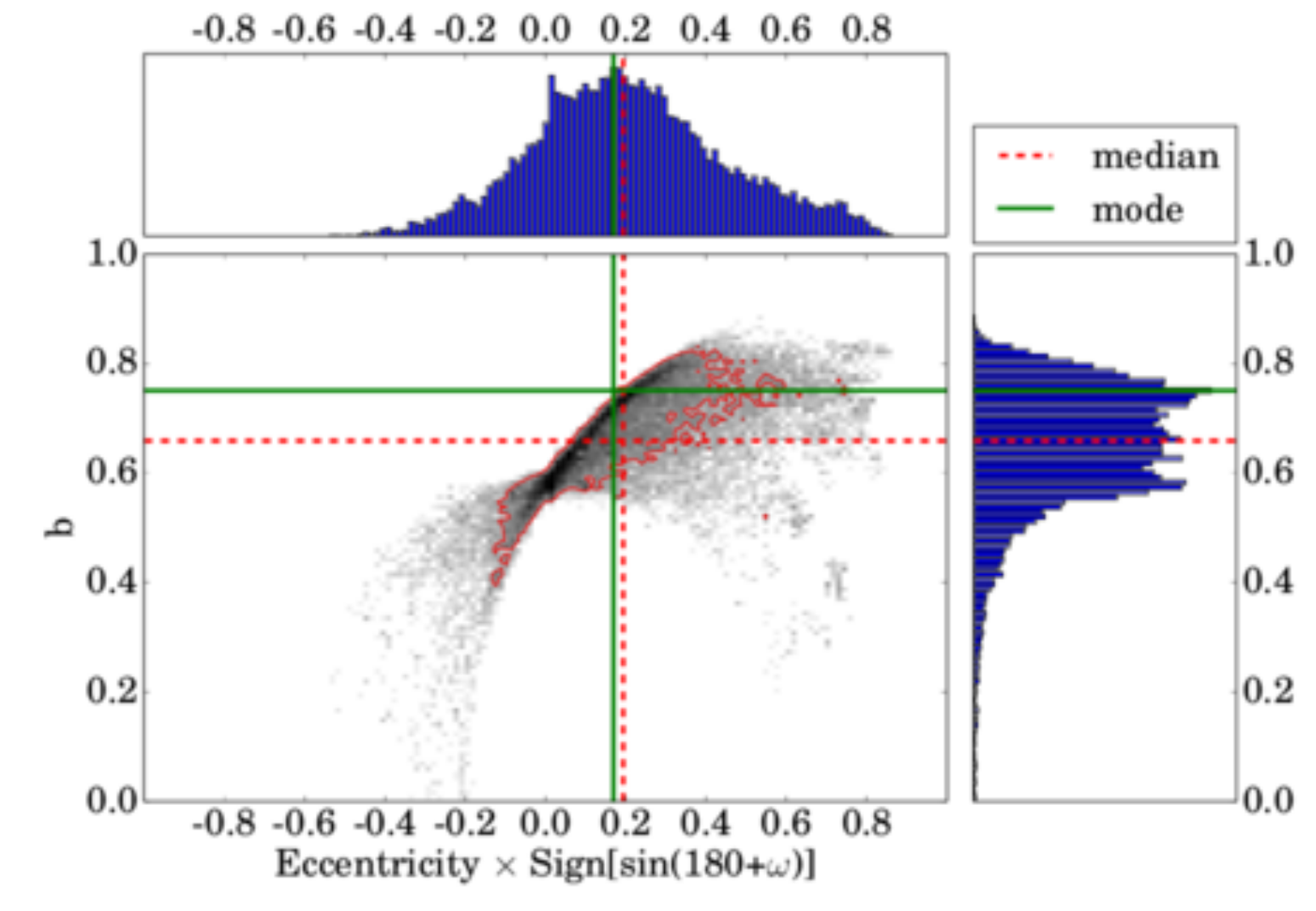}}
  
  \subfloat[Kepler-127c\label{fig:kepler127c}]{\includegraphics[width=0.33\linewidth]{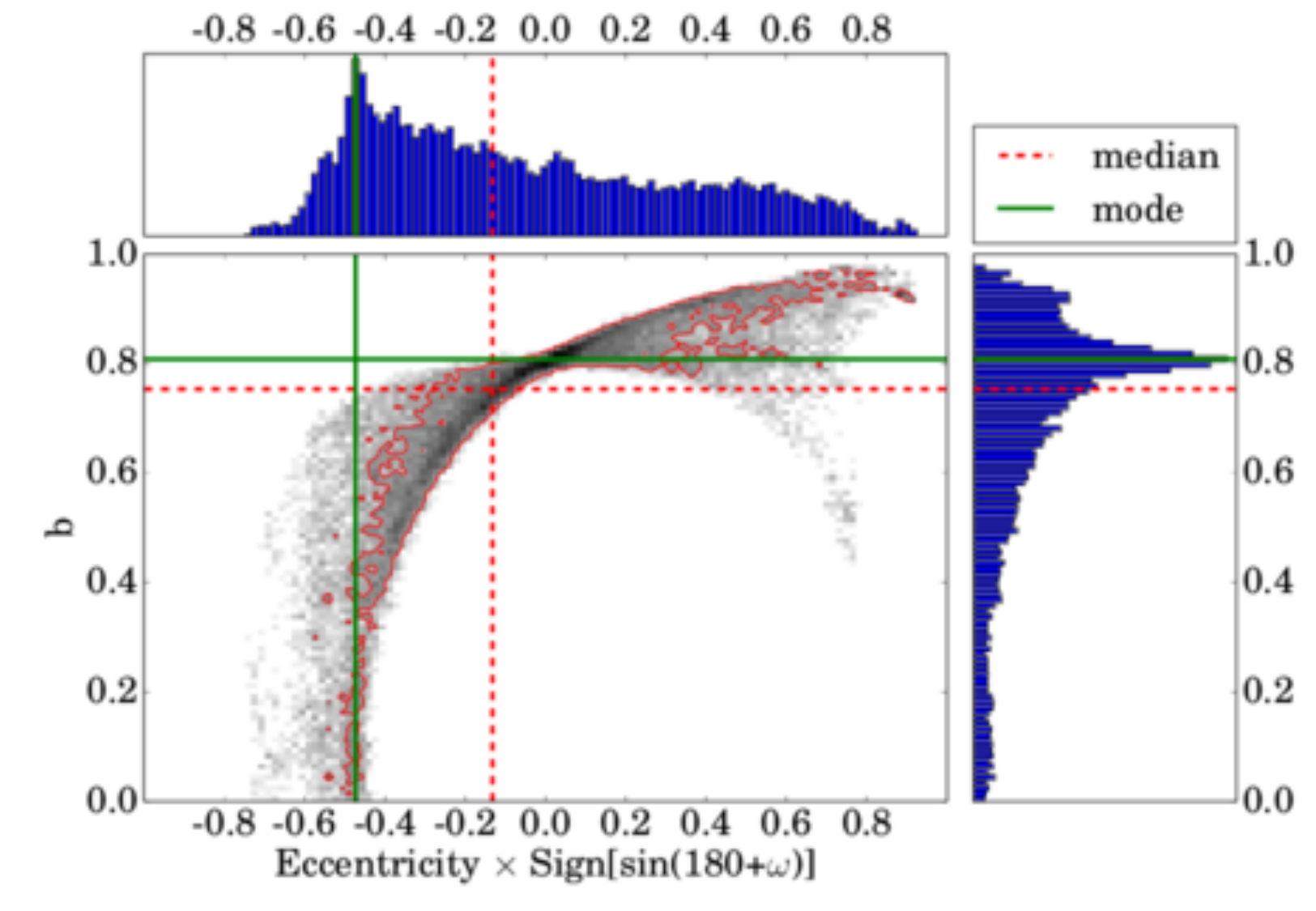}}
  \subfloat[Kepler-127d\label{fig:kepler127d}]{\includegraphics[width=0.33\linewidth]{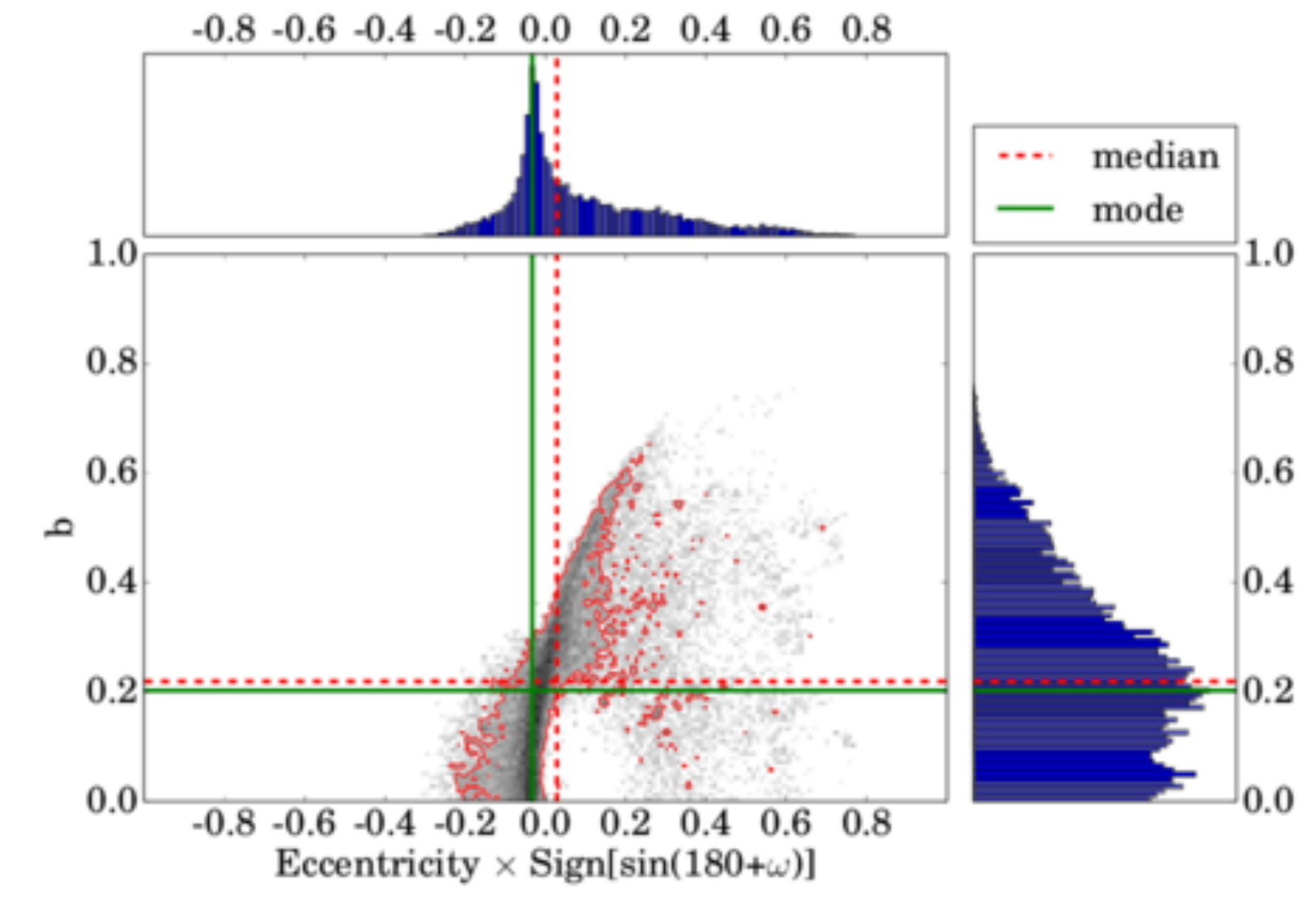}}
  \subfloat[Kepler-129b\label{fig:kepler129b}]{\includegraphics[width=0.33\linewidth]{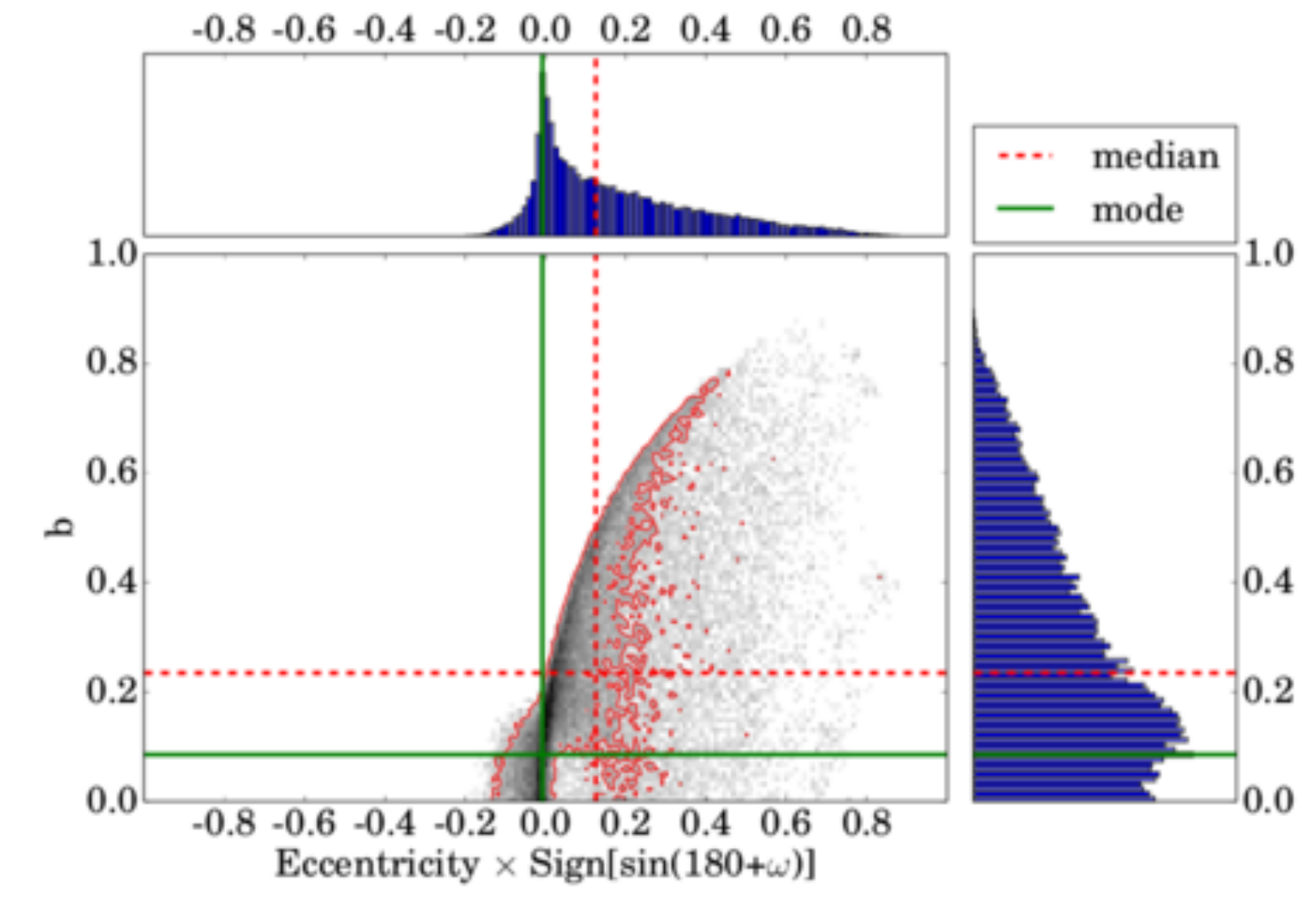}}
  
  \subfloat[Kepler-129c\label{fig:kepler129c}]{\includegraphics[width=0.33\linewidth]{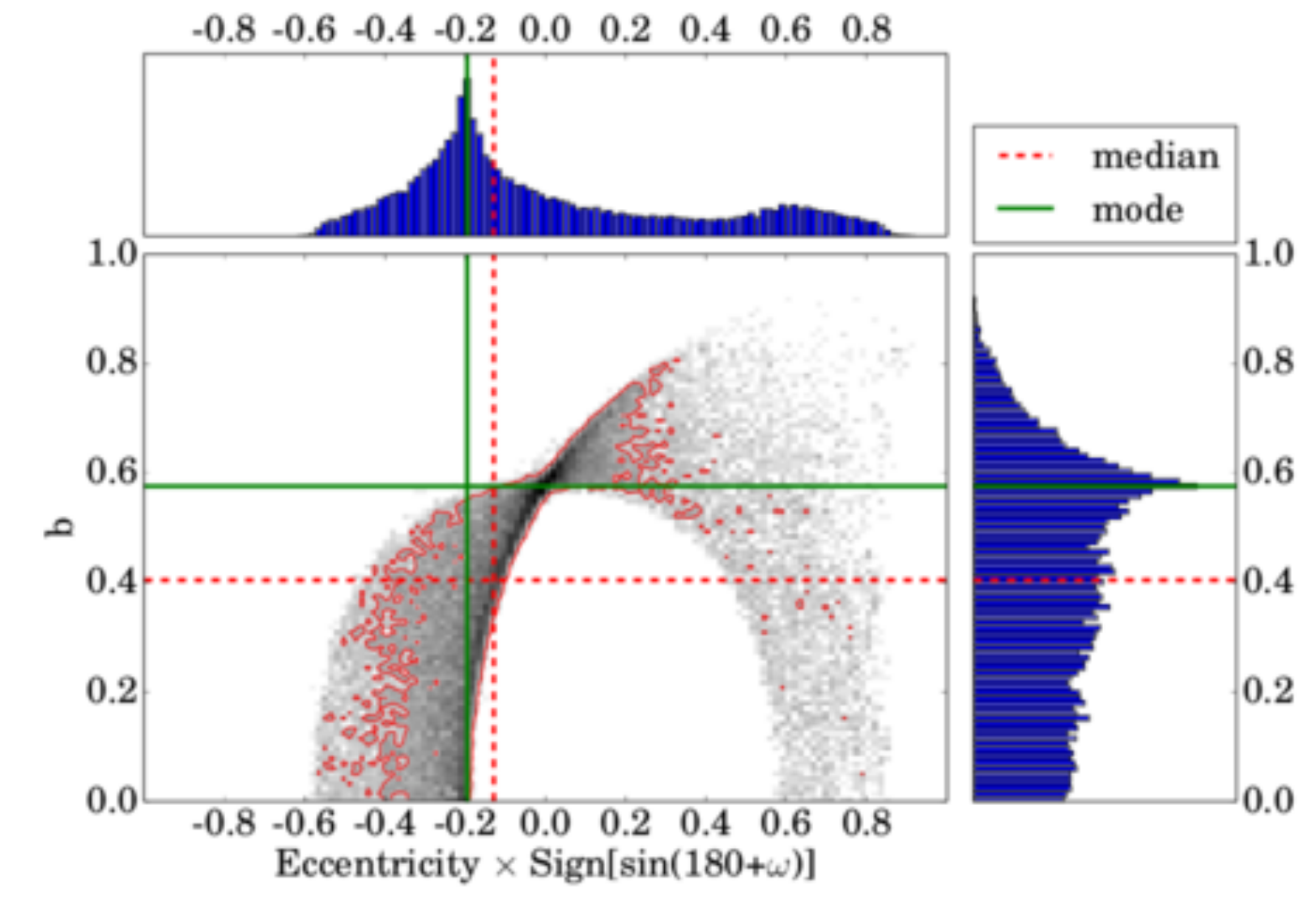}}  
  \subfloat[Kepler-130b\label{fig:kepler130b}]{\includegraphics[width=0.33\linewidth]{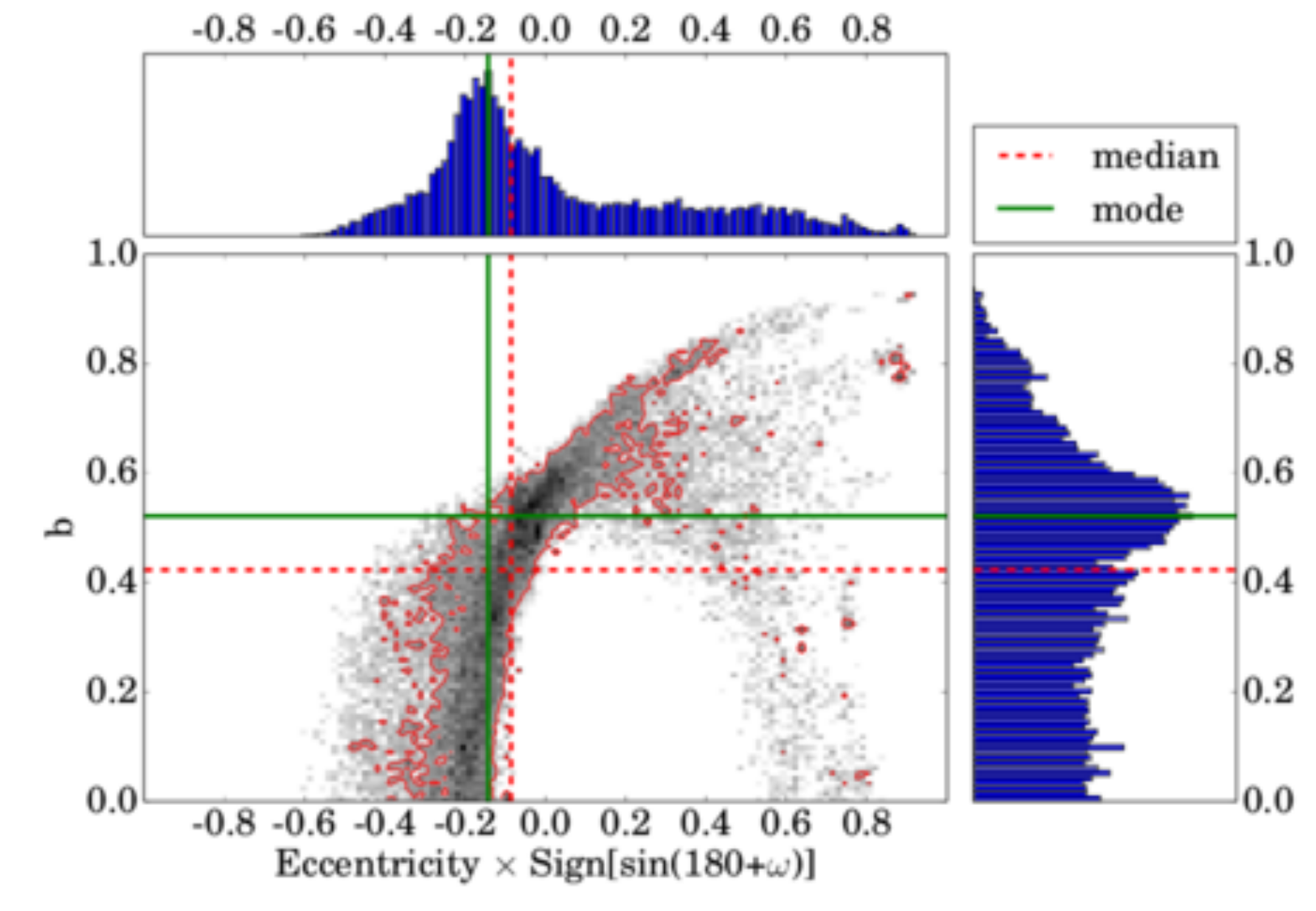}}
  \subfloat[Kepler-130c\label{fig:kepler130c}]{\includegraphics[width=0.33\linewidth]{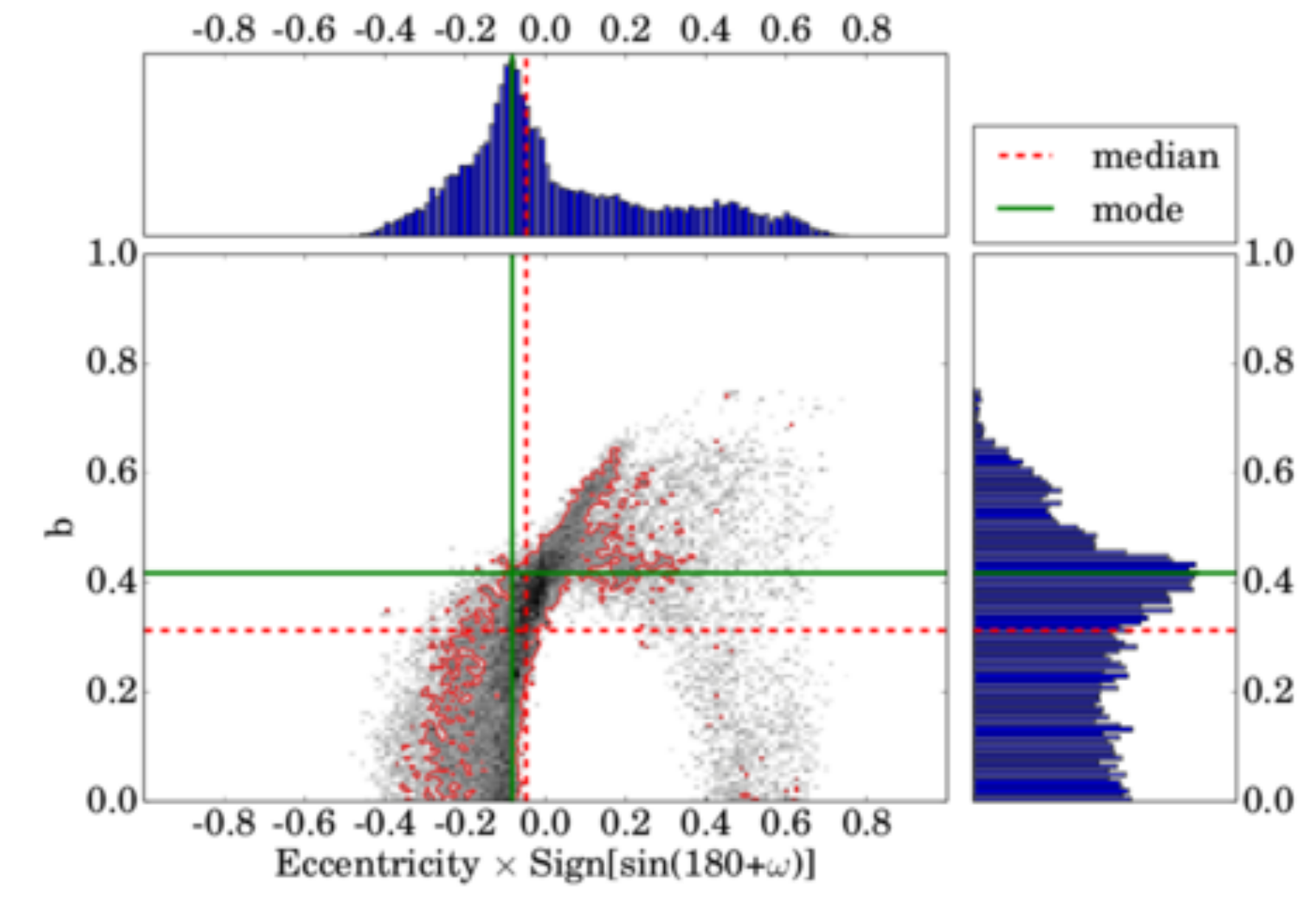}}
  
\end{widepage}
 \caption{(continued) Posterior distributions of individual planets}
\end{figure*}

\begin{figure*}[t]
 \ContinuedFloat
    \begin{widepage}
   \subfloat[Kepler-130d\label{fig:kepler130d}]{\includegraphics[width=0.33\linewidth]{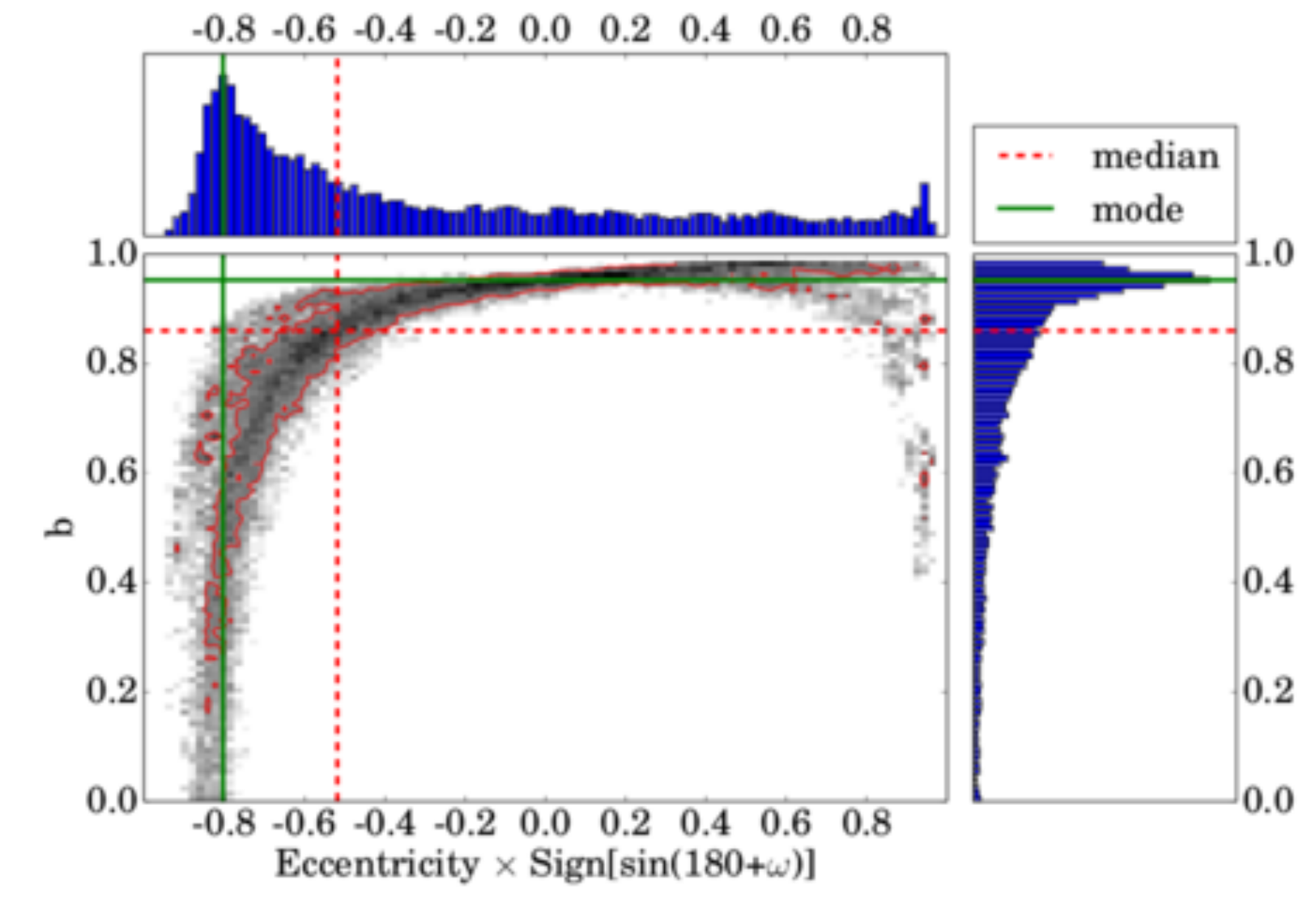}}
  \subfloat[Kepler-145b\label{fig:kepler145b}]{\includegraphics[width=0.33\linewidth]{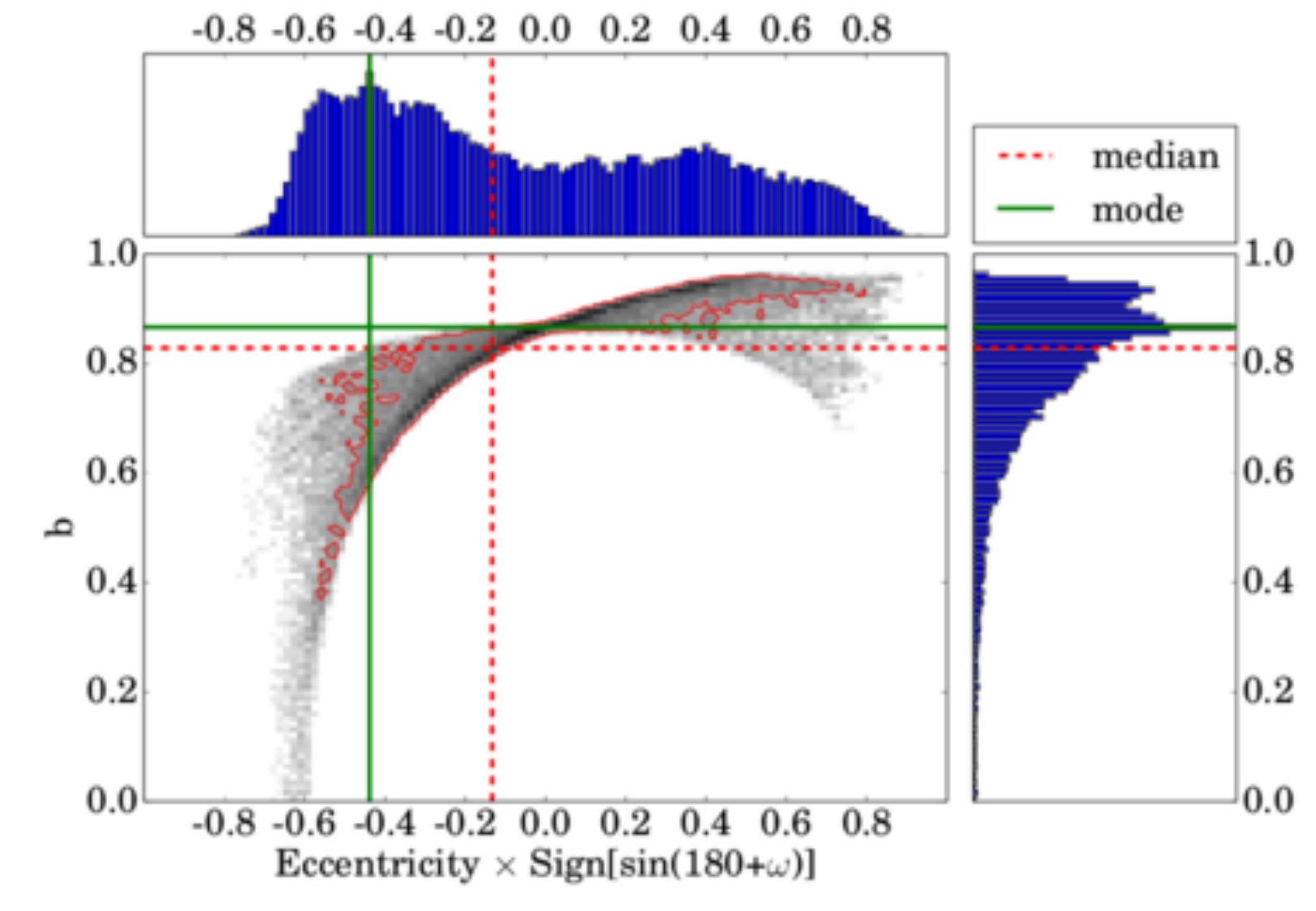}}
  \subfloat[Kepler-145c\label{fig:kepler145c}]{\includegraphics[width=0.33\linewidth]{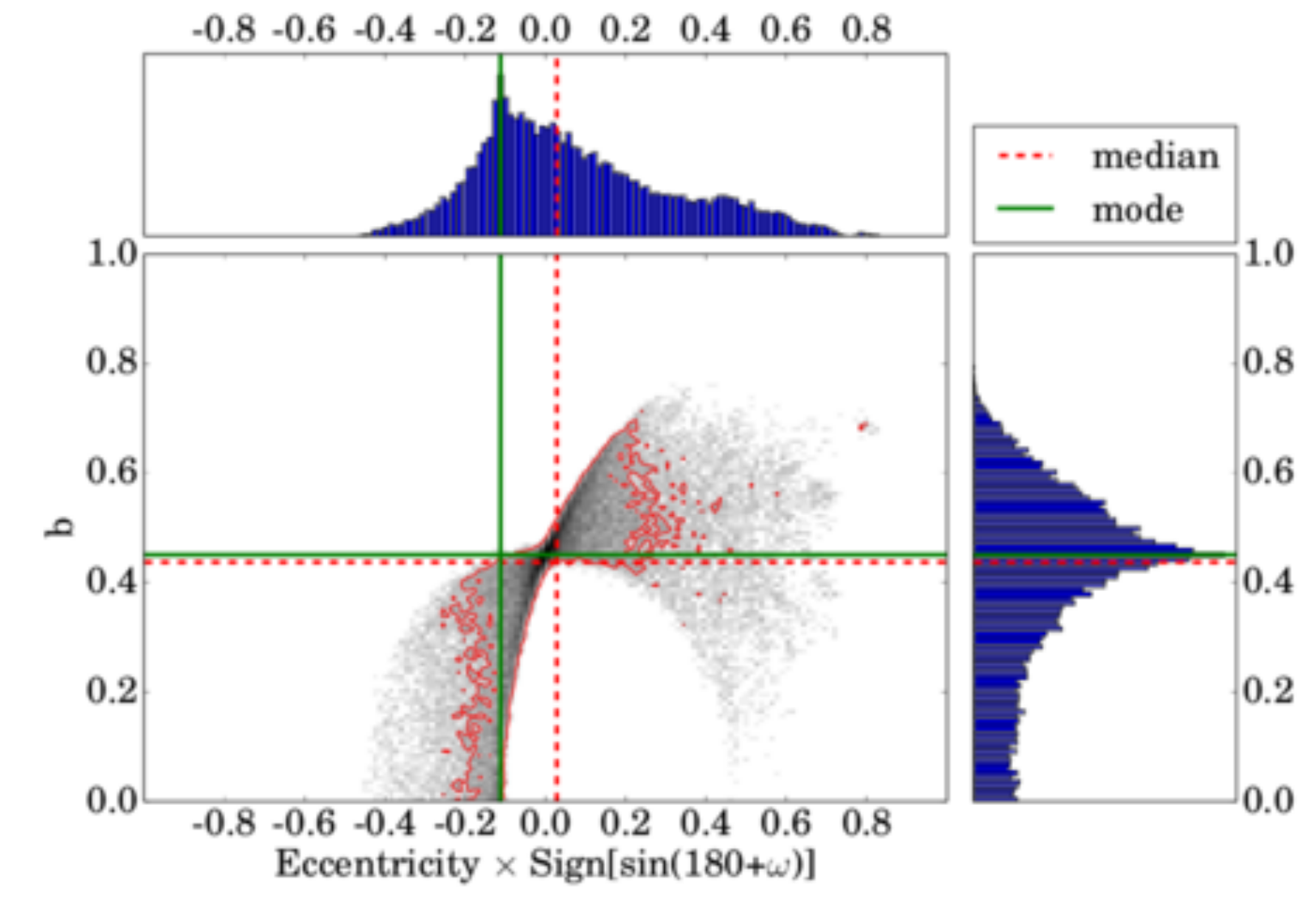}}
  
  \subfloat[Kepler-197b\label{fig:kepler197b}]{\includegraphics[width=0.33\linewidth]{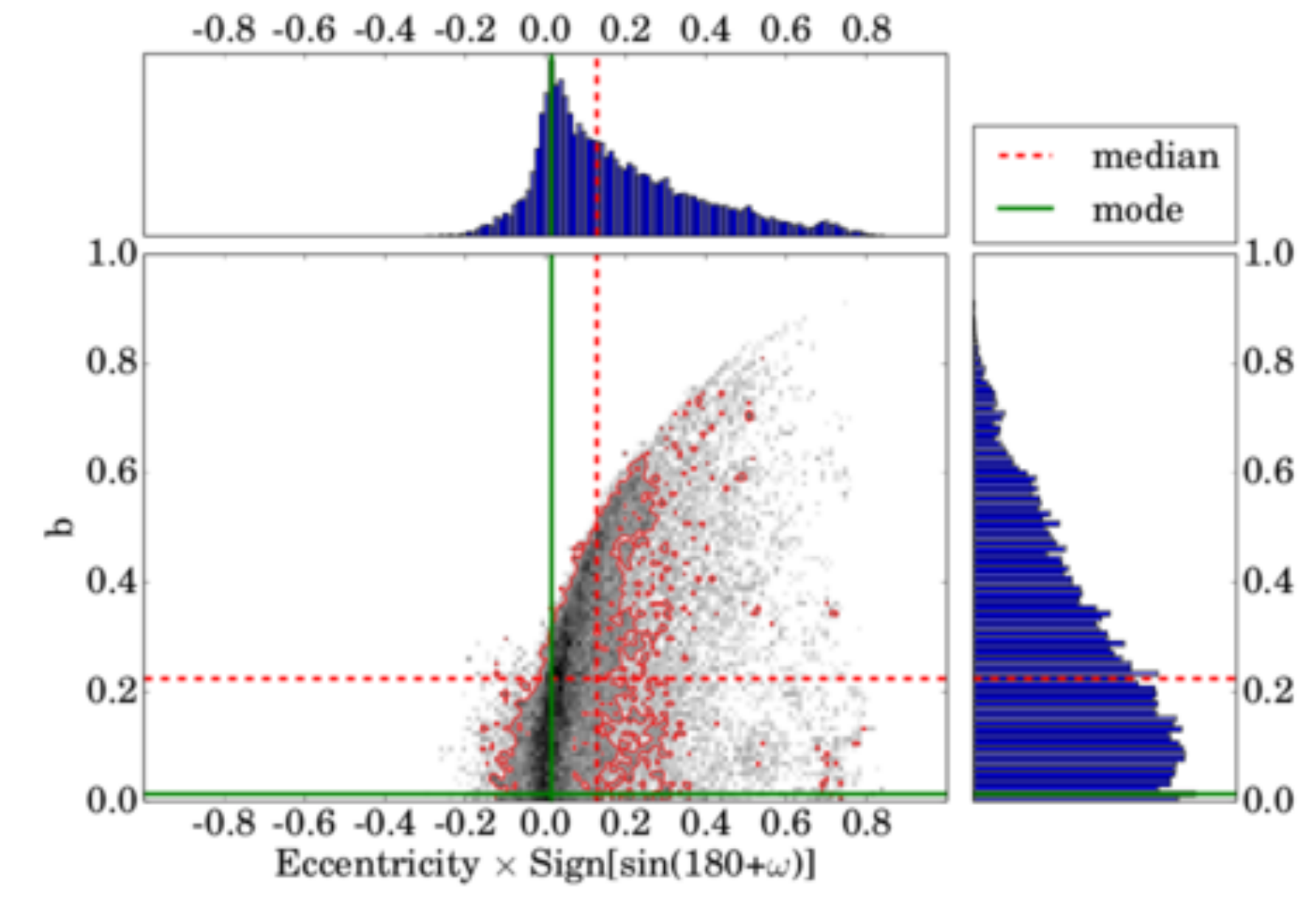}}
  \subfloat[Kepler-197c\label{fig:kepler197c}]{\includegraphics[width=0.33\linewidth]{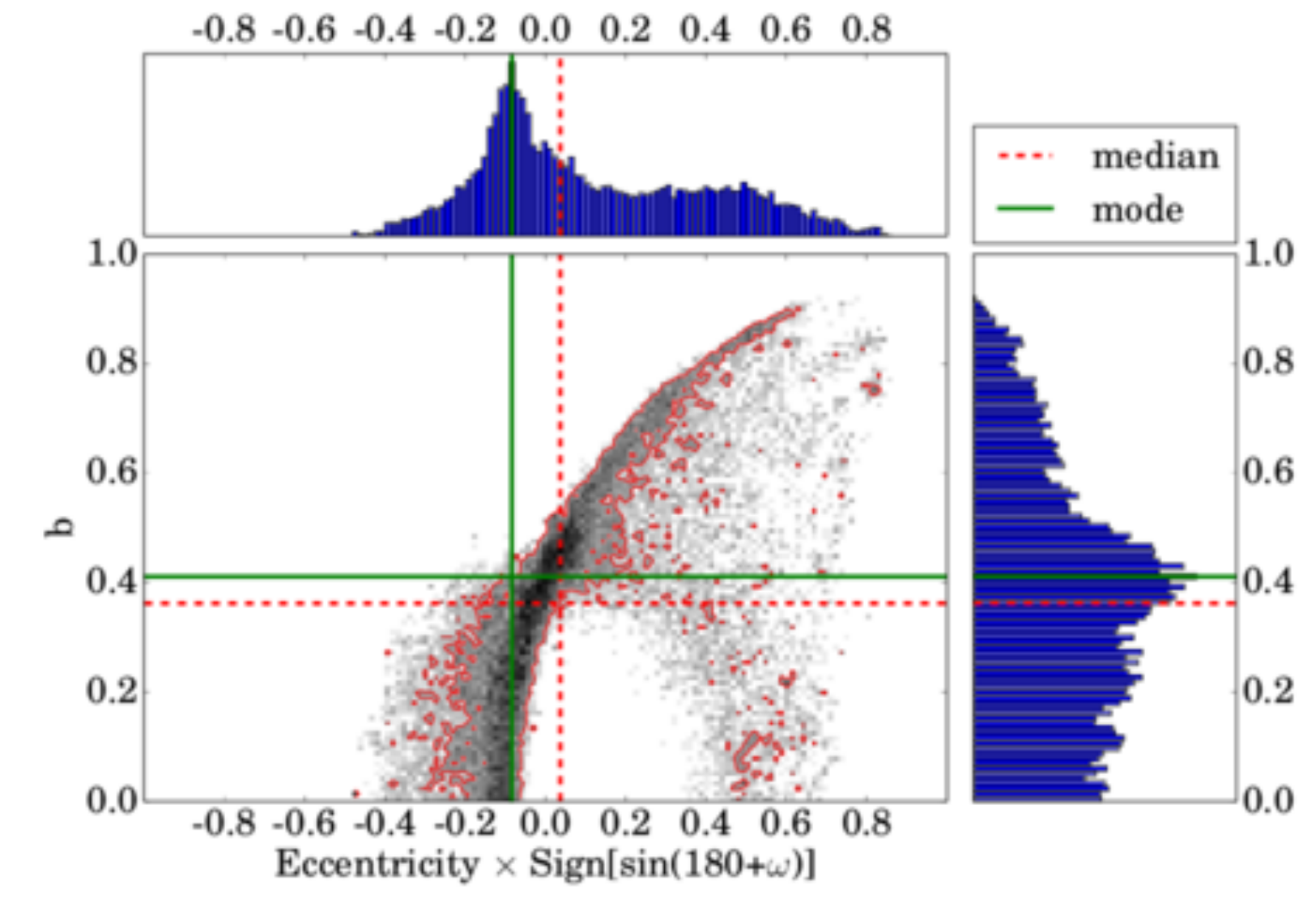}}
  \subfloat[Kepler-197d\label{fig:kepler197d}]{\includegraphics[width=0.33\linewidth]{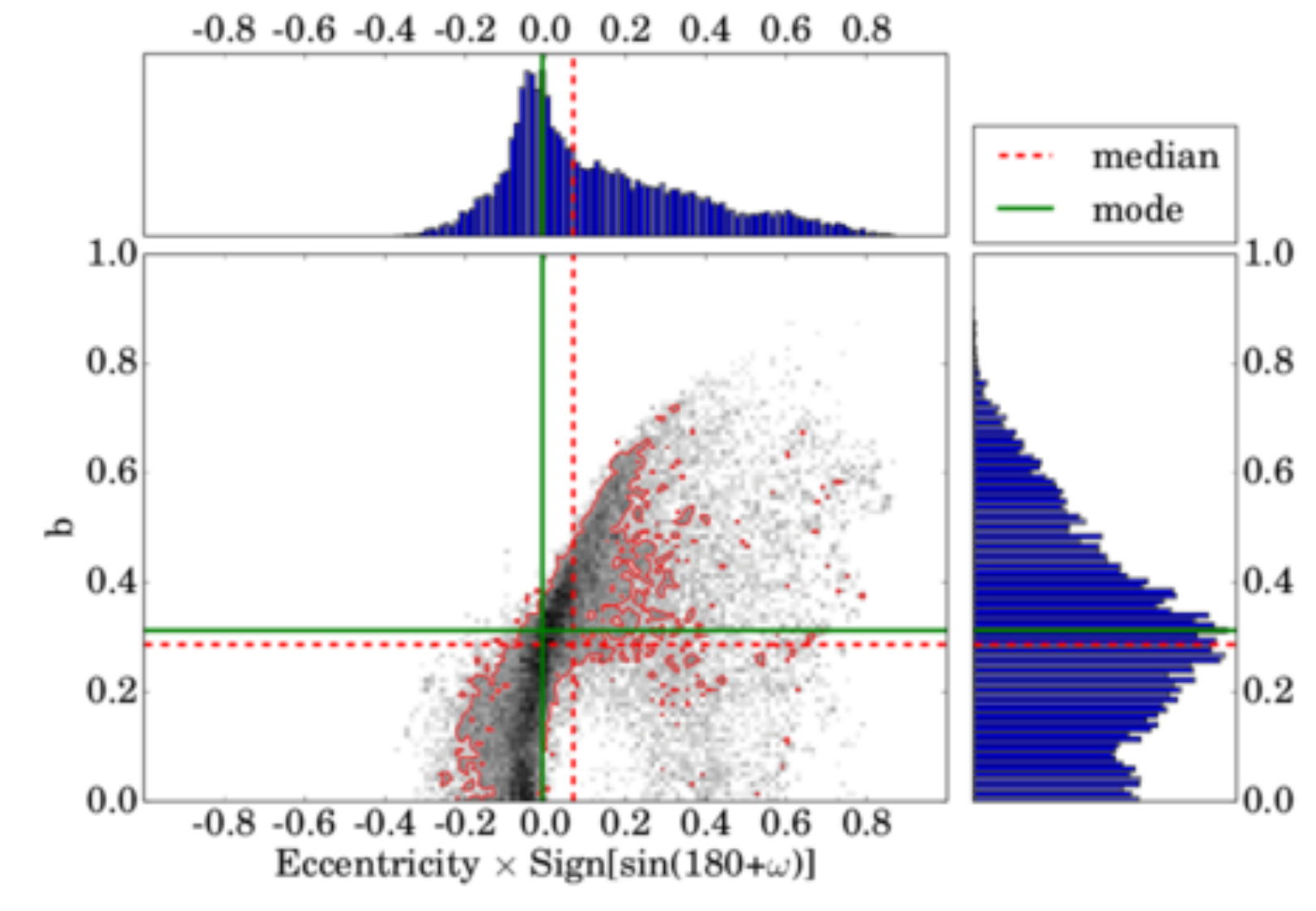}}
  
  \subfloat[Kepler-197e\label{fig:kepler197e}]{\includegraphics[width=0.33\linewidth]{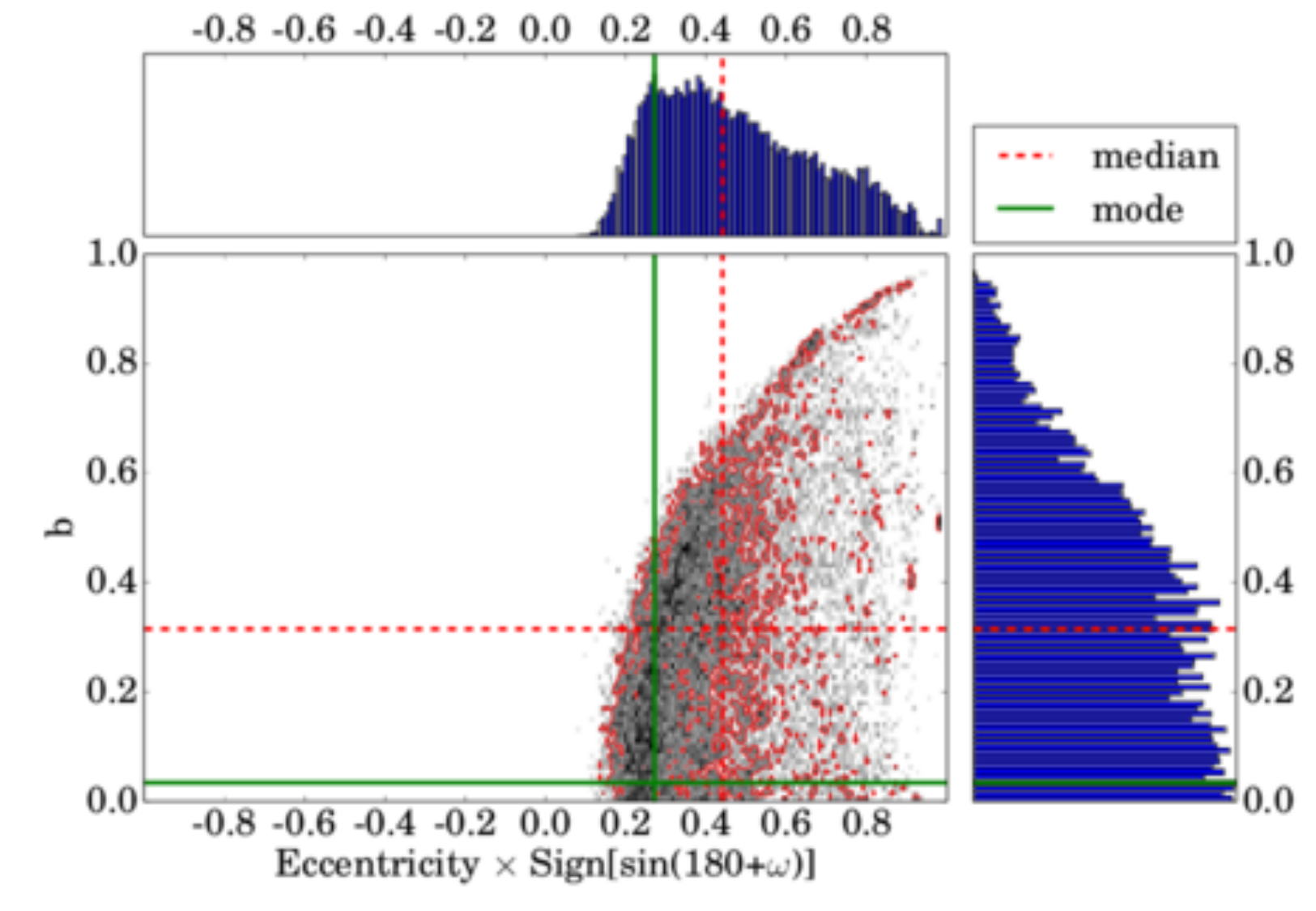}}
  \subfloat[Kepler-278b\label{fig:kepler278b}]{\includegraphics[width=0.33\linewidth]{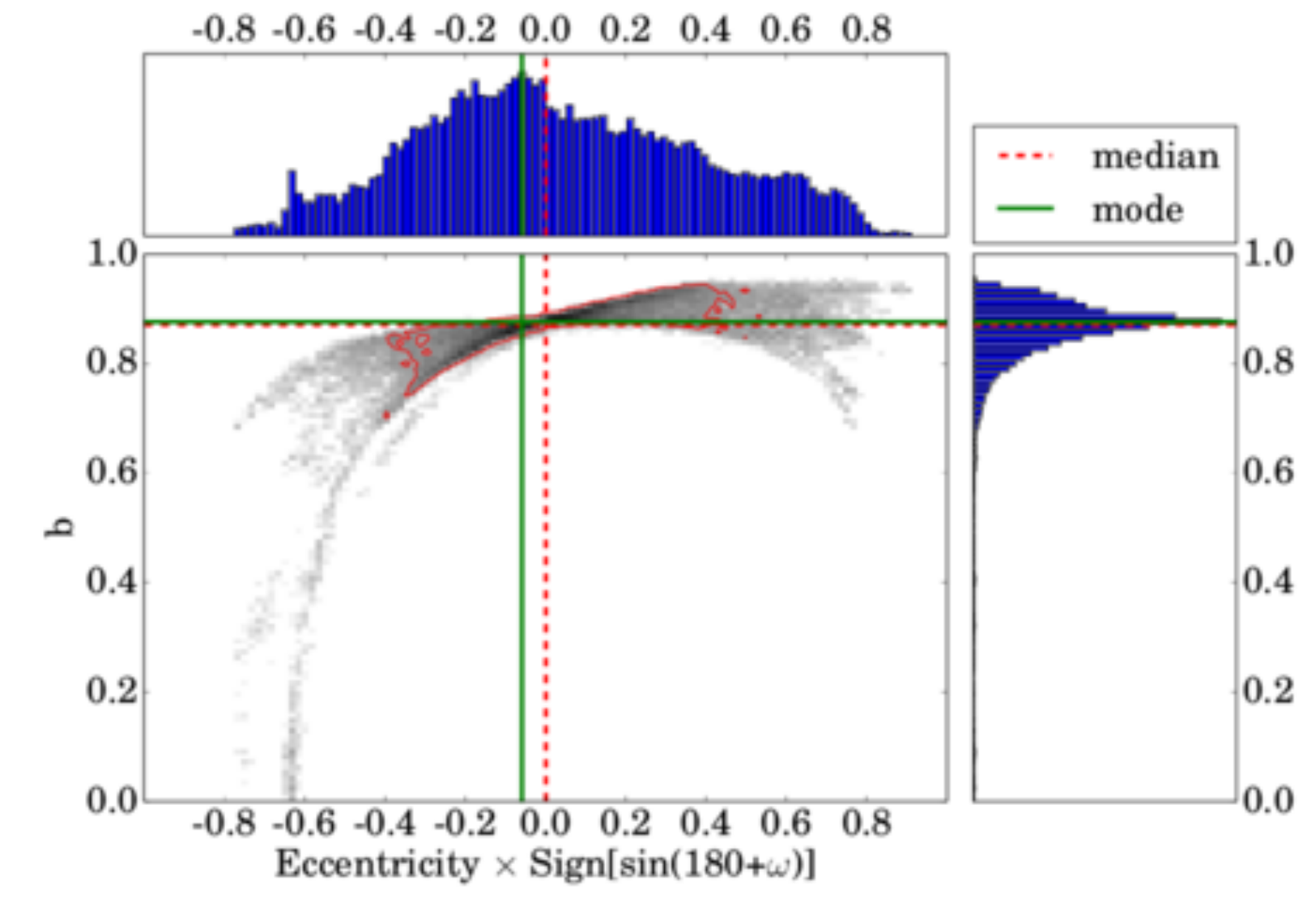}}
  \subfloat[Kepler-278c\label{fig:kepler278c}]{\includegraphics[width=0.33\linewidth]{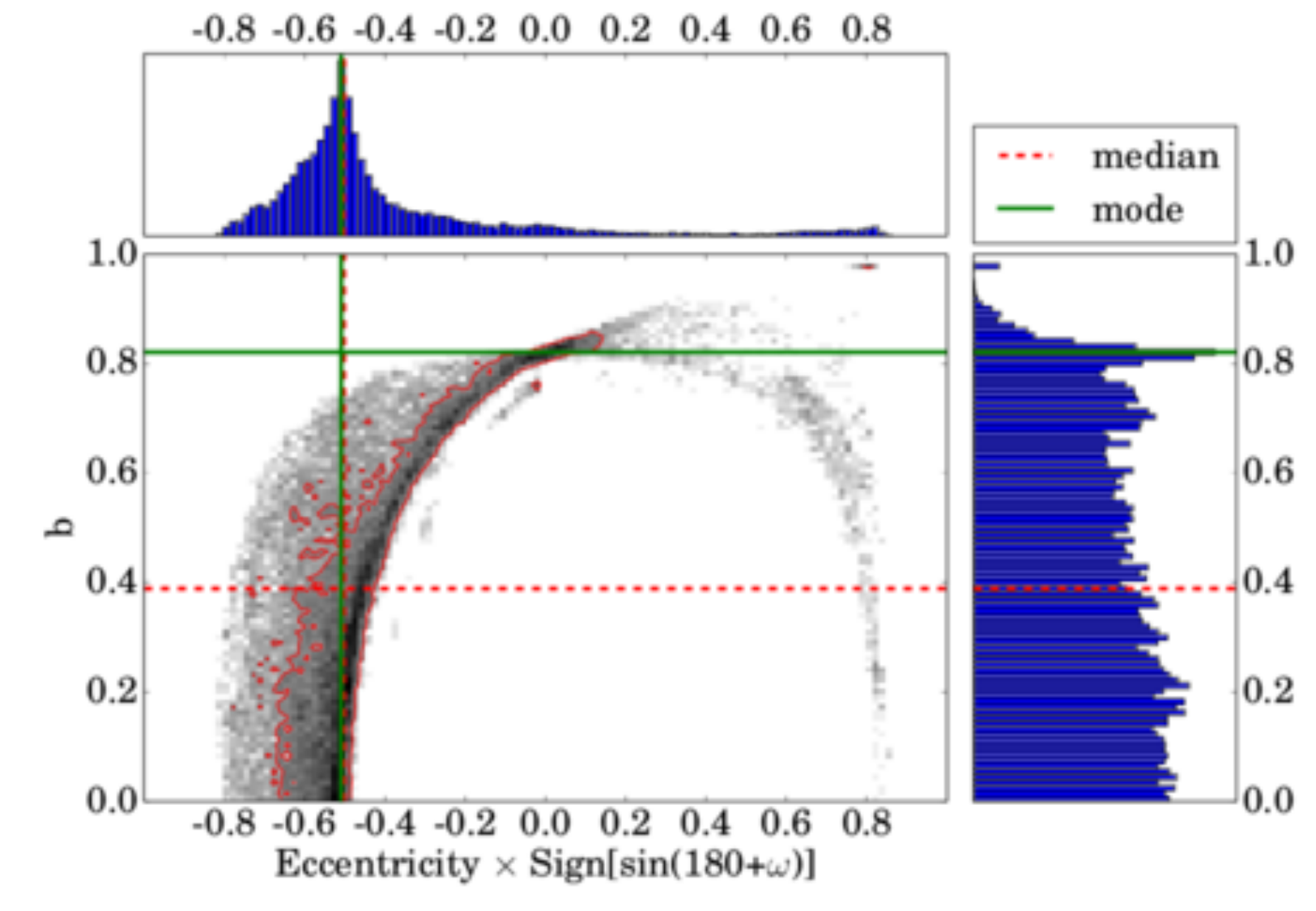}}
  
  \subfloat[Kepler-338b\label{fig:kepler338b}]{\includegraphics[width=0.33\linewidth]{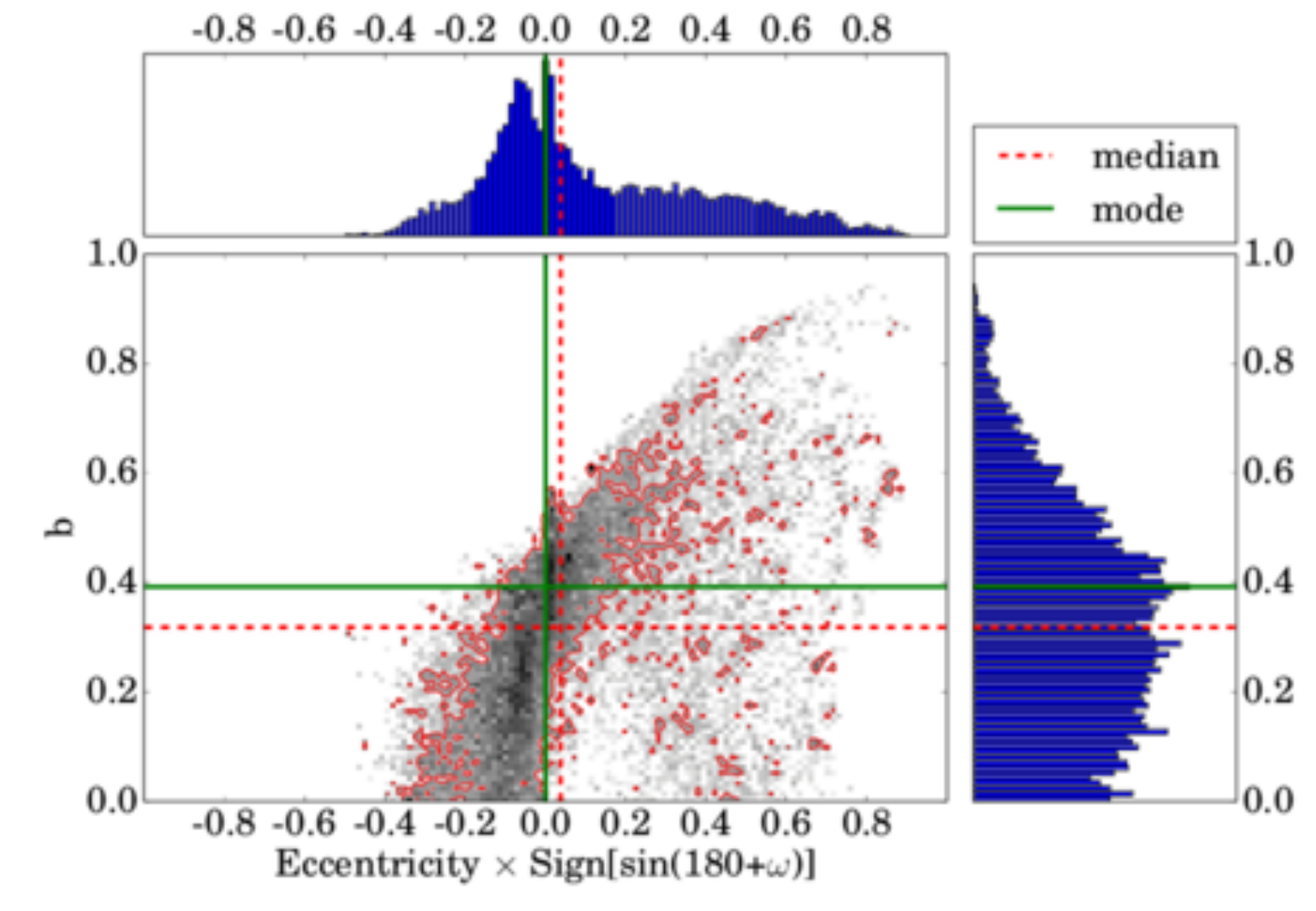}}
  \subfloat[Kepler-338c\label{fig:kepler338c}]{\includegraphics[width=0.33\linewidth]{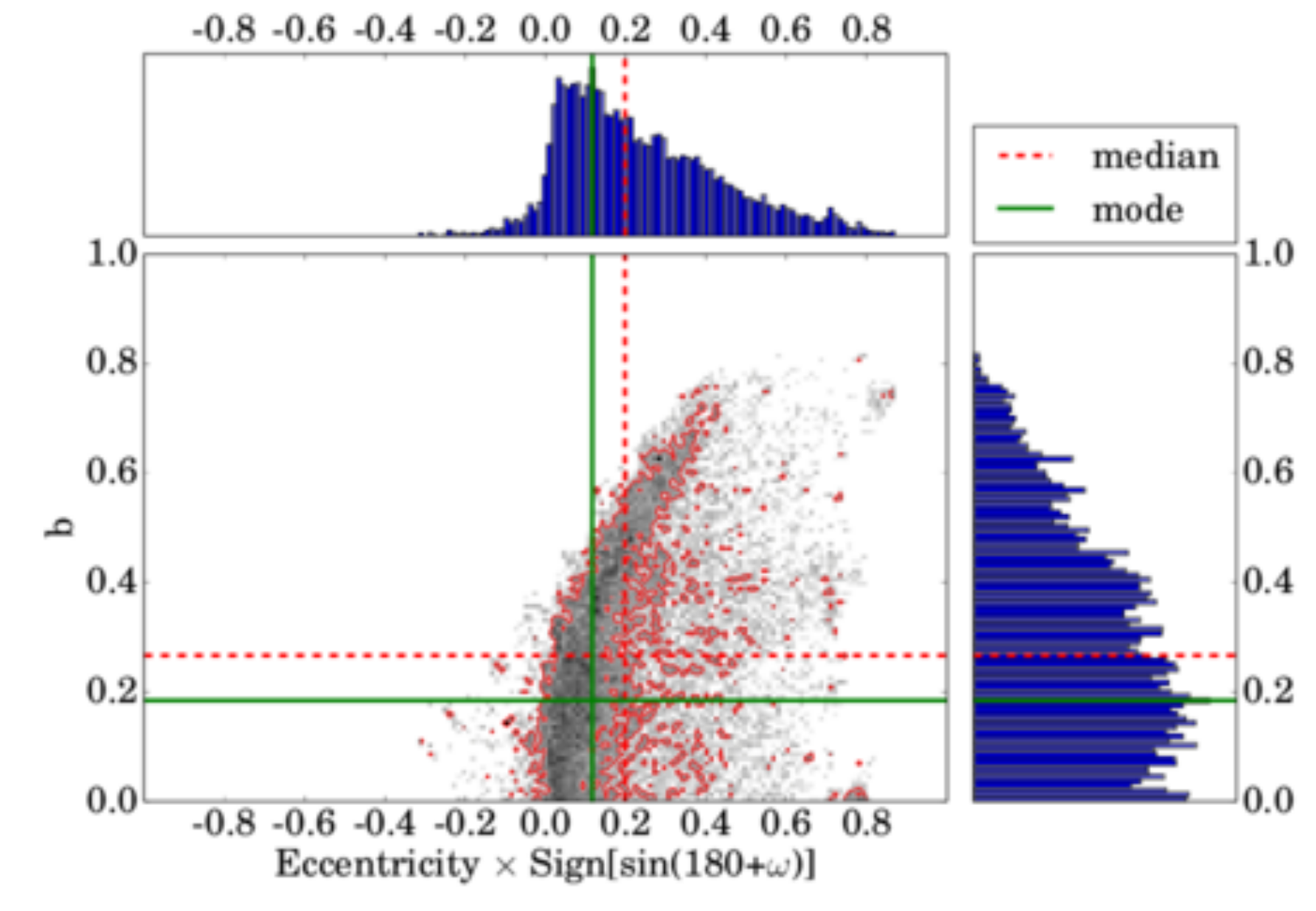}}
  \subfloat[Kepler-338d\label{fig:kepler338d}]{\includegraphics[width=0.33\linewidth]{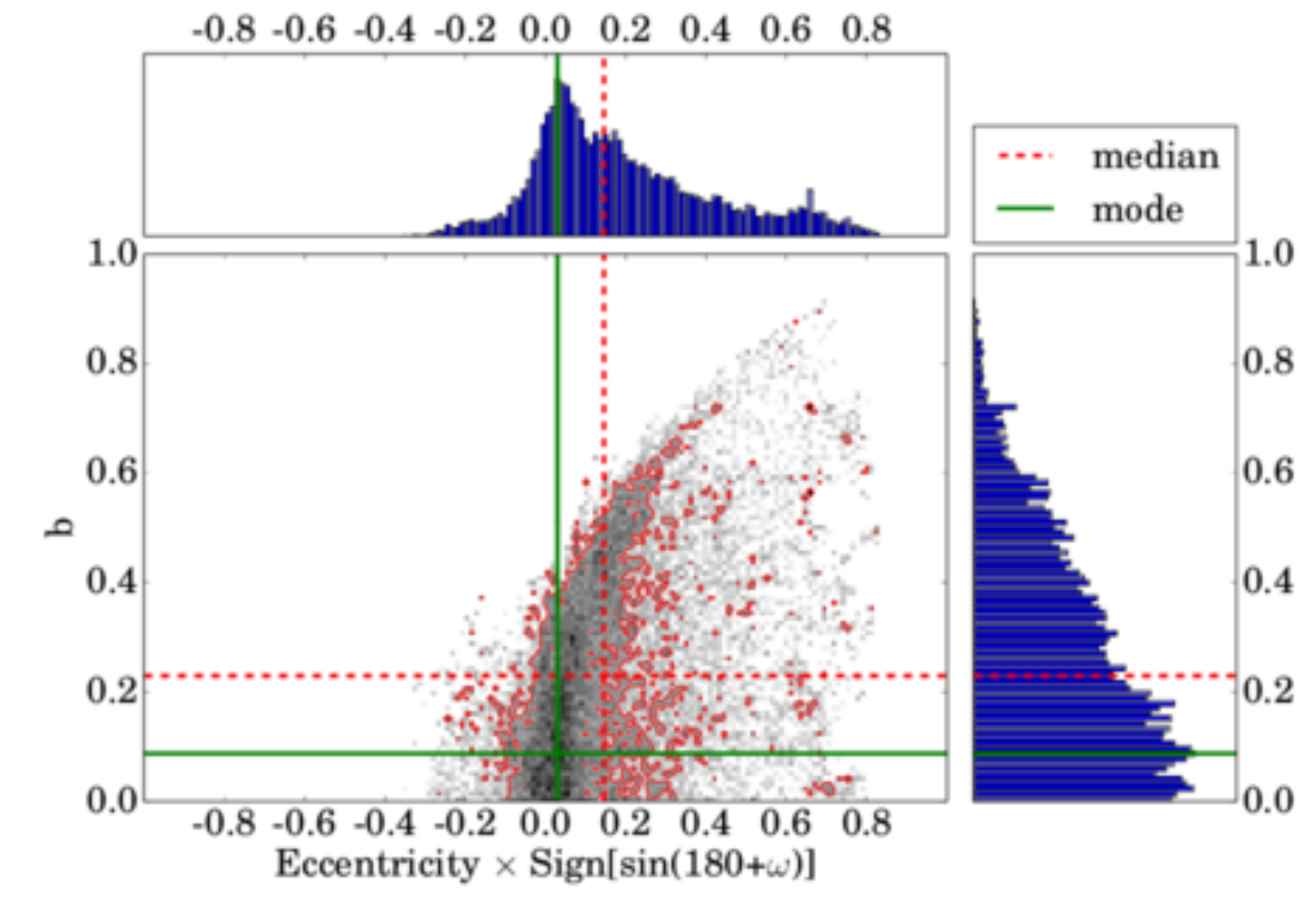}}

\end{widepage}
 \caption{(continued) Posterior distributions of individual planets}

\end{figure*}

\begin{figure*}[t]
 \ContinuedFloat
    \begin{widepage}
    \subfloat[Kepler-338e\label{fig:kepler338e}]{\includegraphics[width=0.33\linewidth]{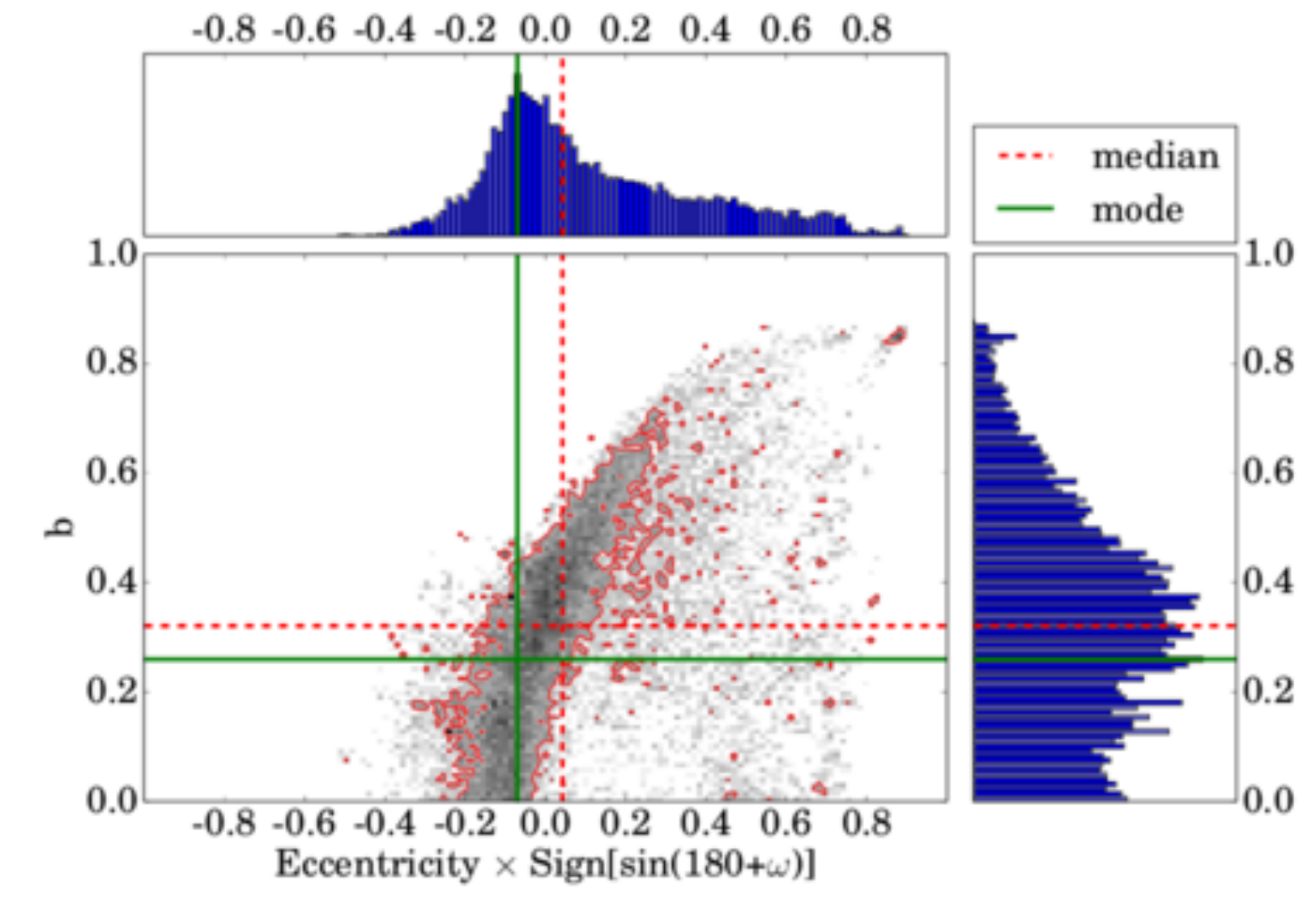}}
  \subfloat[Kepler-444b\label{fig:kepler444b}]{\includegraphics[width=0.33\linewidth]{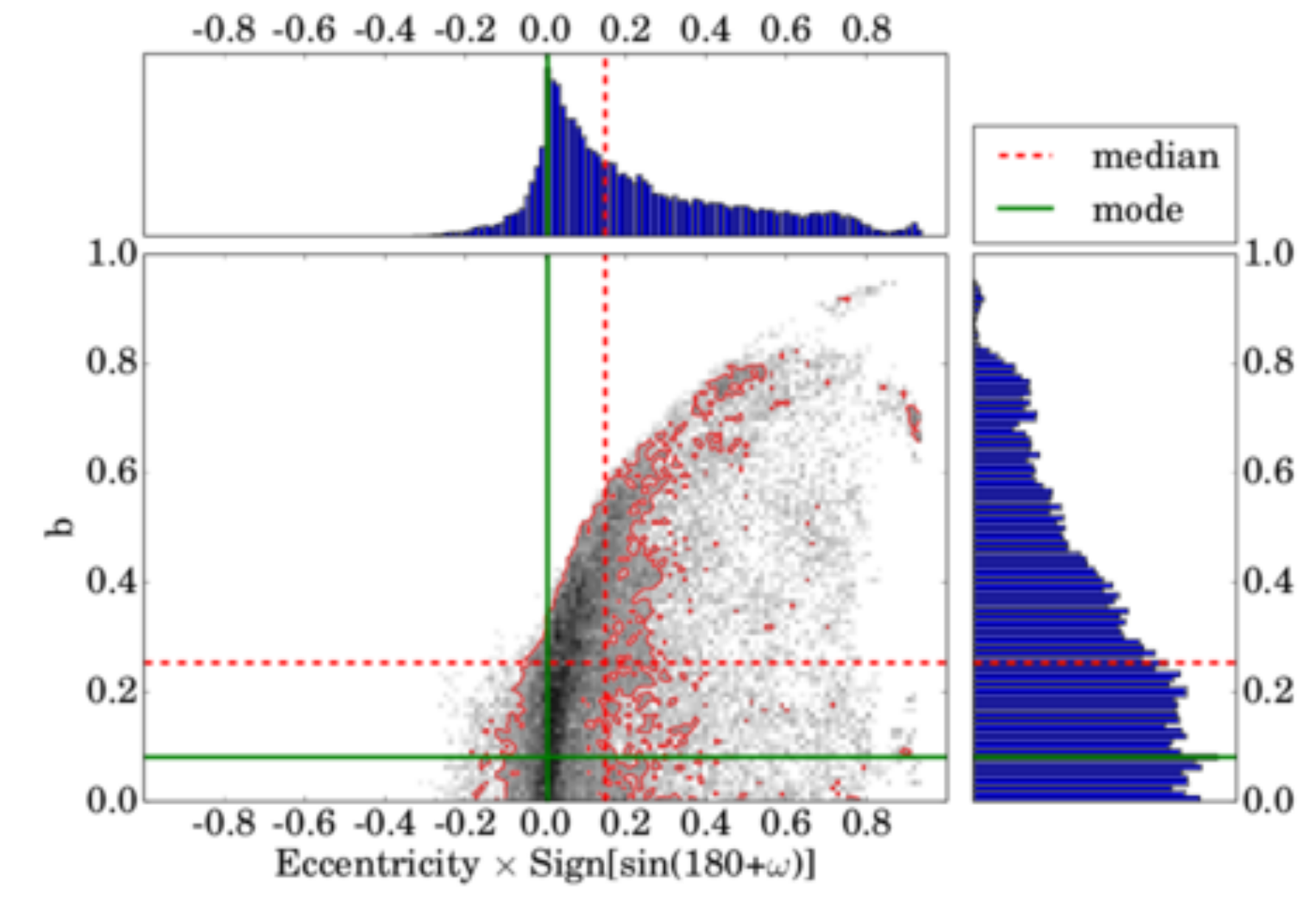}}
  \subfloat[Kepler-444c\label{fig:kepler444c}]{\includegraphics[width=0.33\linewidth]{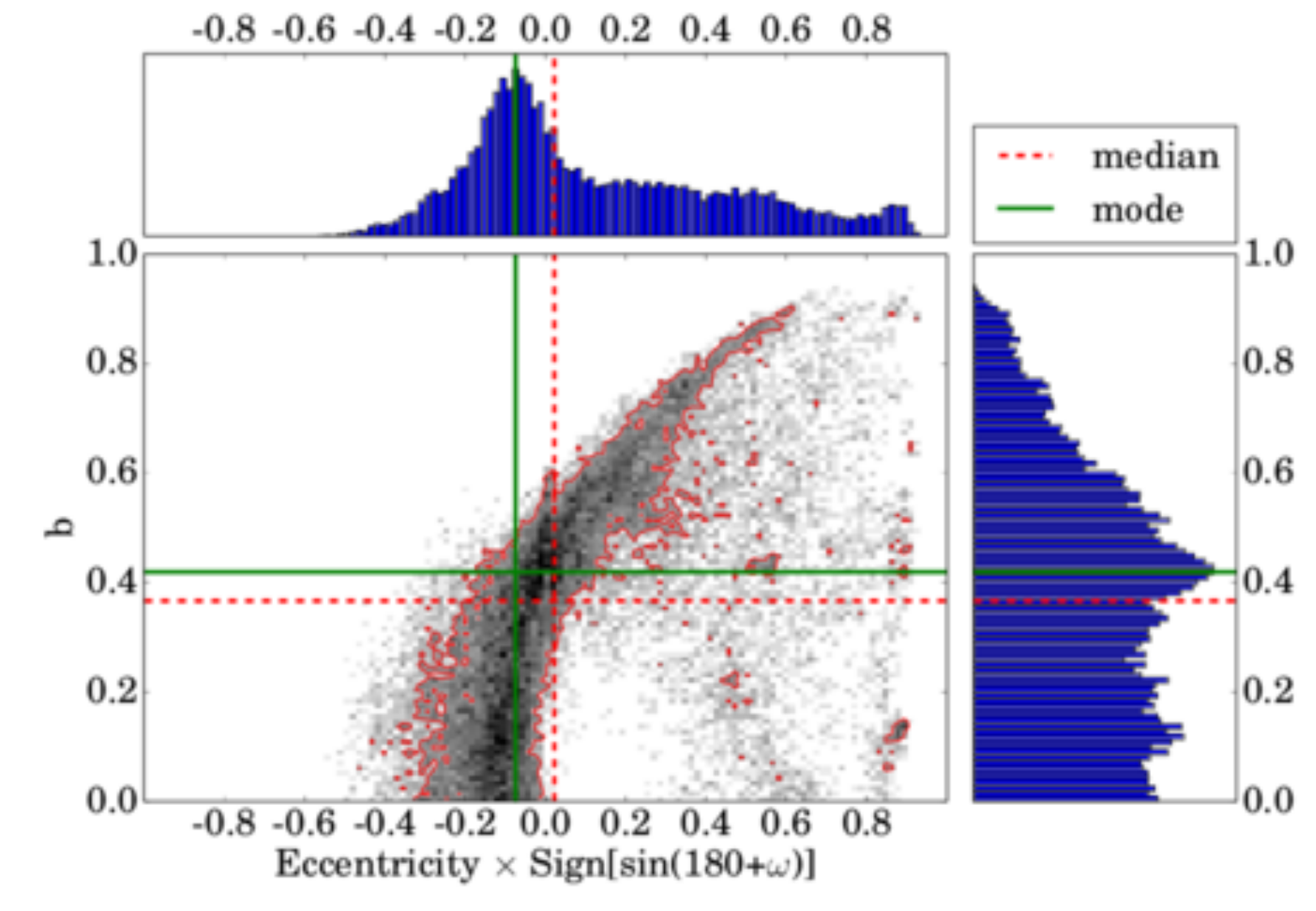}}
  
  \subfloat[Kepler-444d\label{fig:kepler444d}]{\includegraphics[width=0.33\linewidth]{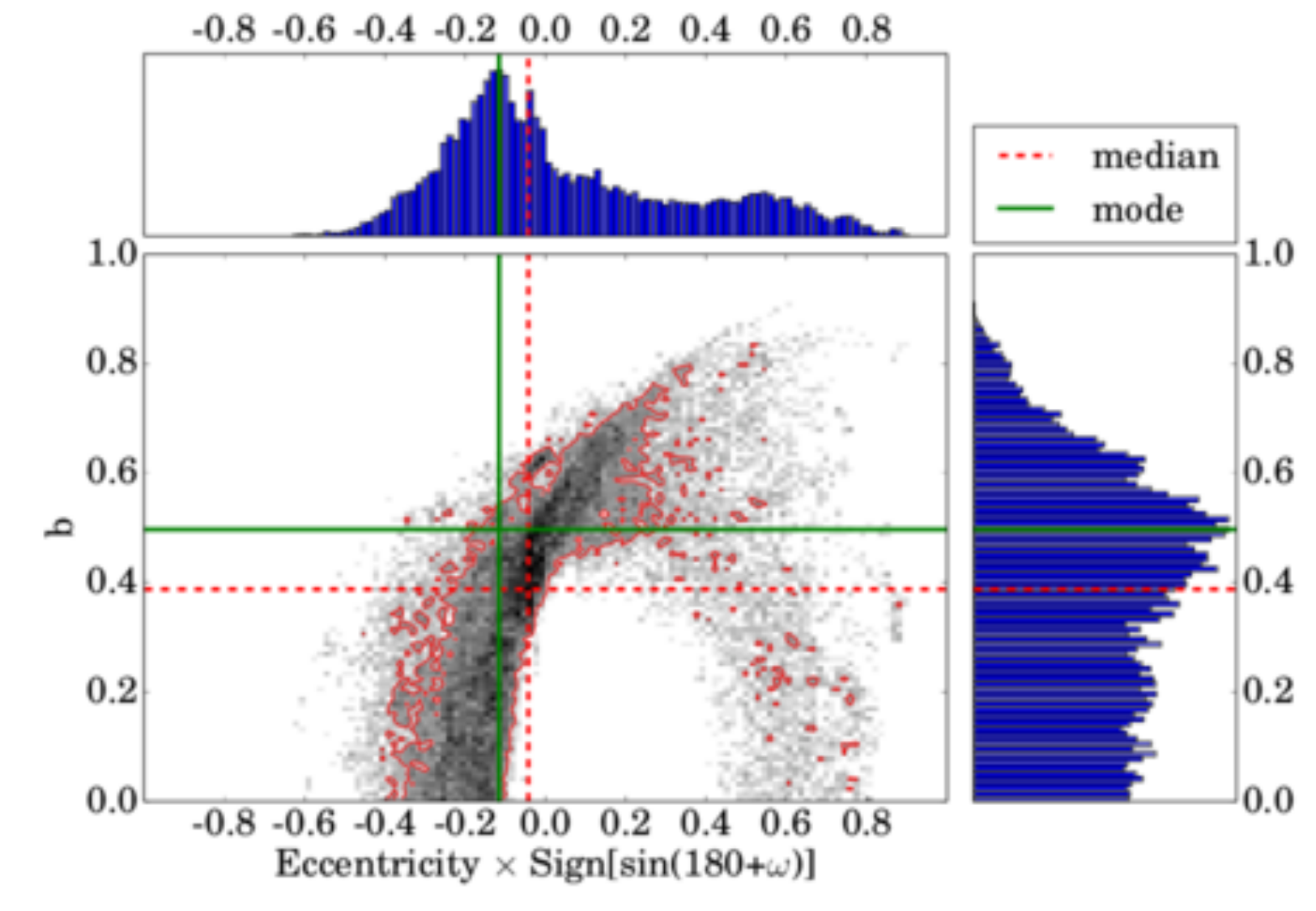}}
  \subfloat[Kepler-444e\label{fig:kepler444e}]{\includegraphics[width=0.33\linewidth]{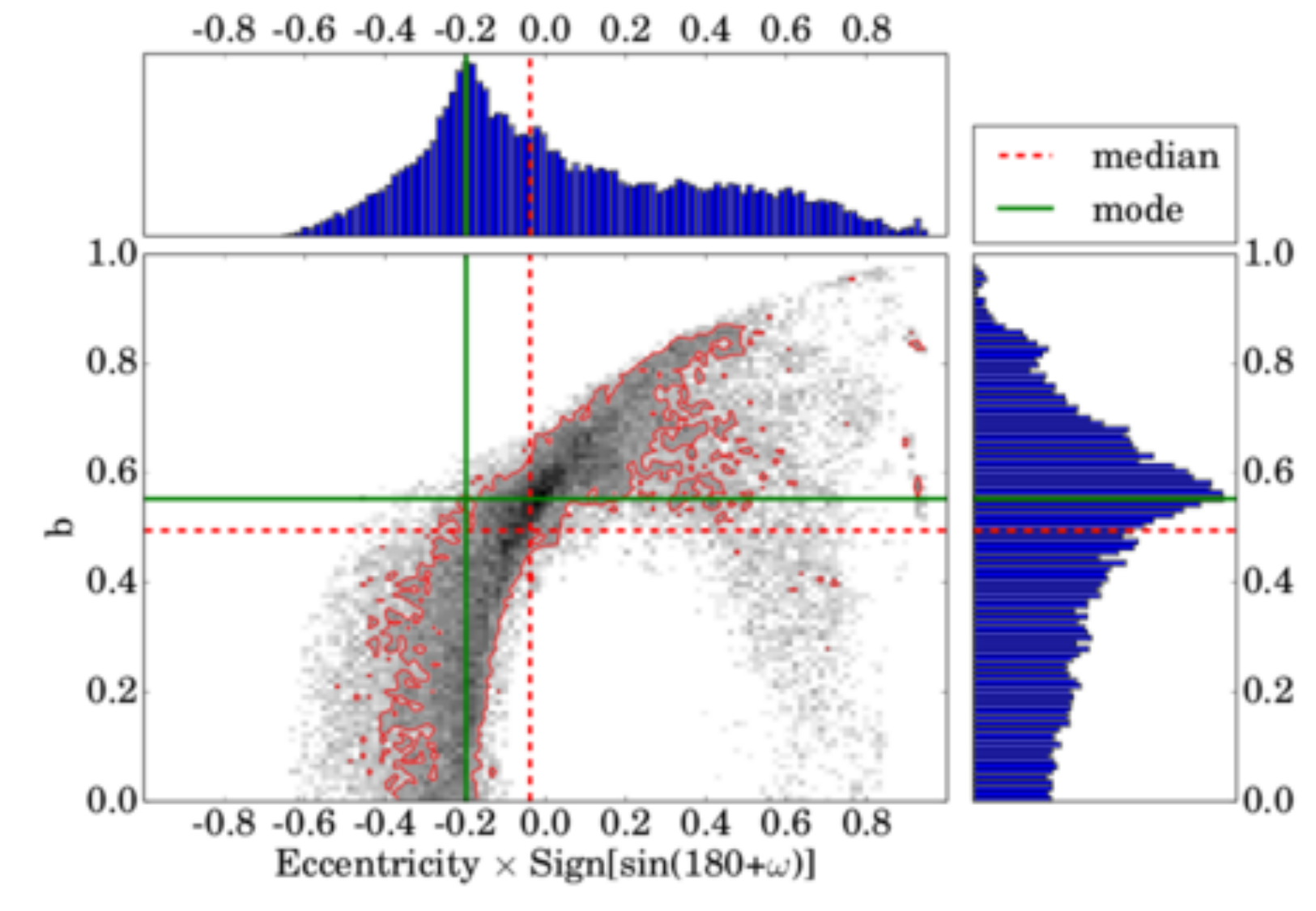}}
  \subfloat[Kepler-444f\label{fig:kepler444f}]{\includegraphics[width=0.33\linewidth]{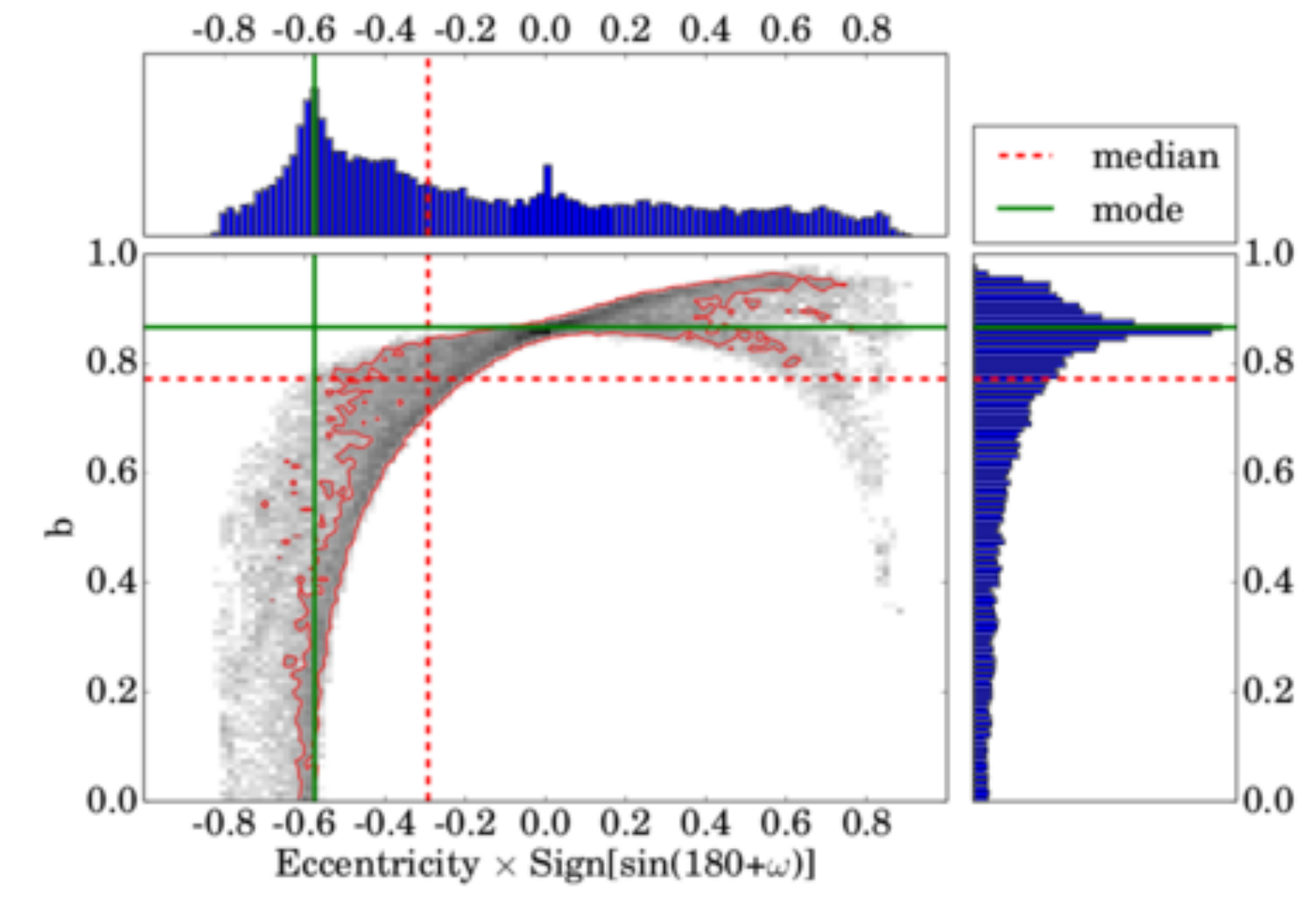}}  
\end{widepage}
 \caption{(continued) Posterior distributions of individual planets}
\end{figure*}

\end{document}